\documentclass[%
reprint,
superscriptaddress,
twocolumn,
amsmath,amssymb,
aps,
pra,
]{revtex4-2}

\usepackage{dcolumn}
\usepackage{bm}
\usepackage{mathtools} 
\usepackage{color}  
\usepackage{braket}
\usepackage{tikz}
\usetikzlibrary{quantikz2}
\usepackage{float}
\usepackage{multirow}
\usepackage{siunitx}
\usetikzlibrary{snakes}
\usepackage{amsmath}
\usepackage{mhchem}
\usepackage{color,soul}
\usepackage[OT1, T1]{fontenc}
\DeclareTextSymbolDefault{\dh}{OT1}
\usepackage{graphicx}
\usepackage[caption=false]{subfig}
\usepackage[export]{adjustbox}
\usepackage{booktabs}

\usepackage{bm}
\usepackage{bbold}

\definecolor{mscolor}{rgb}{0,0.5,0.5}

\definecolor{tgcolor}{rgb}{0.5,0,0.5}

{}
\definecolor{phcolor}{rgb}{0.5,0,0.5}

\newcommand \be{\begin{equation}}
\newcommand \ee{\end{equation}}
\newcommand \bea{\begin{eqnarray}}
\newcommand \eea{\end{eqnarray}}
\newcommand \bse{\begin{subequations}}
\newcommand \ese{\end{subequations}}

\newcommand{\rsub}[1]{\textcolor{black}{#1}}

\newcommand{\UWM}{Department of Physics, University of Wisconsin-Madison, 1150 University Avenue, Madison, WI, 53706 USA}
\newcommand{\NBI}{Niels Bohr Institute, University of Copenhagen, Blegdamsvej 17, 2100 Copenhagen, Denmark}

\def\afterfi#1#2\fi{\fi#1}

\def\periodfl#1{\pflA#1.}
\def\pflA#1#2{\dot#1\ifx.#2\else\afterfi{\pflB#2}\fi}
\def\pflB#1#2{\ifx.#2\dot#1\else#1\afterfi{\pflB#2}\fi}

\def\padotsA#1#2{\dot#1\ifx.#2\else\afterfi{\padotsA#2}\fi}

\def\pmiA#1#2{\ifx.#2\dot#1\else\afterfi{\pmiB#1#2}\fi}
\def\pmiB#1.{\overline{#1}}

\begin{document}

\title{Deterministic carving of quantum states with Grover's algorithm}

\author{Omar Nagib}%
 \email{onagib@wisc.edu}
\affiliation{\UWM}
\author{M. Saffman}%
\affiliation{\UWM}
\author{K. M{\o}lmer}
\affiliation{\NBI}

\date{\today}

\begin{abstract}
We show that iteration of a few ( $\sim N^{1/4}$) unitary steps of Grover's algorithm suffices to perfectly prepare a Dicke state of $N$ atoms in a cavity. We also show that a few subsequent Grover steps can be employed to generate GHZ and Cat states. The Grover iteration is physically realized by global qubit rotations and by the phase shift of single photons  reflected on the cavity. 
Our protocols are deterministic and require no individual addressing of the atoms.  A detailed error analysis accounting for spatial mode matching of the photon to the cavity, spontaneous emission, mirror scattering, and the finite bandwidth of the photon mode is used to predict the fidelity of the prepared states as a function of system parameters and atom-cavity cooperativity. The fidelity can be increased by heralding on detection of the reflected photon. 

\end{abstract}

\maketitle


\section{Introduction} 

Dicke states and their superpositions form an important class of entangled states, with applications in quantum computing, metrology, error correction, and networking \cite{teleclone2012,spin_cat2024,GHZ_clock_2024,Heisenberg_clock_2014,permutation_code_2014,Dicke_network_2009,Dicke_partition_2021,Anonymous_sensing_2022,HL_Dicke_2024,teleclone_1999,Noise_Dicke_2020}. Considering direct physical implementation, various classes of entangled states can be prepared by dissipation \cite{W_Dissipate_2017,Kastoryano2011dissipative,Lin2013Dissipative}, phase estimation \cite{Dicke_phase_estimate_2021}, spin-squeezing \cite{Spin_squeeze_2024}, and pulse sequences with optimal control \cite{optimal_control_2024,Jandura2024, Rydberg_optimal_control_2016, Ion_optimal_control_2013}. Due to their importance, various quantum algorithms have been devised for the preparation of entangled states with a corresponding implementation on quantum circuits. Quantum circuits made from single- and two-qubit gates for deterministic preparation of the Dicke states have been proposed \cite{Bartschi2019Deterministic,Divide_Dicke_2022,Cruz_Dicke_2019,Kaye_Dicke_2001}. For $N$ qubits, both the circuit depth and the gate count grow linearly $O(N)$ with the number of qubits \cite{Bartschi2019Deterministic}, which can be improved to $O(m \log \frac{N}{m})$ for quantum circuits with all-to-all connectivity, where $m$ denotes the number of excitations of the $N$-qubit Dicke state \cite{Bartschi_2022}. When supplemented with ancillas, midcircuit measurements, and feedforward operations, Dicke states can be prepared with an efficient quantum circuit with depth $O(m^{1/4})$ (up to a polylogarithmic correction), using Grover's amplitude amplification \cite{state_prep_shallow_circuits_2024, Cirac_Dicke_2024}. Recently, there have been physical proposals based on Grover's amplitude amplification \cite{Grover1997} in atom-cavity systems \cite{Grover_geometric_2020,Grover_2023}, where the procedure consists of repeated application of a geometric nonlinear phase gate interleaved with global rotations. 

An alternative approach is through probabilistic preparation by projective measurements, where a quantum circuit that counts the number of excitations projects an initial product state into a Dicke state \cite{Childs_2002,Cirac_Dicke_2024,Dicke_phase_estimate_2021}. Physical implementation of this approach, referred to as ``cavity carving'' in atom-cavity systems, has been theoretically proposed \cite{Sorensen2003probabilistic, Chen2015carving, Sorensen2003measurement, ramette2024counterfactual, Davis2018Painting}, and it has been experimentally implemented to successfully produce Bell states and the W state \cite{Welte2017cavity, McConnell2015Entanglement, McConnell2013Generating, Christensen2014Quantum,Dordevic2021Entanglement, Florian2014Entangled}. The success probability of such an approach is equal to the square of the overlap between the initial state and the target state, which could be small for many qubits, necessitating repeating the sequence many times. More precisely, the measurement needs to be repeated $O(m^{1/2})$ times to prepare the Dicke state $m$ \cite{Cirac_Dicke_2024}. This can be improved when coupled with repeated collective measurements and feedforward, where the Dicke state $m=N/2$ can be prepared in $O(\log N)$ trial steps \cite{yu2024efficientpreparationdickestates}. 
  
A challenge to scalability in many physical implementations is that the resources for state preparation typically grow (super) linearly with $N$ \cite{state_prep_shallow_circuits_2024,Ion_optimal_control_2013,Rydberg_optimal_control_2016, Grover_geometric_2020, Grover_2023}. For example, with Rydberg atoms controlled by phase-modulated pulses, at least $O(N)$ phase steps are needed for the state preparation \cite{Rydberg_optimal_control_2016}. In atom-cavity implementations relying on geometric phase gates and amplitude amplification, $O(N^{5/4})$ or $O(N)$ gate sequences as well as $O(N)$ optimization parameters are needed, respectively \cite{Grover_geometric_2020,Grover_2023}. Furthermore, the theoretical fidelity is not guaranteed to be unity, even in the absence of errors and after optimizing the $O(N)$ parameters. This motivates looking for an efficient scheme for physical preparation of the Dicke states, where the resources per step do not scale with the number of qubits. 

\begin{table*}
  \centering
\rsub{
  \begin{ruledtabular}
  \begin{tabular}{l c c c c c c}
    Ref.\ 
    & Depth/Physical resource 
    & Infidelity
    & Ancillas 
    & Interconnectivity 
    & Addressing
    & \parbox[c]{2cm}{\centering Midcircuit\\measurement} \\ \\
    \colrule
    \multicolumn{7}{l}{\bfseries Algorithm}                                                   \\
    a) Dicke $m$:                                                                               \\
    \quad Ref. \cite{Cirac_Dicke_2024}       
      & $O(m^{1/4}l^2_{m,\epsilon})$    
      & $\epsilon$   
      & $O(N + l_{m,\epsilon})$      
      & Nearest‑neighbor      
      & Individual      
      & Yes      \\
    \quad Ref. \cite{yu2024efficientpreparationdickestates}      
      & $O(\ln m + \ln 1/\epsilon)$    
      & $\epsilon$       
      & $O(\ln N)$     
      & All‑to‑all      
      & Global     
      & Yes    \\
    \quad Present     
      & $O(m^{1/4})$*      
      & 0     
      & 0   
      & All‑to‑all   
      & Global      
      & No    \\
    b) GHZ $N$ qubits:                                                                                 \\
    \quad Ref. \cite{QC_LOCC}       
      &  $O(1)$
      &  0
      &  $O(N)$
      &  Nearest‑neighbor
      &  Individual
      &   Yes   \\
    \quad Ref. \cite{Cruz_Dicke_2019}       
      &  $O(\ln N)$
      &  0
      &  0
      &  All-to-all
      &  Individual
      &   No   \\
    \quad Present     
      &  $O(N^{1/4})$*
      &  0
      &  0
      &  All-to-all
      &  Global
      &  No    \\
    \midrule
    \multicolumn{7}{l}{\bfseries Implementations}                              \\
    Dicke $m$:\\
    Cavity carving \cite{Chen2015carving}    
      &  $O(m^{1/2})$ photons transmission**
      &  $O(C^{-1})$
      &  0
      &  All-to-cavity
      &  Global
      &  No    \\
    Grover-based \cite{Grover_geometric_2020} 
      &  $O(N^{5/4})$ phase gates
      &  $O(1-e^{-\pi^2 \kappa /g})$
      &  0
      &  All-to-cavity
      &  Global
      &   No   \\
    Grover-based \cite{Grover_2023}
      &  $O(N)$ phase gates
      &  NA
      &  0
      &  All-to-cavity
      &  Global
      &   No  \\
      Ref. \cite{yu2024efficientpreparationdickestates}    
      &  $O(\ln m)$ photons transmissions***
      &  NA
      &  0
      &  All-to-cavity
      &  Global
      &  Yes    \\
    Present (unheralded)           
      &  $O(m^{1/4})$ photons reflections
      &  $O(C^{-1/2})$
      &  0
      &  All-to-cavity
      &  Global
      &  No    \\
       Present (heralded)           
      &  $O(m^{1/4})$ photons reflections
      &  $O(C^{-2/3})$
      &  0
      &  All-to-cavity
      &  Global
      &  No    \\
  \end{tabular}
  \end{ruledtabular}
  }
  \rsub{\caption{\label{tab:comparison}
    Comparison of various algorithms and physical implementations, including present work, for preparation of the Dicke state with $m$ excitations and $N$ qubits and the $N$‐qubit GHZ state. $\ell_{m,\epsilon}
= \log_{2}\{(1/\ln(4/3))\,[2m(\ln(2m)+9/2)
  + \ln(\mathrm{Poly}(m)/\epsilon^{2})]\}\,$. $C=g^2/\kappa \gamma$ is the atom-cavity cooperativity, where $g$ is the atom-cavity coupling, $\kappa$ is the cavity decay rate, and $\gamma$ is the atom decay rate. *This is under the assumption that one Grover step can be implemented in a constant depth, as we show in the present paper. **The cavity carving scheme, which has success probability $O(m^{-1/2})$, can be converted to a deterministic protocol by repeating the scheme $O(m^{1/2})$ times. ***Each photon is a multi-chromatic photon with $O(N)$ frequency components.
}
  }
\end{table*}

Here, we show that the phase shift applied in the carving scheme in cavity QED can be employed in Grover's algorithm to deterministically prepare a Dicke state in $O(m^{1/4})$ Grover steps, with a reduction in resources compared to previous physical implementations \cite{Grover_geometric_2020, Grover_2023}. In particular, the resource for one Grover step is two photon reflections and global rotations, independent of the number of qubits. We also show that a few subsequent Grover steps $O(N^{1/4})$ can prepare the GHZ and Cat states. The scheme does not require individual addressing and is realizable in other physical systems, e.g., an ensemble of Rydberg atoms, superconducting qubits, or trapped ions. Because of this constant-depth physical implementation of the Grover iteration, we obtain an efficient scaling similar to the one obtained recently with quantum circuits assisted with midcircuit measurement and feedforward \cite{Cirac_Dicke_2024}, but without the need for individual addressing, ancillas, or measurements.

The present work consists of two parts: the first part (Sec. \ref{sec:Grover} to \ref{sec:Grover_Cat}) is strictly on algorithms and the second part (Sec. \ref{sec:physical_implement}) is on the physical implementation. In Sec. \ref{sec:Grover}, we show how Grover's algorithm can be used for perfect state preparation in a minimal number of steps. In Sec. \ref{sec:Dicke_grover}, a Grover's algorithm is presented to prepare a Dicke state with $m$ excitations perfectly in $O(m^{1/4})$ steps, starting from a product state of all the qubits. Unlike previous proposals requiring $O(N)$ optimized angles \cite{Grover_geometric_2020,Grover_2023}, only a single global rotation angle is needed, interleaved with conditional phase inversions, to prepare the state perfectly. A simple equation to easily and numerically find that angle is also given. In Sec. \ref{sec:Grover_GHZ},
a two-step Grover's algorithm to prepare the GHZ state perfectly in $O(N^{1/4})$ steps is presented, starting from a Dicke state and using global rotations. Sec. \ref{sec:Grover_Cat} presents a two-step algorithm for preparation of Cat states. 

Section \ref{sec:physical_implement} describes a physical realization of Grover's algorithm in cavity QED. Dicke states are prepared in atom-cavity systems via single-photon reflections, where the Grover iteration consists of photon-induced phase shifts and global qubit rotations, with no individual addressing required. The scheme requires the reflection of two photons per Grover step for the Dicke states and three photons for the GHZ and Cat states, independent of the number of qubits. Dominant sources of error in the scheme, including spatial mode mismatch, nonideal cavity effects, and photon wavepacket decoherence are analyzed analytically and numerically. The scaling of the errors in state preparation with the cavity parameters is also derived. It is shown that the infidelity scales as $O(C^{-1/2})$ and is improved to $O(C^{-2/3})$ if we herald upon detection of the reflected photon, where $C$ is the cavity cooperativity.  

\rsub{A detailed comparison between the present and the previous work on both the algorithm and the implementation sides can be found in Table \ref{tab:comparison}, and Appendix \ref{appendix:cavity_carve} gives an overview of previous cavity carving schemes. Sec. \ref{sec:discuss} concludes with a discussion.}

For a short high-level summary of the present work, the reader is referred to the companion letter to this paper \cite{Nagib2025a}.

\section{Grover's algorithm}\label{sec:Grover}

\subsection{Introduction}

Grover's algorithm can be used as a state-preparation protocol to prepare a target state, starting from an arbitrary initial state. The initial state $\ket{\psi_i}$ can be decomposed into the sum of two orthogonal states: the target state we wish to prepare $\ket{\psi_t}$ and an orthogonal complement $\ket{\psi_{t,\perp}}$:
\begin{equation}\label{psi_i}
 \ket{\psi_i}=   \sin (\theta/2) \ket{\psi_t} +  \cos (\theta/2)   \ket{\psi_{t,\perp}}
\end{equation}
where $\sin(\theta/2)= \braket{\psi_t|\psi_i}$ is the overlap amplitude between the initial state and the target. A Grover iteration consists of two unitaries applied successively. The first is phase inversion of the target state 
\rsub{\begin{equation}\label{chi_t}
\chi_{t}= \mathbb{1}-2\ket{\psi_t}\bra{\psi_t}
\end{equation}}
i.e., $\chi_{t}\ket{\psi_t}=-\ket{\psi_t}$, and $\chi_{t}$ acts as the identity on all the states orthogonal to the target. The second operation is phase inversion of  $\ket{\psi_i}$\rsub{
\begin{equation}\label{chi_i}
\chi_{i}= \mathbb{1}-2\ket{\psi_i}\bra{\psi_i}
\end{equation}}
which puts a minus sign on the initial state and acts as the identity otherwise. The Grover iteration is then
\begin{equation}
G=\chi_{i}\chi_{t}
\end{equation}
After applying the Grover iteration $k$ times on $\ket{\psi_i}$, we get (up to a global phase):
\begin{equation}\label{Grover_k_steps}
G^k \ket{\psi_i}=  \sin \bigg([2k+1]\frac{\theta}{2}\bigg) \ket{\psi_t} +  \cos\bigg([2k+1]\frac{\theta}{2}\bigg)  \ket{\psi_{t,\perp}}
\end{equation}
The fidelity in preparing the target, $F_{\rm Grover}=|\bra{\psi_t}G^k \ket{\psi_i}|^2$, after $k$ iterations is
\begin{equation}\label{fidelity_k}
F_{\rm Grover}=\sin^2\bigg([2k+1]\frac{\theta}{2}\bigg).
\end{equation}
The target is reached  with high fidelity when $\sin ([2k+1]\frac{\theta}{2}) \approx 1$, i.e., $(2k+1)\frac{\theta}{2} \approx \pi/2$. In other words, the required number of steps $k$ to prepare the target is
\begin{equation}\label{k}
k =\dfrac{\pi}{2\theta}-\dfrac{1}{2}.
\end{equation}
The most well-known application of the Grover iteration is the one pertaining to the equal-amplitude superposition of all classical register product states,  $\ket{\psi_i}= (\ket{0}+\ket{1})^{\otimes N}/2^{N/2} =H^{\otimes N}\ket{0...0}$, where $H^{\otimes N}$ is the global Hadamard gate on $N$ qubits. $\chi_i$ in this case is given by $\chi_{i}=H^{\otimes N} \chi_0 H^{\otimes N}$, where $\chi_0\ket{0...0}=-\ket{0...0}$ and the identity otherwise. This operation is colloquially referred to as the {\it inversion-about-the-mean},  because of its action on the state amplitudes in the basis of classical register product states. Together with the application of a gate that changes the sign of the amplitude on a specific, but unknown, target register product state $\ket{\psi_t}$,  these operations can perform a search for the unknown target state among $n$ states in $k \sim \sqrt{n}$ steps \cite{Grover1997}. Other choices of $\ket{\psi_i}$ and the accompanying phase shift operations $\chi_{i}$ may be appropriate for preparation of other desired target states $\ket{\psi_t}$, as we will see in the following. 

\subsection{Grover's algorithm with a modified phase}\label{subsec:mod_phase}

For $k$ integer steps of the Grover iteration above, the condition to prepare a target state perfectly, i.e., $\sin ([2k+1]\frac{\theta}{2}) = 1$, cannot always be satisfied for an arbitrary initial state. Fortunately, it is possible to modify the phases of the Grover iteration, such that any target state can be prepared with perfect fidelity in integer steps \cite{Roy2022Deterministic, Long2001Grover}. The modified phase inversions are given by
\rsub{\begin{equation}
\chi_{t}(\alpha)= \mathbb{1}-(1-e^{i \alpha})\ket{\psi_t}\bra{\psi_t}
\end{equation}}
\rsub{\begin{equation}
\chi_{i}(\alpha)= \mathbb{1}-(1-e^{i \alpha})\ket{\psi_i}\bra{\psi_i}
\end{equation}}
i.e., $\chi_{t}(\alpha)\ket{\psi_t}=e^{i \alpha}\ket{\psi_t}$ and the identity otherwise and similarly for $\chi_{i}(\alpha)$. The phase $\alpha$ is given by \cite{Long2001Grover}:
 \begin{equation}\label{Long_phase}
\alpha =2 \arcsin \bigg[  \dfrac{1}{\sin( \theta/2)}  \sin\left( \dfrac{\pi}{ 4k+6 }\right)  \bigg ] 
 \end{equation}
where $k$ is an integer equal to or greater than the integer part of $(\pi-\theta)/2\theta$. The modified Grover iteration $G=\chi_{i}(\alpha)\chi_{t}(\alpha)$ is then guaranteed to prepare the target state with perfect fidelity in $k+1$ integer steps. Note that the Grover iteration in the previous subsection corresponds to $\alpha=\pi$. 

\subsection{Exact state preparation in a few Grover steps}\label{sec:exact_prep}

It is possible to prepare a target state with unity fidelity in a few steps, without a special phase $\alpha$, if the initial state is carefully chosen. Starting from the initial state $\ket{\psi_i}$, the action of one Grover step [Eq. \eqref{Grover_k_steps}] is $G \ket{\psi_i} = \sin (3\theta/2) \ket{\psi_t} +  \cos(3\theta/2 ) \ket{\psi_{t,\perp}}$. Suppose that the initial state is chosen such that $\theta/2=\pi/6$, i.e., $\sin (\theta/2)=\sin(\pi/6)=1/2$. Then a single Grover step would prepare $\ket{\psi_t}$ with perfect fidelity since $\sin(3\theta/2)=\sin (\pi/2)=1$, i.e., $G \ket{\psi_i} =\ket{\psi_t}$. More generally, to prepare a target exactly in $k$ integer steps, the overlap amplitude between the target and the initial state needs to be 
\begin{equation}
\braket{ \psi_t|\psi_i}= \sin\left[  \frac{\pi}{2(2k+1)}\right].
\end{equation}
In the next sections, we will show that various multi-qubit entangled states can be rapidly prepared by appropriately rotating an initial  product state to have such an overlap with the target. 

\section{Dicke 
states}\label{sec:Dicke_grover}

The Dicke state $\ket{m}$ with $N$ qubits denotes the \rsub{permutation} symmetric superposition state with $m$ qubits in the state $\ket{1}$ (and $N-m$ in $\ket{0}$), e.g., $\ket{m=0}=\ket{0...0}$ is the state with no qubits in $\ket{1}$, and $\ket{m=1}$ is the Dicke state with only a single qubit in $\ket{1}$ (i.e., the W state), etc. Next, we present several variations of the Grover's algorithm for efficient preparation of the Dicke states. We note that Grover's algorithm for Dicke state preparation has been proposed previously in different settings, e.g., using quantum circuits assisted with ancillas, measurements and feedback \cite{Cirac_Dicke_2024}, and using geometric nonlinear phase gates \cite{Grover_geometric_2020}. Here, we present a systematic and general procedure to prepare any Dicke state perfectly in integer steps, and we extend it to the GHZ and Cat states in the next two sections.

\subsection{Normal Grover's algorithm for Dicke states}\label{subsec:normal_grover}

\begin{figure}[!t]
   \centering
    \includegraphics[width=0.45\textwidth]{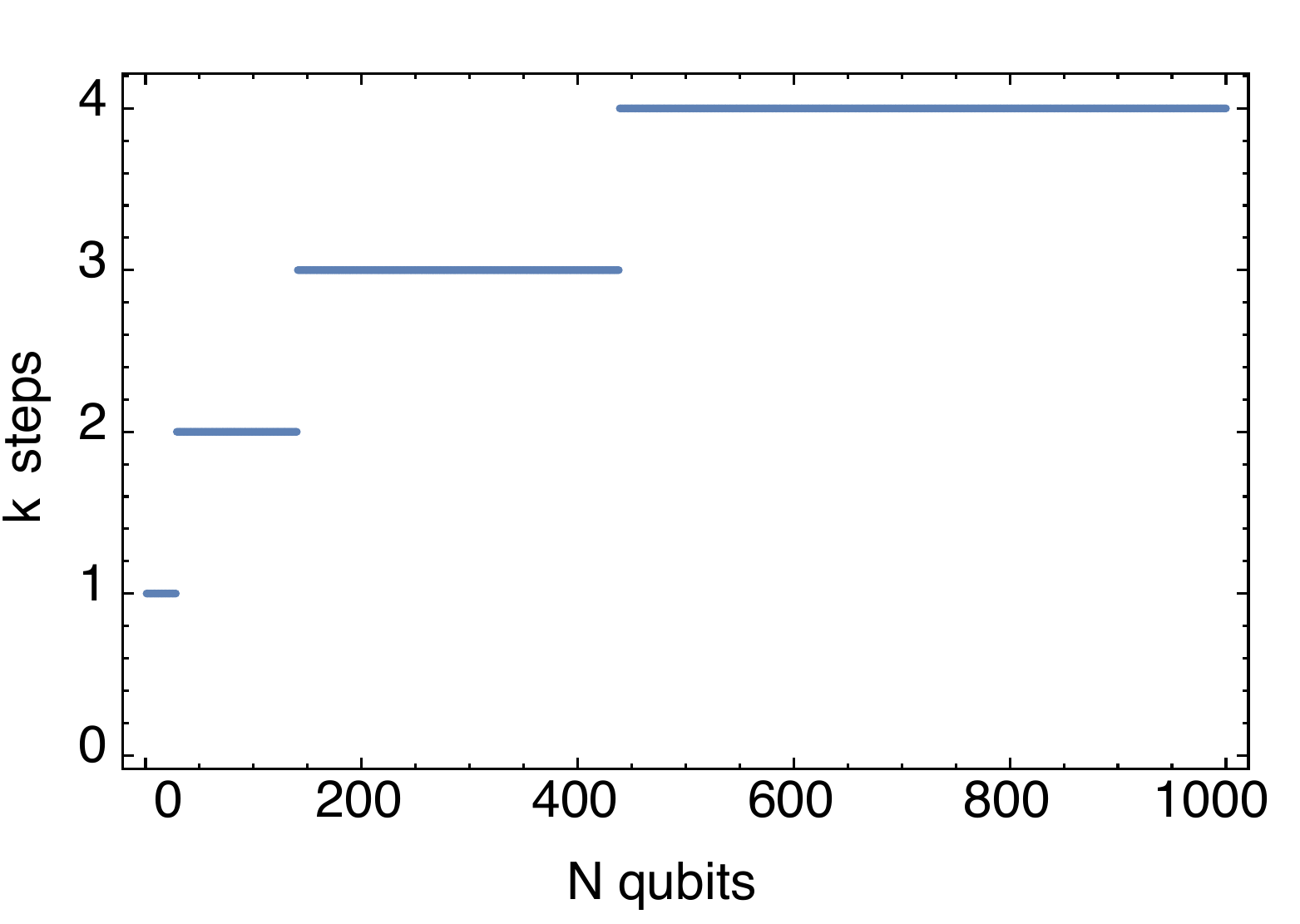}
   \caption{The Grover steps $k$ needed to prepare the Dicke states $m=N/2$ or $(N+1)/2$ using Eq. \eqref{k_N_over_2}. $k$ is rounded to the nearest non-zero integer.} 
 \label{k_mNover2_Dicke_integer}
\end{figure}

To prepare a Dicke state $\ket{m}$ with $N$ qubits, the ``traditional'' Grover iteration would be \cite{Grover1997}
\begin{equation}\label{G}
G= H^{\otimes N} \chi_0 H^{\otimes N} \chi_m
\end{equation}
where $\chi_m$ flips the phase of $\ket{m}$, and does nothing to the rest of the Dicke states $\ket{m' \neq m}$, i.e., $\chi_m\ket{m'} = (1-2\delta_{m,m'})\ket{m'}$. We start from the equal superposition state 
\begin{equation}
 \ket{\psi_i}=   \dfrac{1}{2^{N/2}}(\ket{0}+\ket{1})^{\otimes N}= \dfrac{1}{2^{N/2}} \sum_{n=0}^{N} \sqrt{{N\choose n}}  \ket{n}
\end{equation}
where ${N\choose n}$ is the binomial coefficient. Here, the target is $\ket{\psi_t}=\ket{m}$ with $\sin (\theta/2)= \sqrt{{N\choose m}}/2^{N/2}$ while all the other Dicke states are orthogonal. Using Eq. \eqref{k}, the number of steps $k$ required to prepare $\ket{m}$ is  
\begin{equation}\label{k_N_over_2}
k=\dfrac{\pi}{4 \arcsin(\sqrt{{N\choose m}}/2^{N/2})}-\dfrac{1}{2}
\end{equation}
The fidelity after $k$ steps can be calculated by using Eq. \eqref{fidelity_k}. As an example, Fig. \ref{k_mNover2_Dicke_integer} shows the number of steps $k$ required to prepare $N$ qubits for the even and odd states $m=N/2$ and $m=(N+1)/2$. For $3 \le N \le 1000$ qubits, these states can be prepared within four steps. A good approximation to Eq. \eqref{k_N_over_2} for $m=N/2$ is
\begin{equation}\label{k_N_over_2_approx}
k \approx 0.88N^{1/4}-\dfrac{1}{2}, \ \ m=\dfrac{N}{2}, \dfrac{N+1}{2}
\end{equation}
We note that another Grover iteration that would yield similar results is $G=  R(\pi/2)^{\otimes N} \chi_0  R(-\pi/2)^{\otimes N} \chi_m$, where $R(\phi)^{\otimes N}$ is a global $y$ rotation of all the $N$ qubits. By imparting a chosen phase $\alpha$, as described in Subsec. \ref{subsec:mod_phase}, the Dicke states can be prepared with perfect fidelity in integer steps. $\chi_m$ then would be given by $\chi_m(\alpha)\ket{m'}= e^{i\delta_{m,m'}\alpha}\ket{m'}$ and similarly for $\chi_0(\alpha)$.

\subsection{Scalable Dicke state preparation} \label{subsec:modified_grover_Dicke}

\begin{figure}[!t]
   \centering
    \includegraphics[width=0.45\textwidth]{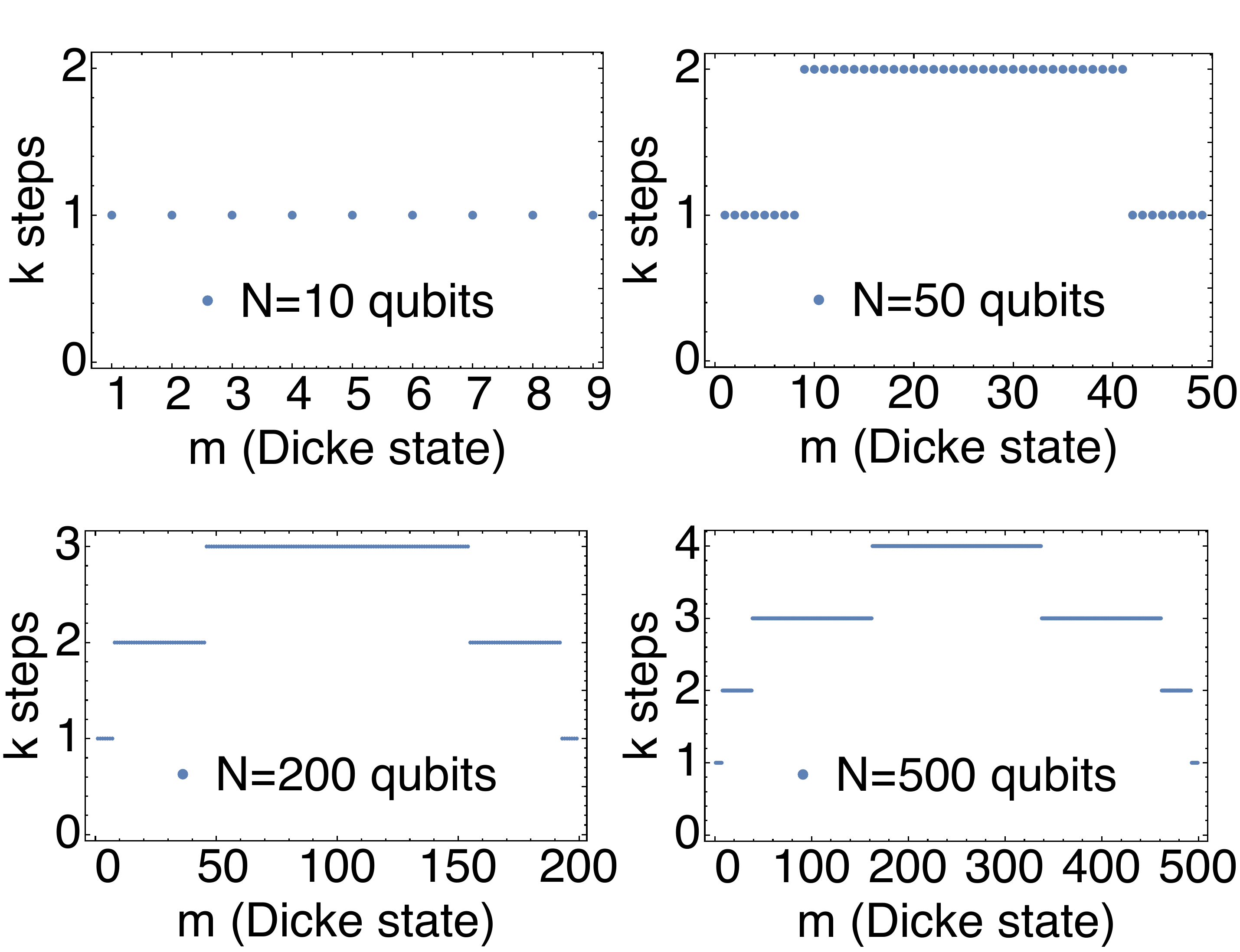}
   \caption{The Grover steps $k$ needed to prepare the Dicke states with high fidelity for different number of qubits $N$. $k$ is rounded to the nearest non-zero integer.}
    \label{fig:kmN}
\end{figure}

The number of steps required under the ``normal'' Grover's algorithm, proposed above, has favorable scaling only for the Dicke states $m\approx N/2$ (i.e., $k \sim N^{1/4}$ steps), but it would require an exponential number of steps in $N$ for Dicke states close to $m = 0$ (e.g., $k\sim 2^{N/2}/N$ for the W state with $m=1$) and $m =N$. Here, we show that a modified Grover's algorithm, with a chosen global rotation different from Hadamard, can prepare $\ket{m}$ with a scaling $k \sim m^{1/4}$ for large $N$. The key idea is to maximize the overlap between the initial state and the target Dicke state and do phase inversion on that initial state. 

To prepare $\ket{\psi_i}$, each qubit $\ket{0}$ is rotated by an angle $\phi$ to the state $ \cos( \phi/2 )\ket{0}+\sin( \phi/2) \ket{1}$. So the initial state is the product state:
\begin{align}\label{psi_initial}
& \ket{\psi_i}=  \bigg(\cos (\phi/2) \ket{0}+\sin (\phi/2)\ket{1}\bigg)^{\otimes N} \nonumber \\&= \sum_{n=0}^{N} \sqrt{{N\choose n}} \cos^{N-n} (\phi/2) \sin^{n} (\phi/2) \ket{n}
\end{align}
Choose $\phi$ such that the overlap between the initial state $\ket{\psi_i}$ and $\ket{m}$ is maximized. This occurs at $\phi=\arccos[(N-2m)/N]$, which corresponds to applying the following single-qubit rotation $R(\phi)$:
\begin{equation}\label{R_phi}
 R(\phi)=\begin{bmatrix}
    \sqrt{\dfrac{N-m}{N}}       & -\sqrt{\dfrac{m}{N}}  \\
    \sqrt{\dfrac{m}{N}}        & \sqrt{\dfrac{N-m}{N}} 
\end{bmatrix}   
\end{equation}
This is a rotation about the $y$ axis of the Bloch sphere of the single qubit by $\phi$, i.e., $ R(\phi)= R_y(\phi)$. Therefore, the initial state is $\ket{\psi_i}=(R(\phi)\ket{0})^{\otimes N}$. The modified Grover iteration is as follows. It consists of phase inversion $\chi_m$, and phase inversion of $\ket{\psi_i}$. A minus sign can be put on $\ket{\psi_i}$ by rotating back to $\ket{m=0}$,  applying $\chi_0$, then rotating back again to $\ket{\psi_i}$, i.e., $\chi_{i}=R(\phi)^{\otimes N}\chi_0R(-\phi)^{\otimes N}$. Therefore, $G$ is
\begin{equation}
G=R(\phi)^{\otimes N}\chi_0R(-\phi)^{\otimes N}\chi_m
\end{equation}
In the limit of many qubits $N \gg 1$, the number of steps required to prepare $\ket{m}$ becomes independent of $N$ (cf. Appendix \ref{appendix:Dicke_prep}):
\begin{equation}\label{k_m_modified}
 k= \dfrac{\pi}{ 4 \arcsin [(1/2 \pi m)^{1/4}]}-\dfrac{1}{2}, \ \ N \gg 1,  \ m \ll N/2 
 \end{equation}
where this can be approximated as $ k \approx 1.24 m^{1/4}-1/2$. By the symmetry of the equations [cf. Appendix \ref{appendix:Dicke_prep} and Eq. \eqref{overlap_max}], the number of steps $k$ to prepare $\ket{m}$ is the same as to prepare $\ket{N-m}$. For values close to $m= N/2$, the number of steps is given by $k \approx 0.88 N^{1/4}-1/2$ [Eq. \eqref{k_N_over_2_approx}]. In the limit $N \gg 1$, for a given $\ket{m}$ the number of steps does not depend on $N$ [Eq. \eqref{k_m_modified}], i.e., the algorithm is scalable in $N$. In Fig. \ref{fig:kmN}, we plot the number of steps required to prepare $\ket{m}$ for different numbers of qubits. $k$ is found by solving $\sin([2k+1]\theta/2)=1$, where $\theta$ is given by Eq. \eqref{overlap_max}. $k$ is analytically given by Eq. \eqref{k_exact_dicke_modified} in Appendix \ref{appendix:Dicke_prep}. Due to the favorable scaling, $k \sim m^{1/4}$, all the Dicke states  with $3 \le N \le 500$ qubits can be prepared with high fidelity within four Grover steps. $k$ is symmetric around $m=N/2$ due to the symmetry of the Dicke states. The states around $m=N/2$ require the largest number of steps to prepare. As before, the Dicke states can be prepared exactly in integer steps by applying $\chi_m(\alpha)$ and $\chi_0(\alpha)$, where $\alpha$ is given by Eq. \eqref{Long_phase} and $\sin (\theta/2)$ by Eq. \eqref{overlap_max}.

\subsection{Preparing Dicke states exactly in a few steps}\label{subsec:Dicke_exact_prep}

In this section, we show how to prepare an arbitrary $N$-qubit Dicke state perfectly in a few steps without using $\alpha$. For exact preparation in $k$ integer steps, we are looking for an initial state prepared by a global rotation $\ket{\psi_i}=R(\phi)^{\otimes N}\ket{0}^{\otimes N}$ such that the overlap angle between this state and the target Dicke state is $\theta/2=\frac{\pi}{2(2k+1)}$ (cf. Sec. \ref{sec:exact_prep}). Therefore, we are looking for the single-qubit rotation $R(\phi)$
\begin{equation} \label{rotation_phi}
 R(\phi)=\begin{bmatrix}
    \cos(\phi/2)     & -\sin (\phi/2)   \\
    \sin(\phi/2)        &  \cos (\phi/2) 
\end{bmatrix}   
\end{equation}
Such that the inner product between $\ket{m}$ and $\ket{\psi_i}$ is 
\begin{align}\label{grover_dicke_overlap_condition}
&\braket{m|\psi_i}\nonumber= \\&  \sqrt{{N\choose m}} \cos^{N-m} (\phi/2) \sin^{m} (\phi/2)=\sin \bigg( \dfrac{\pi}{2(2k+1) }\bigg)
\end{align}
This is a transcendental equation in $\phi$ that can be solved for a given $N$ and $m$. A solution $0 \le \phi \le \pi$ exists if $N$ and $m$ satisfy (see Appendix \ref{overlap_appendix})
\begin{equation}\label{Dicke_condition_solution}
{N\choose m} \bigg(1-\dfrac{m}{N}\bigg)^{N-m}\bigg(\dfrac{m}{N}\bigg)^m\ge \sin^2 \bigg( \dfrac{\pi}{2(2k+1) }\bigg)
\end{equation}
After finding $\phi$, the Grover iteration would then be
\begin{equation}\label{few_step_grover}
G=R(\phi)^{\otimes N}\chi_0R(-\phi)^{\otimes N}\chi_m
\end{equation}
Preparing a Dicke state in one step requires finding a rotation angle that satisfies $\braket{m|\psi_i}= 1/2$. For some $N$ and $m$, an exact preparation in one step is not possible. In this case, we can still try to find $\phi$ and a corresponding $\ket{\psi_i}$ that approximately satisfy the overlap condition, i.e.,  $\braket{m|\psi_i}\approx 1/2$. This would generate the desired Dicke state with fidelity $F \approx 1$ in one step. To prepare a Dicke state in one step with fidelity $F$, we need to find a global rotation angle that satisfies the overlap condition $\braket{m|\psi_i}=\sin ( \frac{1}{3}\arcsin F )$. For example, to prepare $\ket{m}$ with $F=0.999$, we need to find $\phi$ that satisfies $\braket{m|\psi_i}=0.49$. If $\braket{m|\psi_i} = 1/2$ cannot be solved for $\phi$ in the vicinity of $1/2$, then we can relax our condition: we try to prepare $\ket{m}$ in two or more integer steps [Eq. \eqref{grover_dicke_overlap_condition}]. Using Eq. \eqref{Dicke_condition_solution}, we prove the following (see Appendices \ref{W_one_step_appendix} and \ref{N_over_2_steps_appendix}):

\begin{itemize}
\item The $m=1$ Dicke state (W state) with $N$ qubits can always be prepared perfectly in one step for all $N$.

\item The minimum number of steps required to prepare the Dicke state $m=N/2$ is given by $k = 0.88 N^{1/4}-1/2$.

\item We numerically verify that all the Dicke states with $3 \le N \le 500$ qubits can be exactly prepared within four Grover steps.
    
\end{itemize}

\begin{figure}[!t]
    \centering
    \includegraphics[width=0.45\textwidth]{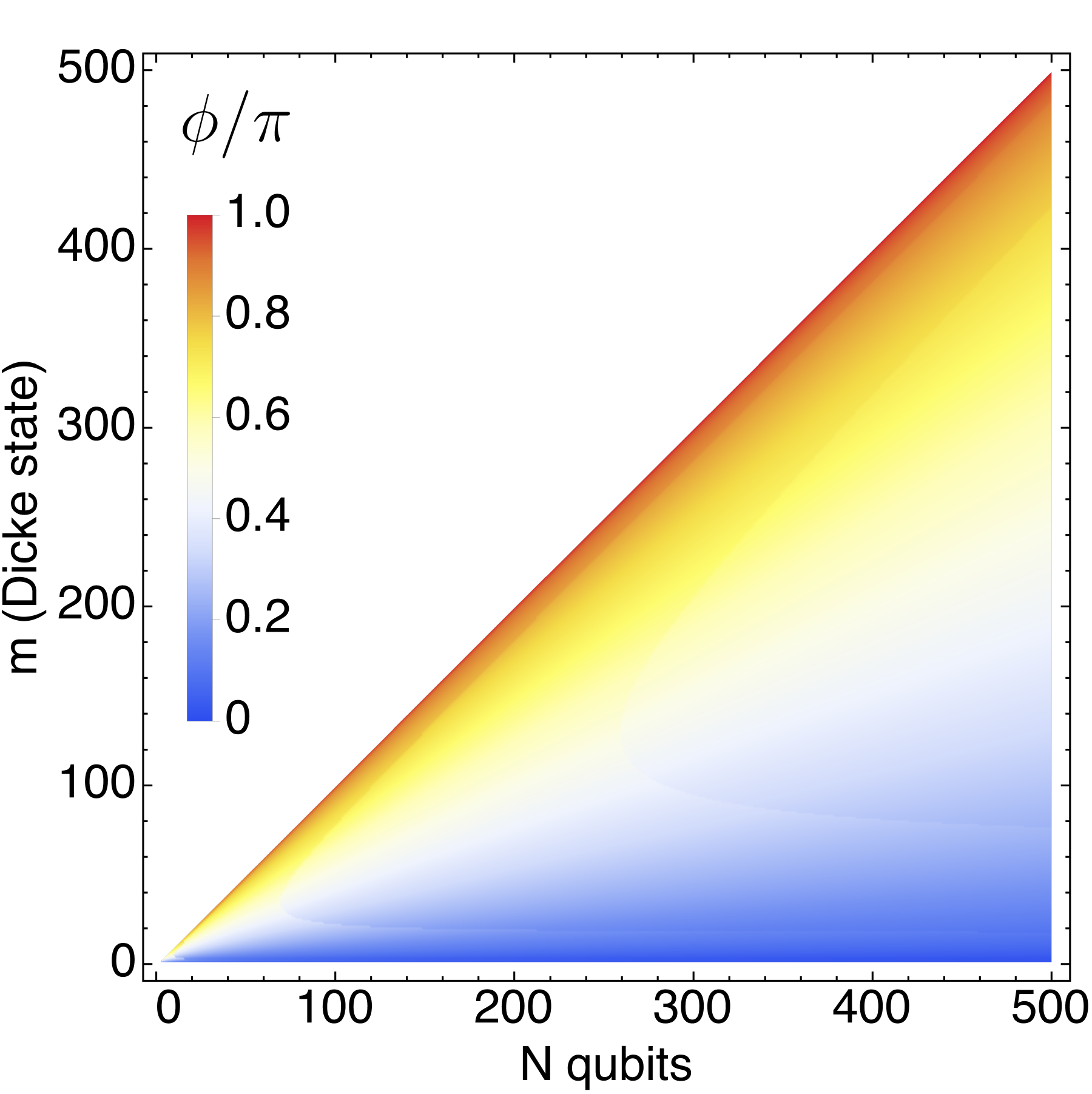}
    \caption{Density plot of the required global rotation angle $\phi$ to prepare a Dicke state exactly with few Grover steps [Eq. \eqref{few_step_grover}].}
    \label{Dicke_phi}
\end{figure}

By inverting Eq. \eqref{Dicke_condition_solution} and solving for $k$, we can find the minimum number of steps $k$ required to prepare $\ket{m}$ perfectly. This gives 
\begin{equation}
 k \ge \dfrac{\pi}{ 4 \arcsin [(1/2 \pi m)^{1/4}]}-\dfrac{1}{2}, \ \ N \gg 1,  \ m \ll N/2 
 \end{equation}
Which is the same scaling found in Eq. \eqref{k_m_modified}. We have numerically identified $\phi$ in Eq. \eqref{grover_dicke_overlap_condition} for all the Dicke states with $3 \le N \le 500$ qubits. In general, Eq. \eqref{grover_dicke_overlap_condition} has two distinct solutions $0 \le \phi \le \pi$ for a given Dicke state. Here, one of the two solutions is shown in Fig. \ref{Dicke_phi}, noting that the other solution has similar values and features. Dicke states with larger (smaller) $m$ require smaller (larger) $\phi$, and the Dicke states $m \approx N/2$ require $\phi \approx \pi/2$ global rotations. 

\section{GHZ states}\label{sec:Grover_GHZ}

\subsection{Two-step Grover's algorithm for the GHZ state}

We show that Grover's algorithm can prepare the GHZ state with a favorable scaling in the number of qubits. The protocol consists of two parts. The first part generates the Dicke state $\ket{N/2}$, which can be efficiently done as was shown previously. After rotating this Dicke state to the $x$-basis, i.e., $\ket{N/2}_x$, it will then be used as the initial state for the second step. The second step applies a different Grover iteration on that initial state to prepare the GHZ state in $k \sim N^{1/4}$ steps. 

Recall that the GHZ state is the superposition of the two extremal Dicke states $\ket{m=0}$ and $\ket{m=N}$
\begin{equation}
\ket{\rm GHZ}=\dfrac{1}{\sqrt{2}}(\ket{0}^{\otimes N} +\ket{1}^{\otimes N})
\end{equation}
For large $N$, the overlap amplitude is maximum between $\ket{m=N/2}_x$ and $\ket{\rm GHZ}$ with $\sim (8/ \pi N)^{1/4}$. An even $m$ close to $N/2$ needs to be chosen as the initial state, so that it has a finite overlap with the GHZ state. If $m=N/2$ is not even, then choose an even $m$ close to $N/2$. After preparing $\ket{N/2}_x$, we apply the following Grover iteration: a phase inversion on the two extremal states $\ket{0}^{\otimes N}$ and $\ket{1}^{\otimes N}$. This flips the phase of the GHZ state, i.e., $\chi_t=\chi_{0}\chi_{N}$. A phase flip can be applied on the initial Dicke state by rotating from the $x$ to the $z$-basis, applying a phase flip on $\ket{N/2}$, then rotating back to the $x$-basis. Thus $G$ is given by 
\begin{equation}\label{G_GHZx}
G=H^{\otimes N}\chi_{N/2}H^{\otimes N}\chi_{0}\chi_{N}
\end{equation}
For a large number of qubits, the number of steps required to prepare the GHZ state is (cf. Appendix \ref{GHZx_appendix})
\begin{equation}
 k= \dfrac{\pi}{ 4 \arcsin [(8/\pi N)^{1/4}]}-\dfrac{1}{2} \approx 0.62 N^{1/4}-\dfrac{1}{2}, \ \ N \gg 1  
 \end{equation}
This shows that the number of steps required to prepare the GHZ state scales in the same way as for preparation of the $\ket{m=N/2}$ Dicke state, i.e., $O(N^{1/4})$. Starting from a product state, the GHZ state needs approximately twice as many steps, since the Dicke state $\ket{N/2}$ is needed as an intermediate state. The algorithm here can also be modified by imparting a phase $\alpha$, given by Eq. \eqref{Long_phase}, to prepare the GHZ state with unit fidelity in an integer number of steps. Instead of using a global Hadamard transformation, one can also take a Dicke state in the $y$-basis as the initial state and apply global $\pi/2$ rotations about the $y$-axis to prepare the GHZ state (cf. Appendix \ref{GHZyz_appendix}). 

\subsection{Preparing the GHZ state exactly in a few steps}
Here, we show how to prepare a GHZ state perfectly in a few steps without resorting to a special phase $\alpha$. The algorithm proceeds in two steps as before. The first step is preparing $\ket{N/2}$ then rotating it by $-\phi$. This rotated Dicke state $\ket{N/2,-\phi}$ will be fed as the initial state in the second step. We look for $\phi$ such that the overlap between that Dicke state and the GHZ state allows for exact preparation within few steps. The overlap condition to prepare the GHZ exactly after $k$ steps is (cf. Appendix \ref{GHZ_condition_solution_appendix})
\begin{align}\label{GHZ_overlap_condition}
&\braket{N/2, -\phi|\rm GHZ}= \\& \nonumber \sqrt{2{N\choose N/2}}  \cos^{N/2} (\phi/2) \sin^{N/2} (\phi/2)=  \sin \bigg( \dfrac{\pi}{2(2k+1) }\bigg)
\end{align}
where both $N$ and $m=N/2$ are even here. This has a solution if $N$ satisfies (cf. Appendix \ref{GHZ_condition_solution_appendix})
\begin{equation}
\dfrac{1}{2^{N-1}}{N\choose N/2} \ge  \sin^2 \bigg( \dfrac{\pi}{2(2k+1) }\bigg)
\end{equation}
By inverting this condition and solving for $k$, we can show that the minimum number of steps $k$ to prepare the $N$-qubit state GHZ state perfectly is 
\begin{equation}
k \ge \dfrac{\pi}{4 \arcsin(\sqrt{{N\choose N/2}}/2^{(N-1)/2})}-\dfrac{1}{2}
\end{equation}
which is approximately $0.62 N^{1/4}-1/2$ for large $N$. Here, the phase inversion on the target state is the same as in the previous subsection. The phase of the initial state can be inverted by rotating from the $\phi$ to the $z$ direction, applying a phase flip on $\ket{N/2}$, and then rotating back. Thus, $G$ simplifies to
\begin{equation}
G=R(-\phi)^{\otimes N} \chi_{N/2}R(\phi)^{\otimes N}\chi_0 \chi_N 
\end{equation}
We conclude this section by stating that it is possible to implement $\chi_N$ in terms of $\chi_0$ using $\chi_N=X^{\otimes N}\chi_0 X^{\otimes N}$, where $X^{\otimes N}$ is a global $X$ gate that flips all the qubits. This might be beneficial in cases when $\chi_N$ cannot be implemented natively or its physical implementation has a lower fidelity compared to $\chi_0$.
\section{Cat states}\label{sec:Grover_Cat}
We propose to use Grover's algorithm to prepare Cat states, which are superpositions of coherent spin states (CSS):
\begin{equation}
\ket{{\rm cat \pm},\phi}=\dfrac{1}{\sqrt{2 \pm 2\cos^N \phi}}(\ket{\phi}^{\otimes N} \pm \ket{-\phi}^{\otimes N})
\end{equation}
where the CSS states $\ket{\pm \phi}^{\otimes N}$ are the Dicke states $\ket{m=0}$ rotated globally by $\pm \phi$. The protocol is similar to that of the two-step Grover's algorithm for a GHZ state: the first part prepares a Dicke state $\ket{m}$ which is chosen to have a favorable overlap with the target Cat state (any Dicke state can be efficiently prepared by Grover as described previously). The second part of the protocol applies another Grover iteration to $\ket{m}$ to prepare the desired Cat state. To prepare a Cat state in $k$ steps, the overlap condition between the initial state $\ket{m}$ and $\ket{{\rm cat \pm},\phi}$ is (cf. Appendix \ref{Cat_appendix})
\begin{align}
&\braket{m|{\rm cat \pm},\phi}\nonumber=\dfrac{2}{\sqrt{2 \pm 2\cos^N \phi}} \\& \times  \sqrt{{N\choose m}} \cos^{N-m} (\phi/2) \sin^{m} (\phi/2)= \sin \bigg( \dfrac{\pi}{2(2k+1) }\bigg)
\end{align}
where we need to solve for $m$ given $k, \ N,$ and $\phi$. The Grover iteration then becomes $G=\chi_m\chi_{\rm cat}$. The phase inversion operator $\chi_{\rm cat}=I-2\ket{\rm cat}\bra{\rm cat}$ is non-trivial to implement in practice. Therefore, we propose to approximate its action by operators that do phase inversion on the two CSS states $\ket{\pm \phi}$ sequentially, i.e., $\chi_{\rm cat} \approx \chi_{\phi} \chi_{-\phi}$. This approximation holds well when the two CSS states are nearly orthogonal. This approximation becomes increasingly more accurate for larger numbers of qubits because the overlap decreases exponentially with the number of qubits, i.e., $|\braket{\phi|-\phi}^{\otimes N}|^2=\cos^{2N} \phi$. Therefore, the approximate Grover iteration becomes
\begin{equation}
G \approx \chi_m \chi_{\phi} \chi_{-\phi}
\end{equation}
$\chi_{\pm \phi}$ is realized by rotating $\ket{\pm \phi}^{\otimes N}$ to $\ket{m=0}$, applying a phase flip, then rotating back to the CSS states, i.e., $\chi_{\pm \phi}=R(\pm \phi)^{\otimes N}\chi_0 R(\mp \phi)^{\otimes N}$. The Grover iteration then becomes
\begin{equation}\label{G_cat}
G \approx \chi_m R(\phi)^{\otimes N} \chi_0 R(-2\phi)^{\otimes N}\chi_0R(\phi)^{\otimes N}
\end{equation}
One can modify the phases here by $\alpha$ to get a unity fidelity as outlined before. 

\section{Physical realization in cavity QED}\label{sec:physical_implement}

\subsection{Idealized case}\label{subsec:ideal_cavity}
\begin{figure}[!t]
   \centering
    \includegraphics[width=0.45\textwidth]{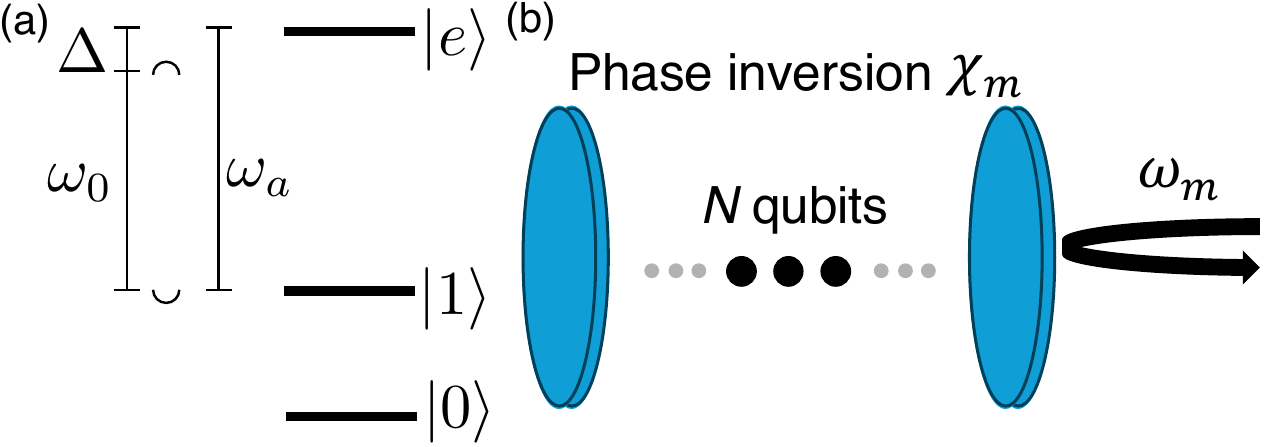}
   \caption{(a) The energy level scheme for the $N$ qubits in a cavity interacting with light. (b) The phase inversion operator $\chi_m$ is realized in cavity QED by reflecting a photon with frequency $\omega_m=\omega_0+m\Omega$.}
    \label{fig:energy_scheme}
\end{figure}
We proceed to show that it is possible to efficiently implement the Grover iteration in the dispersive regime of cavity QED. Consider the setup in Fig. \ref{fig:energy_scheme}: a one-sided cavity with a resonance frequency $\omega_0$. There are $N$ atoms in the cavity with the atomic transition $\ket{1} \leftrightarrow \ket{e}$ with frequency $\omega_a$, whereas $\ket{0}$ is far detuned. \rsub{The cavity is far detuned from the atomic transition such that $|\Delta|=|\omega_0-\omega_a| \gg g$, where $g$ is the atom-cavity coupling strength}. Under these conditions, the interaction Hamiltonian $H$ between the $N$ atoms in the cavity and light is dispersive and given by \cite{Chen2015carving,Circuit_QED_2009}:
\begin{equation}
H=\hbar \Omega \hat{m} \hat{n}_c 
\end{equation}
where $\Omega=\dfrac{g^2}{\Delta}$, and $\hat{n}_c$ is the photon number operator and $\hat{m}$ counts the number of atoms in state $\ket{1}$.  If there are $m$ atoms in $\ket{1}$, i.e., the Dicke state $\ket{m}$, then this shifts the cavity resonance frequency to $\omega_m=\omega_0+m\Omega$, where $\Omega \ll \Delta$. This selective shift of the cavity resonance frequency can be used to implement $\chi_m$ by hitting the cavity with a photon with frequency $\omega_m$ ; this will induce a phase shift of $-1$ only when the atomic state is in $\ket{m}$ (since the photon frequency is resonant with the atom-cavity system). For all other Dicke states, no phase shift occurs after reflection as the photon is off resonant. This realizes the required phase inversion $\chi_m$ [cf. Fig. \ref{fig:energy_scheme}(b)]:
\begin{equation}
\ket{m' } \rightarrow (1-2\delta_{mm'})
\ket{m'}
\end{equation}
Similarly, $\chi_0$ is realized by hitting the cavity with a photon with frequency $\omega_0$ . By going slightly off-resonance, $\chi_m(\alpha)$ can also be implemented. These relations are under ideal cavity conditions. A single Grover step consists of two phase inversions for the Dicke state [Eq. \eqref{few_step_grover}]. Therefore, the scattering of two photons (and global rotations) physically implement a single Grover iteration. For the GHZ and Cat states, three photon scattering events are required [Eqs. \eqref{G_GHZx} and \eqref{G_cat}]. It is important to emphasize that the resources to implement $\chi_m$ do not scale with the number of qubits in this scheme. Moreover, the scheme does not require individual addressing of the qubits. We remark that a similar physical implementation has been proposed previously in the context of using Grover's algorithm to solve subset sum problems \cite{Dicke_partition_2021}. \rsub{We contrast the present approach with previous probabilistic carving schemes relying on a two-sided symmetric cavity, where the photon is probabilistically transmitted or reflected, conditioned on the state of the atomic qubits \cite{Chen2015carving} (see Appendix \ref{appendix:cavity_carve} for an overview).}

\subsection{Effect of the spatial mode mismatch}\label{subsec:spatial_mismatch}

 \begin{figure}[!t]
    \centering
    \includegraphics[width=0.45\textwidth]{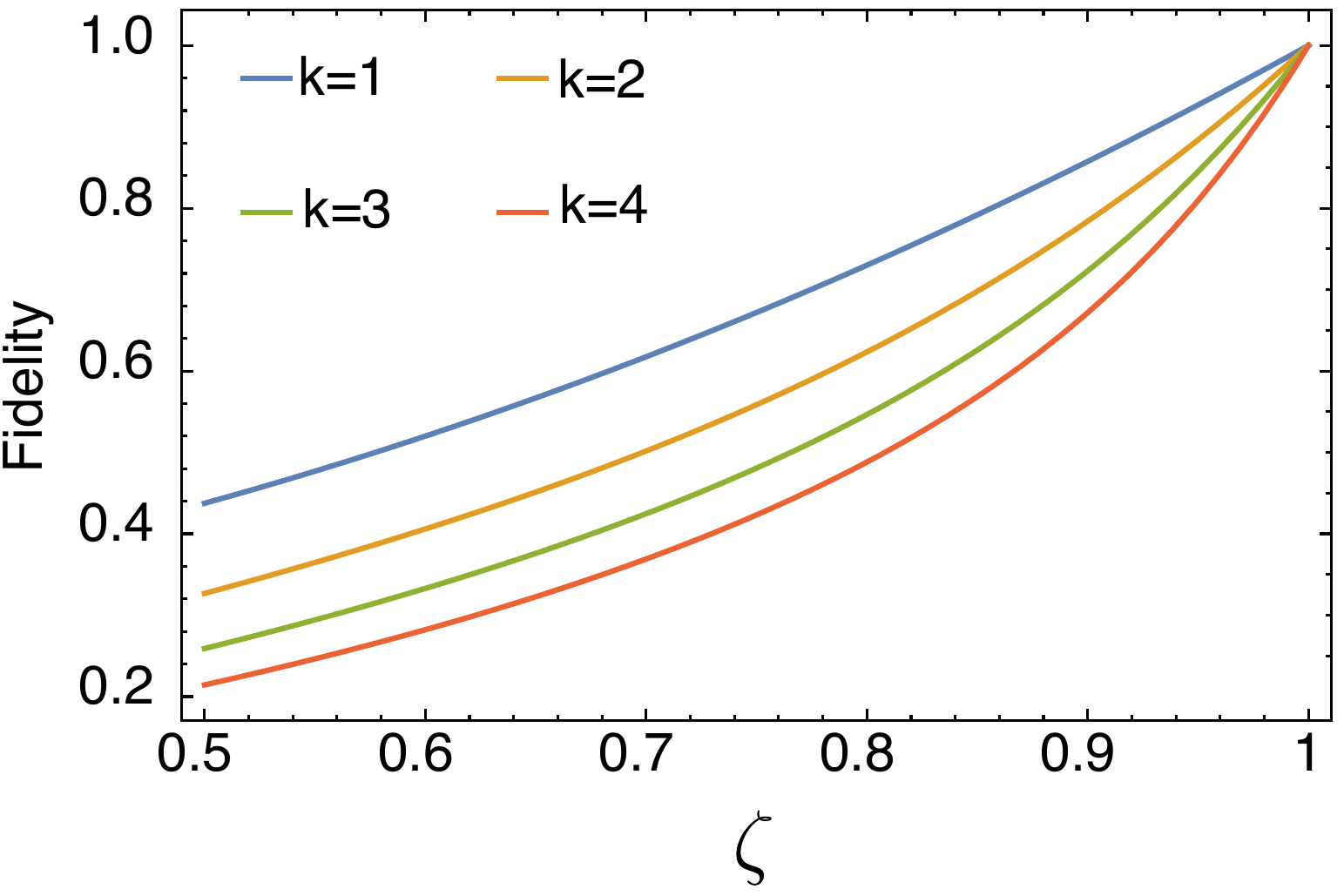}
    \caption{Fidelity after applying Grover's algorithm versus the spatial mode matching efficiency $\zeta$ for the four initial states with $\theta/2=\pi/6, \ \pi/10, \ \pi/14,$ and $\pi/18$, which can prepare a desired target in $k=1, \ 2, \ 3,$ and $ 4$ steps, respectively. The fidelity is defined as the overlap between the output of Grover's algorithm and the target state.}
    \label{F_mismatch}
\end{figure}

Here, we discuss the effect of spatial mode mismatch between the incoming photon and the cavity mode on Grover's algorithm.  It was shown above that the phase inversion operator $\chi$ on the atomic qubits $\rho$ can be physically implemented by reflecting a photon off the cavity. If the spatial mode matching efficiency is $\zeta$, then the mismatched part of the photon mode $1-\zeta$ is reflected off the cavity without interacting with the atomic qubits \cite{Cavity_Rempe_review_2015}, i.e., after reflection the atomic state becomes
\begin{equation}
 \rho \rightarrow  \zeta \chi \rho  \chi^{\dag}+(1-\zeta)\rho
\end{equation}
Since all the Dicke states with $3 \le N \le 500$ qubits can be prepared with $ 1\le k \le 4$ scattering events, we restrict the analysis to these cases, noting that a similar analysis can be done for the GHZ and Cat states.

In Appendix \ref{appendix:mismatch}, we develop an error model and analytically compute the fidelity, resulting from applying the Grover iteration in the presence of spatial mode mismatch, as a function of $\zeta$. The results of those calculations [Eqs. \eqref{F_mismatch1}, \eqref{F_mismatch2}, \eqref{F_mismatch3}, and \eqref{F_mismatch4}] are shown in Fig. \ref{F_mismatch}. The figure shows the fidelity, i.e., the overlap between the target and  the output $\rho^{(k)}_{\rm out}$ after $k$ Grover steps, versus $\zeta$ given the four initial states $\theta/2=\pi/6, \ \pi/10, \ \pi/14,$ and $\pi/18$. Target states that require more steps to prepare are more sensitive to spatial mode mismatch, as expected. For $|1-\zeta| \ll 1$, the infidelity after applying $k$ steps is found to the lowest order to be (cf. appendix \ref{appendix:mismatch})
\begin{equation}
1-F(k) \approx \dfrac{2k+1}{2}(1-\zeta), \ \ \dfrac{\theta}{2}=\frac{\pi}{2(2k+1)}
\end{equation}
i.e., the infidelity grows linearly with both the number of steps and the amount of mismatch. Achieving a fidelity of $F\ge 0.99$ requires a corresponding mode matching of $\zeta\ge (0.993,0.996,0.997, 0.998)$ for the states that require $k=(1,2,3,4)$ Grover steps, respectively. 

\subsection{Nonideal cavity and photon wavepacket}

It is possible to implement $\chi_m$ exactly only in the idealized case of a resonant monochromatic photon hitting an ideal cavity. In practice, the finite width of the photon wavepacket entangles the photonic and atomic state, leading to decoherence. An additional source of decoherence is from the nonideal cavity, causing photon losses through spontaneous emission, transmission, and scattering by the cavity mirror. In this subsection, we develop an error model taking these effects into account, using input-output theory \cite{Input_output_1984,Input_output_1985}, Kraus operators \cite{Kraus_2012}, and the Liouville (superoperator) formalism \cite{optically_pumped_atoms}. An incoming photon after reflection will be in the following four outgoing modes: reflection, transmission, spontaneous emission, and mirror scattering. The complex amplitudes for these modes are \cite{Daiss2019Single}
\rsub{
\begin{subequations}
\begin{align}
r_n(\omega)=&1-\dfrac{2 \kappa_r(i \Delta+i\omega + \gamma)}{n g^2+(i \Delta+i\omega + \gamma)(i \omega + \kappa)} \label{r_n} \\ 
t_n(\omega)=&\dfrac{2 \sqrt{\kappa_r \kappa_t }(i \Delta+i\omega + \gamma)}{n g^2+(i \Delta+i\omega  + \gamma)(i \omega + \kappa)}  \label{t_n} \\
a_n(\omega)=&\dfrac{2 \sqrt{\kappa_r \gamma }\sqrt{n}g}{n g^2+(i \Delta+i\omega  + \gamma)(i \omega + \kappa)} \label{a_n}\\
m_n(\omega)=&\dfrac{2 \sqrt{\kappa_r \kappa_m }(i \Delta+i\omega  + \gamma)}{n g^2+(i \Delta+i\omega  + \gamma)(i \omega + \kappa)}  \label{m_n}
\end{align}
\end{subequations}
}
where \rsub{$\omega=\omega_p-\omega_0$ is the detuning between the incoming photon frequency $\omega_p$} and the bare cavity frequency $\omega_0$, and $n$ is the number of atoms coupled  to the cavity (i.e., in $\ket{1}$). $\gamma$ is the atom decay rate, and $\kappa=\kappa_r+\kappa_t+\kappa_m$ is total cavity decay rate, which consists of cavity decay rates into the three outgoing modes of reflection, transmission, and mirror scattering, respectively. \rsub{For $n$ coupled atoms, the shifted cavity resonance frequency $r_n(\delta \omega_n)=-1$ [Eq. \eqref{r_n}] occurs at } \rsub{\begin{align}\label{cavity_shift_freq}
\nonumber &\delta \omega_n = {\rm Re} \bigg \{ \dfrac{1}{2}\Bigl(
  i\gamma - \Delta 
  + i(\kappa - \kappa_r)
 \\ & + \sqrt{4g^{2}n - \bigl(\gamma + i\Delta - \kappa + \kappa_r\bigr)^{2}}
\Bigr) \bigg \}
\end{align}}
\rsub{The real part gives the resonance frequency while the imaginary component is due to dissipation, which causes losses, i.e., $|r_n(\delta \omega_n)|<1$, as well as the phase inversion being approximate as $r_n(\delta \omega_n) \approx -1$. Assuming the cavity QED dispersive regime, i.e., $ng^2/\Delta^2 \ll 1$ or $n \Omega \ll \Delta$, the shifted resonance frequency of the cavity is $\delta\omega_n\approx n g^2/\Delta$ to a leading order}. Thus a phase inversion occurs on resonance, $r_n(ng^2/\Delta) \approx -1$, while off resonance we get $r_n(|\omega| \gg ng^2/\Delta) \approx 1$. A photon pulse of finite duration will have different frequency components and will hence not be reflected with these  exact phase factors. 

In Appendix \ref{appendix:kraus}, we develop a Kraus operator description of cavity QED experiments under various settings, and Appendix \ref{appendix:error_analysis} applies this formalism to model the errors in implementing $\chi_m$ and $G$. Here, we introduce the Kraus operators for reflection, transmission, spontaneous emission, and cavity mirror scattering acting on the Dicke states $\ket{k}$
\begin{subequations}\label{Kraus_operators}
\begin{align}
K_r(\omega)=&\sum_k r_k(\omega) \ket{k}\bra{k} \label{K_r} \\
K_t(\omega)=&\sum_k t_k(\omega) \ket{k}\bra{k} \label{K_t}\\
K_a(\omega)=&\sum_k a_k(\omega) \ket{k}\bra{k} \label{K_a}\\
K_m(\omega)=&\sum_k m_k(\omega) \ket{k}\bra{k} \label{K_m}
\end{align}
\end{subequations}
where the coefficients are given by Eqs. \eqref{r_n}-\eqref{m_n}. There are different experimental settings one can consider: 1) implementing $\chi_m$ by hitting the cavity with a photon and not heralding on it reflecting back or 2) conditioning on the photon being detected after reflection (otherwise the experiment is deemed a failure and needs to be restarted). To make the error analysis analytically and numerically tractable, we vectorize the density matrix $\rho$, where $\bm{\rho}$ is the ``vectorized'' version of the density matrix. A vectorization ${\rm vec}(\rho)$ maps the density matrix $\rho=\sum_{ij} \rho_{ij} \ket{i} \otimes \bra{j}$ into the following vector $\bm{\rho}={\rm vec}(\rho)= \sum_{ij} \rho_{ij} \ket{i}\otimes\ket{j}$ \cite{optically_pumped_atoms}. Under this superoperator (Liouville) formalism, density matrices are acted on by superoperators, e.g., the action of $\chi_n$ on the vectorized density matrix is modeled by $(\chi_n \otimes \chi_n^{*})\bm{\rho}$. In what follows, we will apply the error model from Appendices \ref{appendix:kraus} and \ref{appendix:error_analysis} to analytically and numerically calculate the effect of the dominant error sources and their scaling with the cavity parameters.

\begin{figure*}[htp!]
    \centering
    \includegraphics[width=\textwidth]{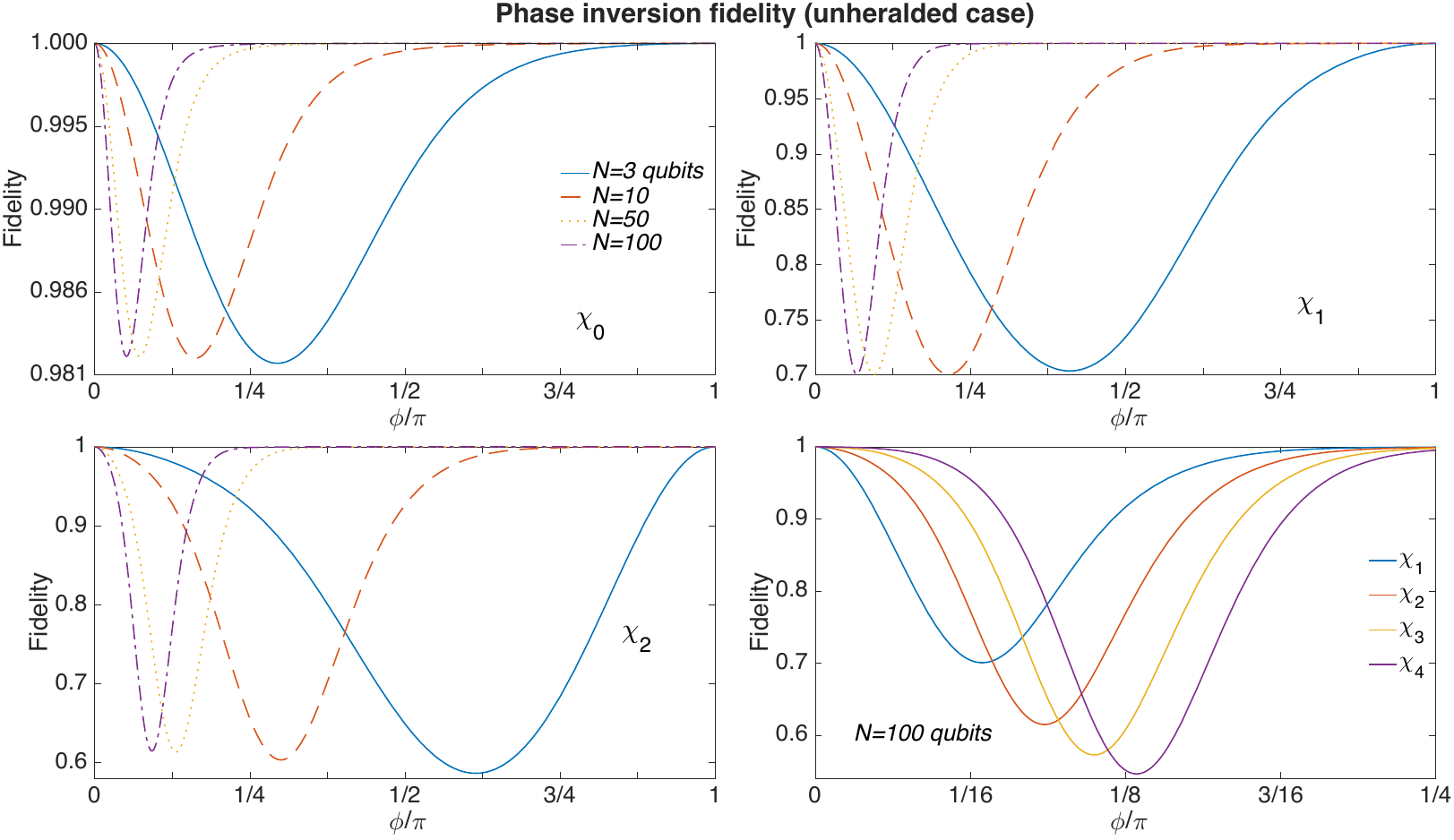}
    \caption{The fidelity in implementing the phase inversion operator $\chi_m$ on various CSS states $\ket{\phi}^N=R^N(\phi)\ket{m=0}$ as function of the global rotation angle $\phi$, for the case of not heralding \rsub{[Eqs. \eqref{vec_rho_unherald} and \eqref{superF}]}. The first three panels shows $F(\chi_m) (m=0,1,2)$ for the qubit number $N=3,10,50,$ and $100$. The last panel shows $F(\chi_m)$($m=1,2,3,4$) for $N=100$ qubits. The system parameters are $\kappa=\kappa_r=1$, $\gamma=1$, $g=10$, $C=100$, $w=0.1$, $d=(C/m)^{1/4}$ for $m=1,2,3,4$ and $\Delta=0 (d=\infty)$ for $m=0$.}
    \label{fig:F_chi_phi_unherald}
\end{figure*}

\subsubsection{Phase inversion: not heralding on the photon reflecting back}

Consider here the case of a single photon hitting the cavity and not heralding on the photon reflecting back. The interaction between the photon and the atom-cavity system can be captured by the following three dimensionless quantities: $w=\sigma/\kappa$, $C=g^2/\gamma \kappa$, and $d=g^2/\Delta\kappa=\Omega/\kappa$. $w$ measures the relative magnitude of the wavepacket and cavity bandwidth. Ideally, $w$ needs to be small so that the photon only inverts the phase of the desired Dicke state. The atom-cavity cooperativity $C$ quantifies the relative magnitude of coherent interaction to dissipation, where $C=\infty (0)$ implies that the coherent interaction (dissipation) dominates. $d$ is the ``resolution'' in frequency needed for the scattered photon to distinguish between two neighboring Dicke states $m$ and $m\pm 1$.

In Appendix \ref{appendix:error_analysis}, we derive the scaling of the infidelity in $\chi_m$ with $C, \ d,$ and $w$ by explicit calculations. Here we will give a physical argument for the scaling. First, consider $\chi_0$, implemented by a photon hitting the cavity with a cavity-photon detuning $\omega=0$. We would like to engineer the atom-cavity interaction such that there is a phase shift when there are no atoms coupled to the cavity, and no phase shift when there is at least a single coupled atom or more, i.e., $r_n=-1$ for $n=0$ and $r_n=1$ when $n>0$. After reflection, the atomic and photonic states are entangled as $\sum_{n} \ket{n}\otimes \ket{\psi_n}_p$, and tracing over the photonic degree of freedom reduces the coherence between the various Dicke states, causing infidelity. The coherence between the Dicke states is related to the overlap between the corresponding photonic states as ${\rm Re}(\braket{\psi_n|\psi_l})\sim{\rm Re}(r_n r^*_l)$. To a leading order in $C \gg 1$ and $d \gg 1$, and assuming $w \ll 1$, we have ${\rm Re}(r_0 r^{*}_n) \sim -1 +2/d^2n^2+2/nC$, where ${\rm Re}(r_0 r^*_n)=-1$ is the ideal value. The first error term $1/d^2n^2$ captures the cavity's ability to resolve between no or $n$ coupled atoms. It has the form $1/(nd)^2=(\kappa/n\Omega)^2$, since $n\Omega=\omega_{n}-\omega_0$ is the detuning between the resonance frequencies of the Dicke states $m=n$ and $m=0$, and it is quadratic because the lineshapes of the atoms and the cavity are Lorentzian-like in the detuning. The second error term $1/nC$ is due to spontaneous emission. For a given $C$, the desired goal is achieved by increasing the Dicke states resolution $d$ as much as possible, i.e., being on resonance $\Delta=0$. Dropping prefactors, and setting $\Delta=0(d=\infty)$, this implies the scaling (cf. appendix \ref{appendix:error_analysis} for details):
\begin{equation}\label{F0_C}
1-F(\chi_0)\sim \dfrac{1}{C}+w^2, \ \Delta=0
\end{equation}
Intuitively, $F(\chi_0)$ is maximized when the cavity and atom frequencies are the same, and the atom-cavity interaction is maximized, so that we are able to easily distinguish between ``no'' or ``at least one or more'' coupled atoms.

Now consider $\chi_m$ with $m\neq 0$, implemented by a photon detuned from the cavity by $\omega \approx mg^2/\Delta$. Here, the goal is more stringent, since the cavity needs to count the exact number of atoms coupled, and only invert the phase when there are $m$ atoms coupled. Suppose we try to decrease the first error $1/d^2$ by increasing $d$.  Since increasing $d=g^2/\kappa \Delta$ is equivalent to decreasing $\Delta$, so now all the shifted cavity frequencies $\omega_n$ become closer to the atomic resonance $\omega_a$. Thus, this will increase the probability of spontaneous emission, which goes like $|a_m(mg^2/\Delta)|^2\sim md^2/C$. Therefore, there is a tradeoff between the the two errors, $1-F \sim 1/d^2+m d^2/C$, which will achieve a minimum at a certain $d$, namely $d\sim (C/m)^{1/4}$ in this case. Plugging in that value in the infidelity gives the scaling (cf. Appendix \ref{appendix:error_analysis}):
\begin{equation}\label{Fm_C}
1-F(\chi_m) \sim \dfrac{1}{\sqrt{C}}+w^2, \ m \neq 0, \ d \sim (C/m)^{1/4}
\end{equation}
i.e., to maximize the fidelity $F(\chi_m)$ for a given $C$, we set the detuning to $\Delta=g^2/d\kappa$ with the choice $d\approx(C/m)^{1/4}$.

In appendix \ref{appendix:error_analysis}, we develop an analytic and numerical error model that calculates the final output state, for any qubit number and cavity parameters interacting with a finite-bandwidth wavepacket. Given an input atomic state $\bm{\rho}_{\rm in}$ and a photon wavepacket $\Phi(\omega)$, the atomic output state $\bm{\rho}_{\rm out}$ after the photon reflection is 
\begin{align}\label{vec_rho_unherald}
\bm{\rho}_{\rm out}= \bigg[ \int d \omega |\Phi(\omega)|^2\bigg (K_r(\omega)\otimes K_r^{*}(\omega)+K_t(\omega)\otimes K_t^{*}(\omega) \nonumber\\ +K_a(\omega)\otimes K_a^{*}(\omega)+K_m(\omega) \otimes K_m^{*}(\omega)\bigg ) \bigg] \bm{\rho}_{\rm in}
\end{align}
I.e., the atomic state is in a mixed state of the cases where the photon reflected, transmitted, scattered with the atom(s) or the cavity mirror. The integration comes from tracing over the photon wavepacket, which causes decoherence if the bandwidth of the photon is large compared to the atom-cavity decay width. We consider a Gaussian wavepacket $|\Phi(\omega,\Omega,\sigma)|^2=e^{ -(\omega-\Omega_c)^2/2 \sigma^2}/\sqrt{2 \pi} \sigma$, with a bandwidth $\sigma$ and a central frequency $\Omega_c$. To invert the phase of the Dicke state $\ket{n}$, the central frequency needs to be on resonance with the cavity shifted frequency, i.e., \rsub{ $\Omega_c = \delta \omega_n$ [Eq. \eqref{cavity_shift_freq}]}. Under this Liouville superoperator approach, the input $\bm{\rho}_{\rm in}$ has been separated from the Kraus operators and the integration. We shall denote $K(\omega) \otimes K^*(\omega)$ as the Kraus superoperators. Let's denote the sum of the averaged Kraus superoperators [the terms in the square bracket in Eq. \eqref{vec_rho_unherald}] as $\mathcal{K}_{\rm avg}$. The matrix elements of $\mathcal{K}_{\rm avg}(\Omega_c)$ are \rsub{computed numerically, with approximate analytic expressions given by Eqs. \eqref{Kr_avg_approx}-\eqref{Ka_avg_approx}, cf. Appendix \ref{appendix:error_analysis} for more details}. Now the output atomic state, for an arbitrary input atomic state and wavepacket, is generated by acting on the input with the averaged Kraus superoperator
\begin{equation}\label{vec_rho_unherald2}
\bm{\rho}_{\rm out}= \mathcal{K}_{\rm avg}(\Omega_c)\bm{\rho}_{\rm in}.
\end{equation}
Therefore, the physical implementation of the ideal operator $\chi_n$ is described by setting \rsub{$\Omega_c= \delta\omega_n$} above. The averaged Kraus superoperator \rsub{$\mathcal{K}_{\rm avg}(\delta\omega_n)$} is the physical inversion superoperator approximating the corresponding ideal superoperator $\mathcal{X}_n=\chi_n \otimes \chi_n^{*}$. The ideal output after applying $\chi_n$ is $\bm{\rho}_{\rm out}({\rm ideal})=\mathcal{X}_n\bm{\rho}_{\rm in}$. The fidelity $F$ in implementing $\chi_n$ is the overlap between the ideal (pure) output state and the nonideal (mixed) output state, which is given by the dot product between these two vectors in the Liouville formalism \cite{optically_pumped_atoms}:
\begin{equation}\label{superF}
F=\bm{\rho}_{\rm out}({\rm ideal})\cdot \bm{\rho}_{\rm out}
\end{equation}
\begin{figure}[!t]
    \centering
    \includegraphics[width=0.49\textwidth]{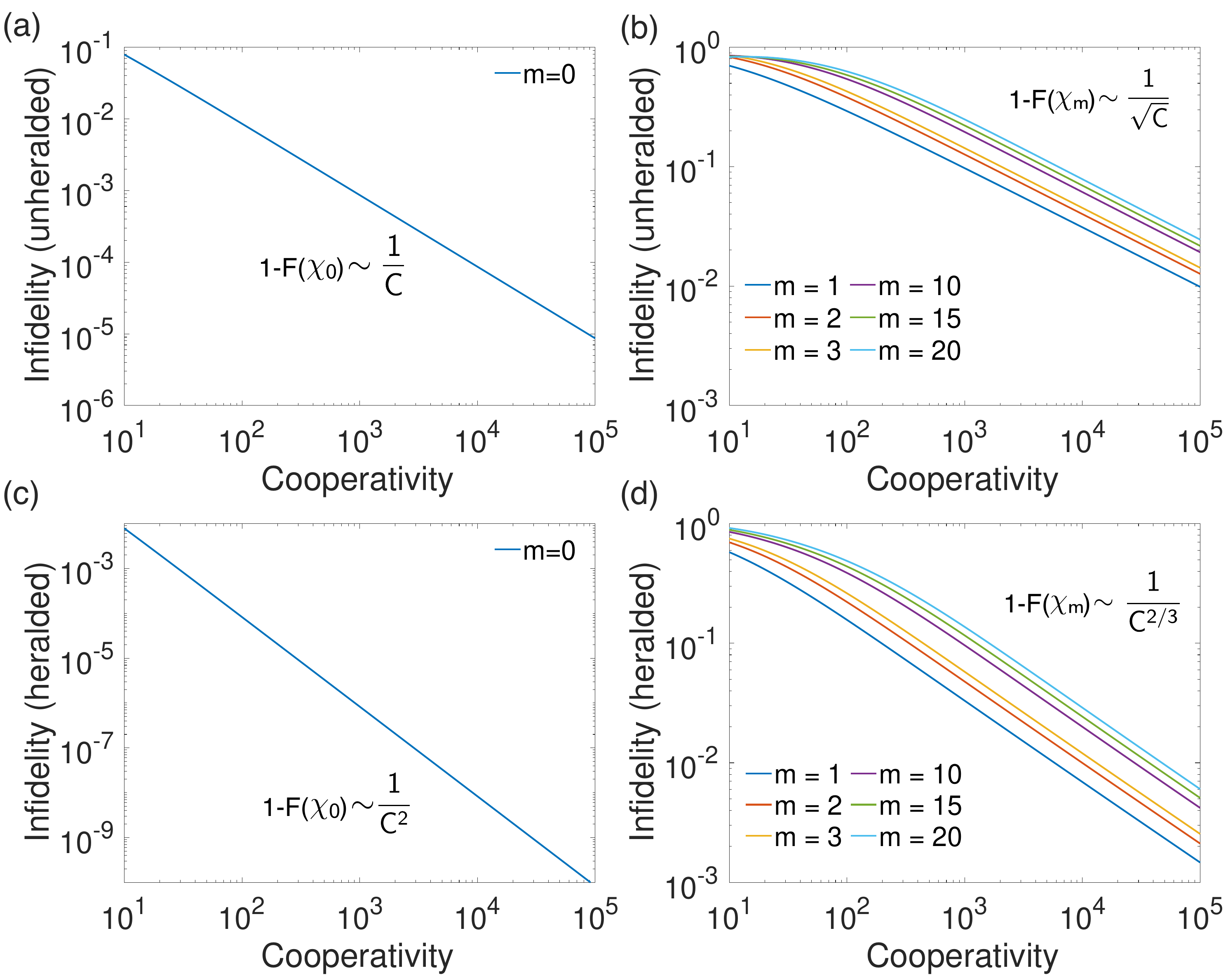}
    \caption{The infidelity of the phase inversion $\chi_m$ for $N=40$ qubits versus the atom-cavity cooperativity $C$ for different $m$, for the case of not heralding [(a) and (b)] and heralding [(c) and (d)]. The initial CSS states considered are:  $\phi_{N,0} \approx 1.5/\sqrt{N}$ for $m=0$, and $\phi_{N,m}=\arccos[(N-2m)/N]$ for for $m \neq 0$. $w=0$, $d=\infty$ for $m=0$, and $d=(C/m)^{1/4}$ ($d=(C/m)^{1/3}$) for $m \neq 0$ for the case not heralding (heralding). }
    \label{fig:F_chi_C}
\end{figure}
In Fig. \ref{fig:F_chi_phi_unherald}, we apply the preceding error model to numerically simulate the fidelity $F(\chi_m)$ in implementing $\chi_m$ on various initial CSS states $\ket{\phi}^N=R^N(\phi)\ket{m=0}$. $F(\chi_m)$ is shown as a function of $\phi$ for realistic cavity parameters \cite{High_C}, for the case of not heralding. There are several important observations to make. The fidelity approaches unity for the cases $\phi=0,\pi$ since these correspond to the Dicke states $\ket{m=0}(\ket{m=N})$, which are not in superposition with any other Dicke states, so there is no loss of coherence. For a given qubit number $N$ and system parameters, $F(\chi_m)$ has a single  minimum in $0 \le \phi \le \pi$, which occurs at $\phi_{N,m}=\arccos[(N-2m)/N]$ for $m\neq 0$. This $\phi_{N,m}$ corresponds to the rotation angle that maximizes the overlap between the $N$-qubit Dicke state $\ket{m}$ and the CSS state $R^N(\phi)\ket{m=0}$. The reason this state achieves a minimum is because $\chi_m$ is an operation that inverts $\ket{m}$ and applies the identity on the rest of the Dicke states; thus the greater the overlap between $\ket{m}$ and $\ket{\phi}^N$, the more sensitive the state will be to the phase inversion, causing greater infidelity \cite{Klaus_Cohen_network_2018}. A particularly important consequence of this remark is that the point of minimum fidelity (at $\phi_{N,m}$) is weakly dependent on $N$ for a given $\chi_m$, as can be seen in the figures. In Appendix \ref{appendix:Dicke_prep}, we show that the overlap between $\ket{m}$ and the CSS state with $\phi_{N,m}$ converges to $\sim (1/2 \pi m)^{1/4}$ in the limit $N \gg 1$ and $m \ll N/2$, which is independent of $N$. Since the infidelity in $\chi_m$ depends on the overlap between the Dicke state $\ket{m}$ and the CSS state, and since this overlap is independent of $N$ for $\phi_{N,m}$, then this implies that the minimum point of $F(\chi_m)$ is approximately independent of $N$ for small $m$.  This makes the proposed implementation of $\chi_m$ scalable in the qubit number, which is an important requirement for quantum state preparation protocols. Conversely, for large $N$, only a small range of $\phi$, around $\phi_{N,m}$, has significant overlap with $\ket{m}$. For $\phi$ much different from $\phi_{N,m}$, the overlap is greatly suppressed in $N$. This makes the fidelity higher for these states, since they are less sensitive to phase inversion errors. This explains the behavior of $F(\chi_m)$ getting narrower in shape as $N$ increases.

We focus our attention on the set of CSS states with minimum fidelity, given by $\phi_{N,m}=\arccos[(N-2m)/N]$, since they give a lowerbound on $F(\chi_m)$. For $m =0$, we numerically find that the CSS states with $\phi_{N,0} \approx 1.5/\sqrt{N}$ have the minimum fidelity for $N$ qubits. \rsub{In Figs. \ref{fig:F_chi_C}(a) and (b)}, we plot the scaling of the infidelity in $\chi_m$ against the atom-cavity cooperativity $C$ on a logarithmic scale for $N=40$ qubits, \rsub{assuming a monochromatic photon (i.e., $w=0$)}. $1-F(\chi_0)$ displays a different power law ($1/C$) than for the other $\chi_m$ ($1/\sqrt{C}$), in agreement with Eqs. \eqref{F0_C} and \eqref{Fm_C}. The different power scaling implies that $\chi_0$ will always have much higher fidelity than other $\chi_m$. In the limit of large $C$, the infidelity increases monotonically and nonlinearly with $m$ for these states. We note that the particular form of the $m$-dependence depends on the class of CSS states studied.

\subsubsection{Phase inversion: heralding on detection of the reflected photon}

\begin{figure*}[!t]
    \centering
    \includegraphics[width=\textwidth]{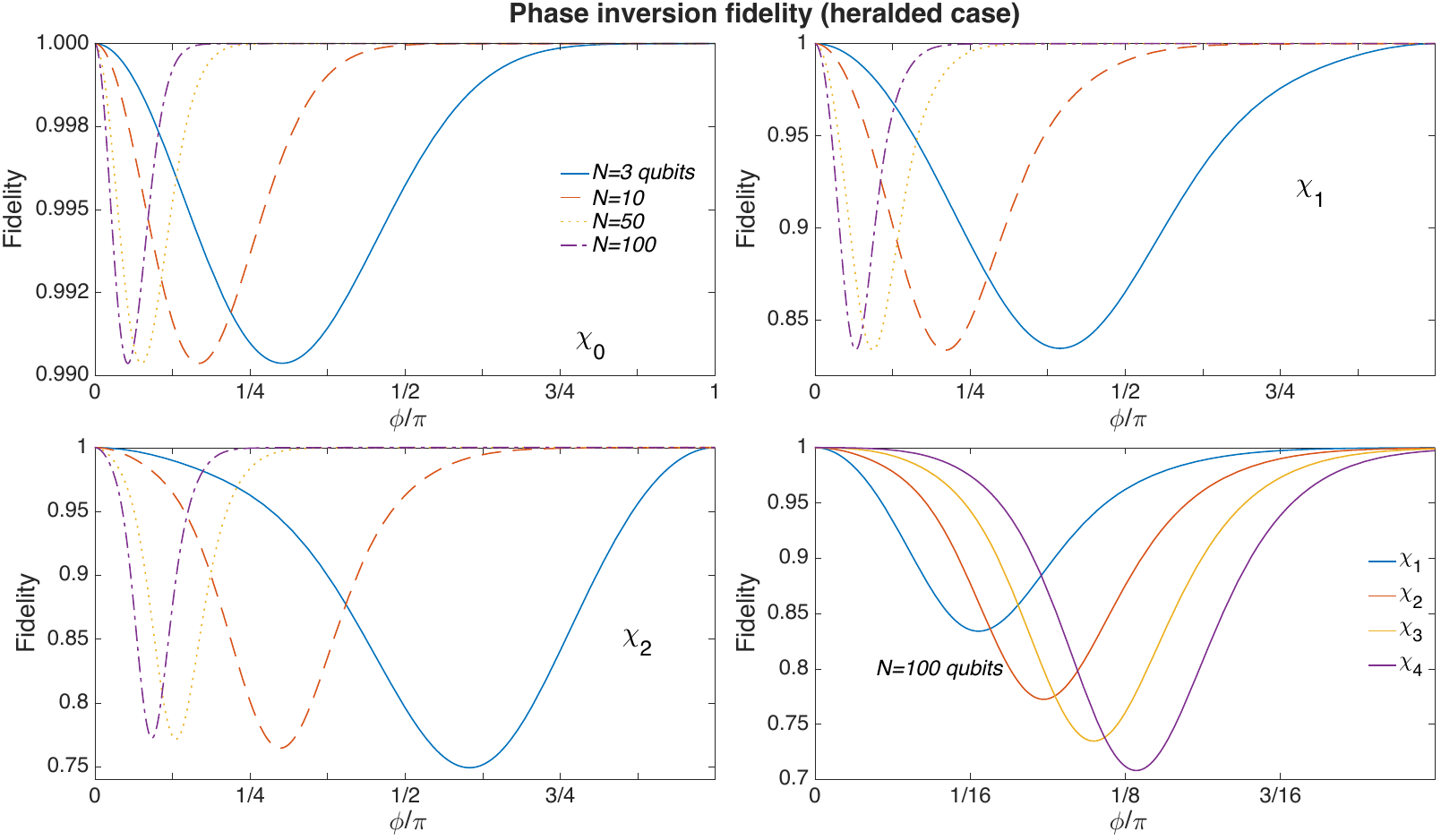}
    \caption{The fidelity in implementing the phase inversion operator $\chi_m$ on various CSS states $\ket{\phi}^N=R^N(\phi)\ket{m=0}$ as function of the global rotation angle $\phi$, for the case of heralding \rsub{[Eqs. \eqref{rho_out_Kr2} and \eqref{superF_norm}]}. The first three panels shows $F(\chi_m) (m=0,1,2)$ for the qubit number $N=3,10,50,$ and $100$. The last panel shows $\chi_m$($m=1,2,3,4$) for $N=100$ qubits. The system parameters are $\kappa=\kappa_r=1$, $\gamma=1$, $g=10$, $C=100$, $w=0.1$, $d=(C/m)^{1/3}$ for $m=1,2,3,4$ and $\Delta=0 (d=\infty)$ for $m=0$. }
    \label{fig:F_chi_phi_herald}
\end{figure*}

Given the same system parameters, we can increase the fidelity of the phase inversion further by heralding on detection of the reflected photon as this eliminates the error due to spontaneous emission. First, consider $\chi_0$. Following a similar analysis as before, the fidelity is maximized at zero atom-cavity detuning with the scaling (cf. Appendix \ref{appendix:error_analysis}):
\begin{equation}\label{F0_herald_C}
1-F(\chi_0)\sim \dfrac{1}{C^2}+w^2, \ \Delta=0
\end{equation}
For $\chi_m$ with $m \neq 0$, there is a trade-off between the sum of two error sources, which go like $1-F \sim 1/d^2+m^2d^4/C^2$. The fidelity is then maximized with the choice $d \approx (C/m)^{1/3}$, which gives the scaling (cf. Appendix \ref{appendix:error_analysis}):
\begin{equation}\label{Fm_herald_C}
1-F(\chi_m) \sim \dfrac{1}{C^{2/3}}+w^2, \ m \neq 0, \ d \sim (C/m)^{1/3}
\end{equation}
i.e., $\Delta=g^2/d\kappa$ with the choice $d\approx(C/m)^{1/3}$ maximizes the fidelity. Therefore, heralding improves the scaling of the fidelity with the atom-cavity cooperativity.

In the case of heralding, the (unnormalized) output atomic state is generated by acting on the input with the averaged reflection Kraus superoperator \rsub{$\mathcal{K}_{\rm r, avg}(\delta \omega_m)$} (cf. Appendix \ref{appendix:error_analysis}):
\begin{equation}\label{rho_out_Kr2}
\bm{\rho}_{\rm out}= \mathcal{K}_{\rm r, avg}\bm{\rho}_{\rm in}, \ \mathcal{K}_{\rm r, avg}= \int d \omega |\Phi(\omega)|^2    K_r(\omega)\otimes K_r^{*}(\omega) 
\end{equation}
The norm of the atomic state gives the success probability, i.e., ${\rm Tr}(\rho_{\rm out})$ measures the efficiency of implementing $\chi_m$. In the superoperator form, the trace is given by the dot product of the vectorization of the identity matrix and the output, i.e., ${\rm Tr}(\rho_{\rm out})={\rm vec}(\mathbb{1}).\bm{\rho}_{\rm out}$ \cite{optically_pumped_atoms}. 
In Appendix \ref{appendix:error_analysis}, we derive the scaling of the success probability in implementing $\chi_m$ with $C$, given by the probability that the photon reflects back and is successfully detected after the atom-cavity interaction:  
\begin{equation}
P_{\rm success}(\chi_m)=1-O(C^{-1/3}), \ d \sim C^{1/3}
\end{equation}
which approaches unity for $C\gg 1$. The fidelity is the overlap between the ideal and nonideal output, normalized by the trace:
\begin{equation}\label{superF_norm}
F=\dfrac{\bm{\rho}_{\rm out}({\rm ideal})\cdot \bm{\rho}_{\rm out}}{{\rm vec}(\mathbb{1}).\bm{\rho}_{\rm out}}
\end{equation}
Fig. \ref{fig:F_chi_phi_herald} shows the numerically simulated fidelity $F(\chi_m)$ of implementing $\chi_m$ on various initial CSS states $\ket{\phi}^N=R^N(\phi)\ket{m=0}$, as a function of $\phi$ for the case of heralding. It shares the same quantitative and qualitative features of the case of not heralding (cf. Fig. \ref{fig:F_chi_phi_herald}), with higher fidelities for all $\chi_m$. For $m \neq 0$, the fidelity increases by about an extra $15 \%$ for the system parameters considered ($C=100$). Fig. \ref{fig:F_chi_C}(c) and (d) shows the scaling of the infidelity in $\chi_m$ against the atom-cavity cooperativity $C$ for $N=40$ qubits, for the CSS states with minimum fidelity. The figure verifies the improved scaling of $F(\chi_m)$ versus $C$ that is provided by heralding.

\subsubsection{Error analysis of the Grover iteration: case of not heralding}

\begin{figure*}[!t]
    \centering
    \includegraphics[width=\textwidth]{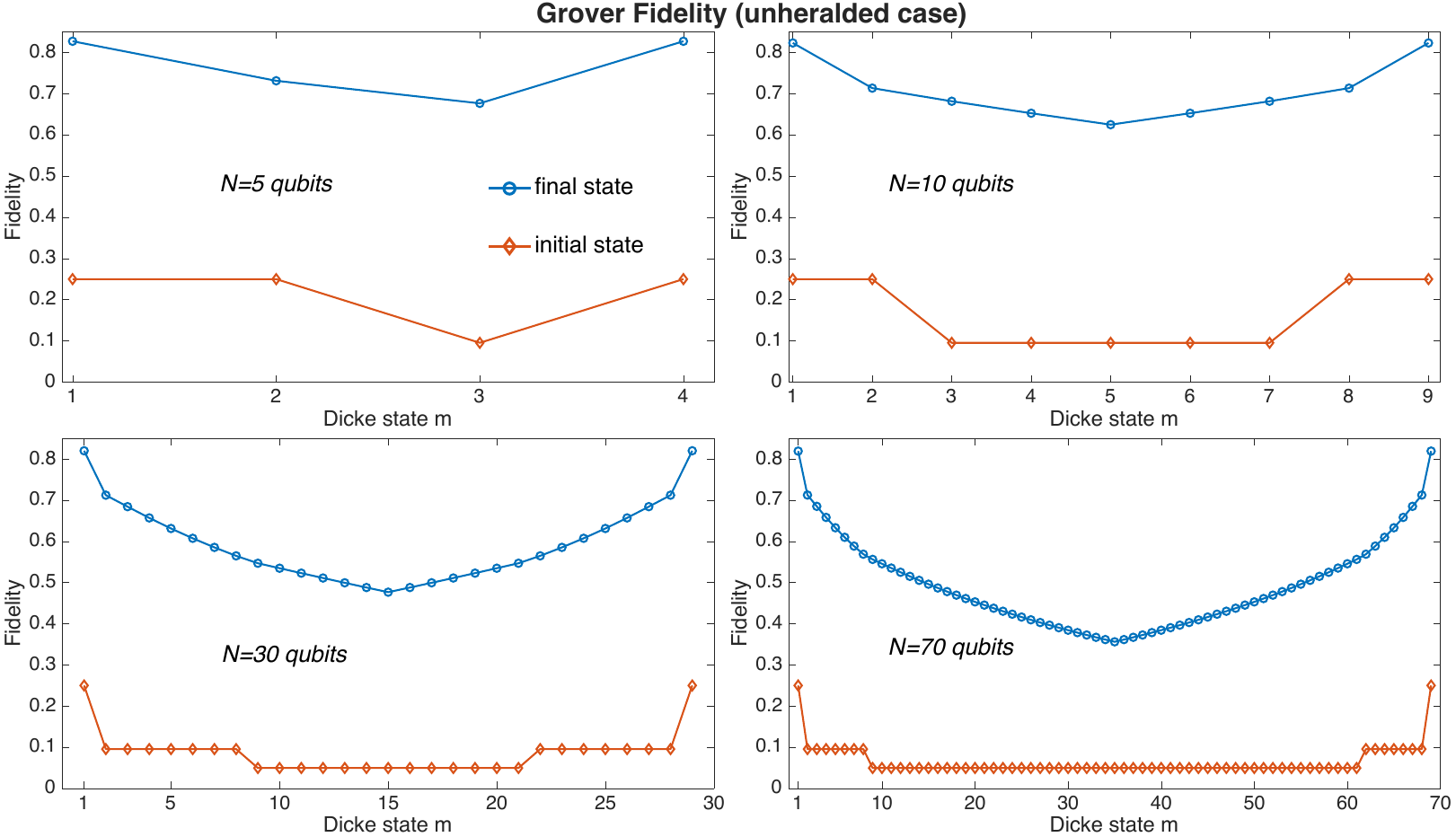}
    \caption{The fidelity in preparing the Dicke state $\ket{m}$ using Grover's algorithm [Eqs. \eqref{G_Dicke_physical} and \eqref{rho_out_Grover}] versus $m=1,2,...,N-1$, for the case of not heralding. Blue (red) curve shows the final (initial) states' fidelity after (before) applying Grover's algorithm. Each figure shows the fidelity for a given qubit number $N$. The system parameters used are $C=100,$ $w=0.1$, and $\kappa_t=\kappa_m=0$. For the implementation of $\mathcal{K}_{\rm avg}(0)$ in Eq. \eqref{G_Dicke_physical}, we set $ d=\infty\leftrightarrow \Delta=0 $. For \rsub{$\mathcal{K}_{\rm avg}(\delta \omega_m)$}, $d$ is numerically optimized (around $\sim (C/m)^{1/4}$) to maximize the fidelity. The number of steps $k$ (and the initial state) has been optimized to maximize the fidelity.}
    \label{fig:F_Grover_Dicke_m_unherald}
\end{figure*}

The ideal unitary Grover iteration to prepare the Dicke state $\ket{n}$ is $G=R(\phi)^{\otimes N}\chi_0R(-\phi)^{\otimes N}\chi_n$. Through the process of vectorization, we construct the corresponding ``physical'' Grover iteration (cf. Appendices \ref{appendix:kraus} and \ref{appendix:error_analysis}):
\rsub{\begin{equation}\label{G_Dicke_physical}
\mathcal{G}=  \mathcal{R}(\phi)  \mathcal{K}_{\rm avg}(0) \mathcal{R}(-\phi) \mathcal{K}_{\rm avg}(\delta \omega_n)
\end{equation}}
where \rsub{$\mathcal{K}_{\rm avg}(\delta \omega_n)$} is the physical approximation of $\mathcal{X}_n=\chi_n \otimes \chi_n$, implemented by a photon wavepacket with a photon-cavity detuning \rsub{$\Omega_c=\delta\omega_n$} hitting the cavity, and $\mathcal{K}_{\rm avg}(0)$ implements $\mathcal{X}_0$ with $\Omega_c=0$. $\mathcal{R}(\phi)=R(\phi)^{\otimes N} \otimes R(\phi)^{\otimes N}$ is the rotation superoperator in the Liouville space. The matrix elements of the global rotation in the Dicke basis is given by the Wigner small $d$-matrix, i.e., $\bra{m}R(\phi)^{\otimes N} \ket{n}=d^{j=N/2}_{m-N/2,n-N/2}(\phi)$. The output after $k$ steps can be obtained by applying the physical Grover superoperator $k$ times on the input
\begin{equation}\label{rho_out_Grover}
\bm{\rho}^{(k)}_{\rm out}=\mathcal{G}^k\bm{\rho}_{\rm in} 
\end{equation}
where the cases of heralding versus not heralding can be handled by using the Kraus operators of Eqs. \eqref{rho_out_Kr2} or \eqref{vec_rho_unherald}, respectively.

\begin{figure*}[!t]
    \centering
    \includegraphics[width=\textwidth]{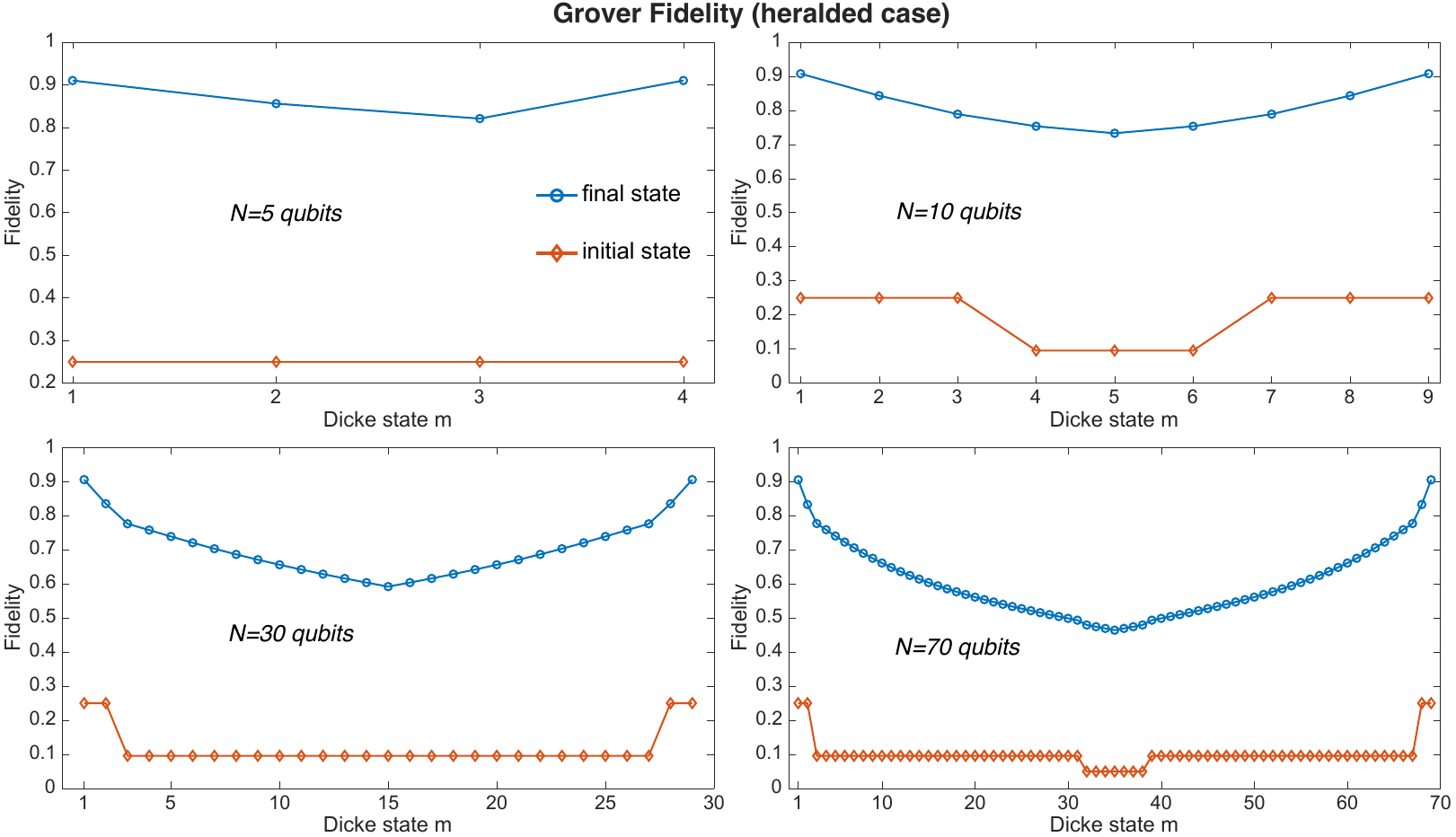}
    \caption{The fidelity in preparing the Dicke state $\ket{m}$ using Grover's algorithm [Eqs. \eqref{G_Dicke_physical} and \eqref{rho_out_Grover}] versus $m=1,2,...,N-1$, for the case of heralding. Blue (red) curve shows the final (initial) state's fidelity after (before) applying Grover's algorithm. Each figure shows the fidelity for a given qubit number $N$. The system parameters used are $C=100$ and $w=0.1$, and $\kappa_t=\kappa_m=0$. For the implementation of $\mathcal{K}_{\rm avg}(0)$ in Eq. \eqref{G_Dicke_physical}, we set $ d=\infty\leftrightarrow \Delta=0 $. For \rsub{$\mathcal{K}_{\rm avg}(\delta \omega_m)$}, $d$ is numerically optimized (around $\sim (C/m)^{1/3}$) to maximize the fidelity. The number of steps $k$ (and the initial state) has been optimized to maximize the fidelity.}
    \label{fig:F_Grover_Dicke_m_herald}
\end{figure*}

\begin{figure*}[!t]
    \centering
    \includegraphics[width=\textwidth]{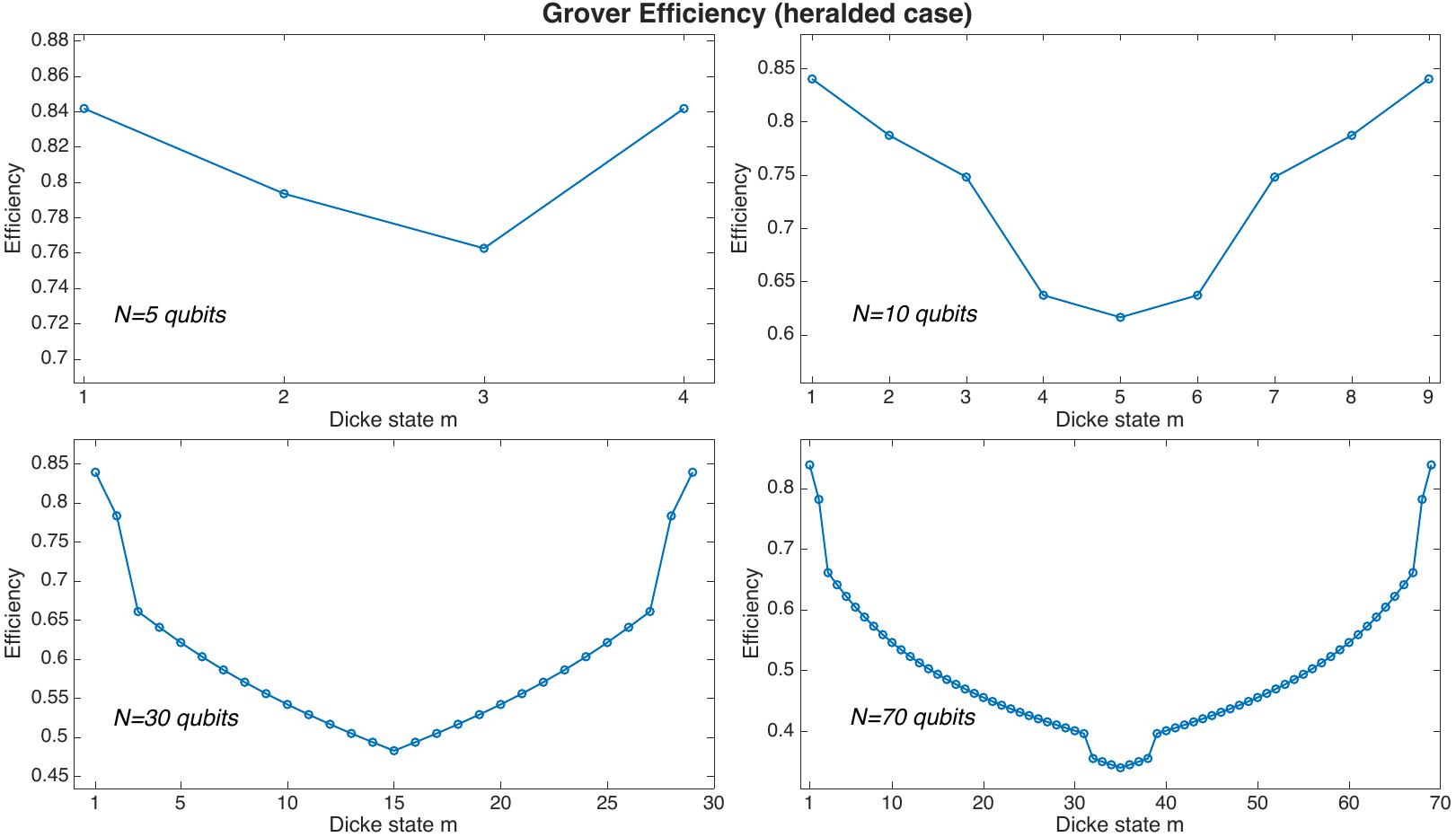}
    \caption{The efficiency (success probability) in preparing the Dicke state $\ket{m}$ using Grover's algorithm [Eqs. \eqref{G_Dicke_physical} and \eqref{rho_out_Grover}] versus $m=1,2,...,N-1$, for the case of heralding. same system parameters as Fig. \ref{fig:F_Grover_Dicke_m_herald}.}
    \label{fig:P_Grover_Dicke_m_herald}
\end{figure*}

In Sec. \ref{subsec:Dicke_exact_prep}, where no errors were assumed, the optimal strategy was to prepare the desired Dicke state in the least number of steps $k$ with perfect fidelity. In this section, where errors exist, we optimize for $k_{\rm opt}$ and $\Delta_{\rm opt}$ (or equivalently $d_{\rm opt}$) that maximizes the fidelity for every Dicke state $\ket{m}$. The corresponding initial CSS state will be given by $\phi$ that satisfies Eq. \eqref{grover_dicke_overlap_condition} with $k=k_{\rm opt}$. One can further optimize for $\phi$ independent of $k$, which is not done here since we found that it gives similar results to optimizing for $k$ alone. Dicke states with $m<N/2$ are prepared just as outlined. For any Dicke state with $m>N/2$, we prepare the Dicke state $\ket{N-m}$, and then apply a global $X^{\otimes N}$ gate that flips all the qubits to the desired target $\ket{m}$. This is advantageous, compared to directly preparing the Dicke states $m>N/2$, because the infidelity of $\chi_m$ is large for large $m$.

First, we study the case of not heralding. Fig. \ref{fig:F_Grover_Dicke_m_unherald} shows the fidelity in preparing the Dicke states $\ket{m}$ versus $m$, for realistic cavity parameters. The Blue (red) curves show the fidelity after (before) applying Grover's algorithm. For the Dicke states $\ket{m}$ that require $k=1,2,$ and $3$ steps, the global rotation angle $\phi$ has been chosen to make the initial states have an overlap $\sin^2(\pi/(2[2k+1])) = 0.25, 0.096,$ and $0.0495$ with $\ket{m}$. In the error-free case, this will lead to unit fidelity after applying Grover's algorithm (see Sec. \ref{sec:exact_prep}). Dicke states with larger $m$ have lower fidelity, primarily owing to the fidelity of $\chi_m$ decreasing with $m$, as well as error accumulation from the application of more than a single Grover step. We find that $k_{\rm opt}$ that maximizes the fidelity is not necessarily the smallest $k$ that prepares a given Dicke state. 

\begin{figure}[!t]
    \centering
    \includegraphics[width=0.45\textwidth]{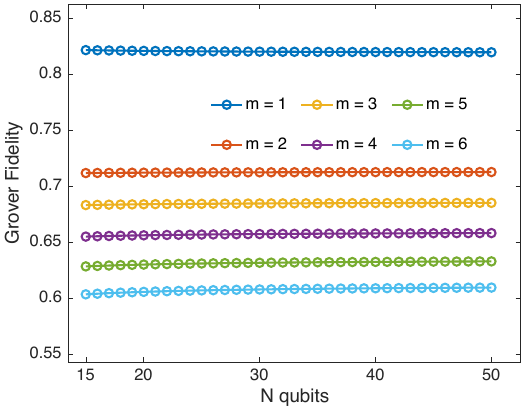}
    \caption{Fidelity to prepare a given Dicke state, for small $m$, versus the number of qubits $15\le N\le 50$, for the case of not heralding. Same parameters as Fig. \ref{fig:F_Grover_Dicke_m_unherald}.}
    \label{fig:F_Grover_Dicke_N}
\end{figure}

Previously, we have remarked that the fidelity of $\chi_m$ only depends weakly on $N$ in the limit $N \gg 1$ and $m \ll N/2$. In Fig. \ref{fig:F_Grover_Dicke_N}, we plot the fidelity in preparing the Dicke states as a function of the number of qubits $N$, for the Dicke states from $m=1$ to $m=6$. We see that, indeed, the Grover fidelity does not change with the number of qubits for a given system parameter. This comes from the fact that the for a given Dicke state $\ket{m}$, the corresponding initial CSS state has essentially the same Dicke states distribution, independent of $N$.

\begin{figure}[!t]
    \centering
    \includegraphics[width=0.45\textwidth]{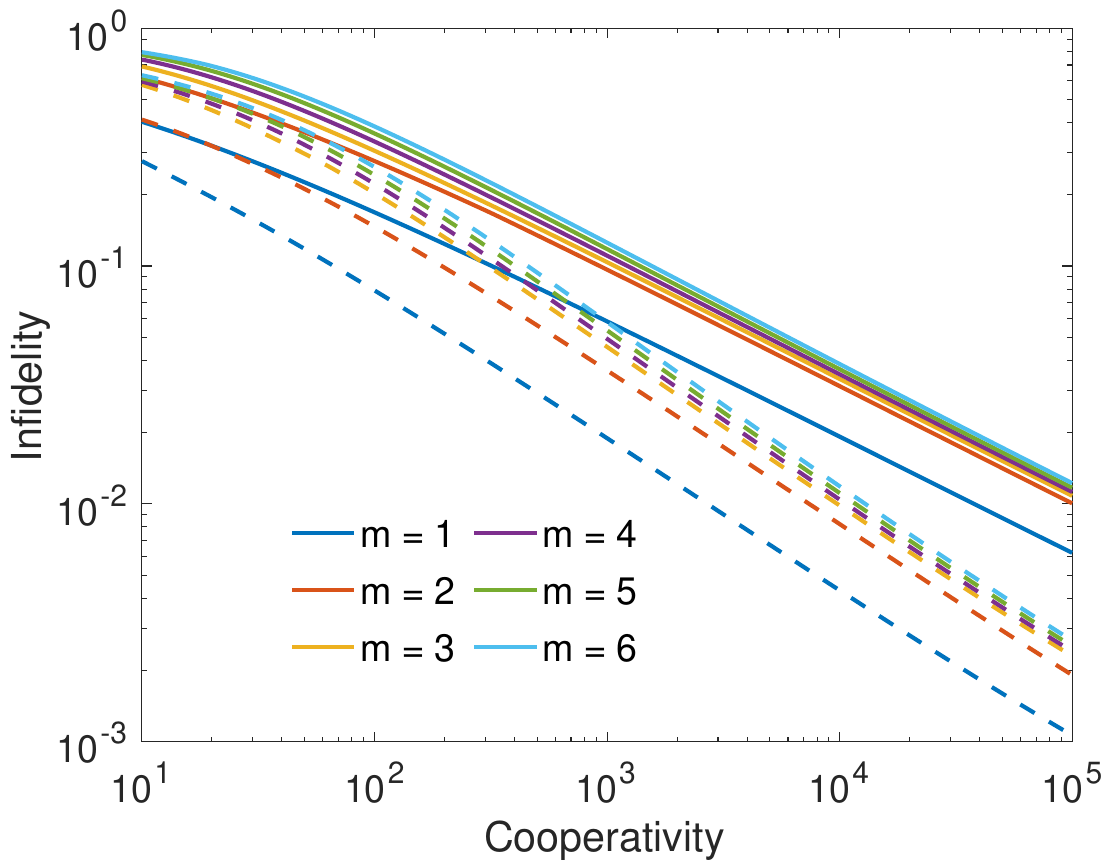}
    \caption{Scaling of the infidelity with the atom-cavity cooperativity in preparing a given Dicke state for $N=15$ qubits, for the cases of heralding (dashed) and not heralding (solid). Same parameters as Fig. \ref{fig:F_Grover_Dicke_m_unherald} but with $w=0.01$.}
    \label{fig:F_Grover_Dicke_C}
\end{figure}

\subsubsection{Error analysis of the Grover iteration: case of heralding}

In the case of heralding, the Grover iteration will be given only by the averaged reflection Kraus operator
\rsub{
\begin{equation}\label{G_Dicke_physical_herald}
\mathcal{G}=  \mathcal{R}(\phi)  \mathcal{K}_{\rm r, avg}(0) \mathcal{R}(-\phi) \mathcal{K}_{\rm r, avg}(\delta \omega_n)
\end{equation}}
where the unnormalized output after $k$ steps is  $\bm{\rho}^{(k)}_{\rm out}=\mathcal{G}^k\bm{\rho}_{\rm in}$. If any photons are not detected due to spontaneous emission or loss, then the protocol fails and needs to be repeated. Thus the norm of the output state gives the efficiency or success probability of Grover's algorithm.

Fig. \ref{fig:F_Grover_Dicke_m_herald} shows the fidelity in preparing the Dicke states $\ket{m}$ versus $m$, for the case of heralding. We see that heralding can significantly boost the fidelity compared to the case of not heralding (cf. Fig. \ref{fig:F_Grover_Dicke_m_unherald}), increasing the fidelity by an extra $10\%$. Since this is a heralding scheme, we also compute its efficiency (i.e., success probability). Fig. \ref{fig:P_Grover_Dicke_m_herald} shows that, for the system parameters considered ($C=100$), the success probability can still be significant ($\sim 50 \%$) for Dicke states with small $m$. As $m$ increases, the probability of spontaneous emission increases, decreasing the efficiency. Moreover, Dicke states requiring more Grover steps will have lower success probability. Both the success probability and fidelity  increase with $C$. Therefore, heralding allows for boosting the fidelity without sacrificing too much efficiency.

Finally, we study the scaling of the infidelity with $C$ in Grover's algorithm. Assuming no errors coming from the global rotations, the Grover iteration inherits the infidelity coming from $\chi_m$ and $\chi_0$. For the not heralded scheme, $\chi_m$ and $\chi_0$ scale as $1/\sqrt{C}$ and $1/C$, and for the heralded scheme as $1/C^{2/3}$ and $1/C^2$, respectively. For $C \gg 1$, the errors due to $\chi_m$ dominate that of $\chi_0$. Fig. \ref{fig:F_Grover_Dicke_C} shows the scaling of the infidelity with the cooperativity for $N=15$ qubits for the not heralded (solid) and heralded (dashed) cases. The figure confirms numerically the scaling behavior, i.e., as the cooperativity changes by four orders of magnitude from $10$ to $10^{5}$, the error drops by two [$(10^4)^{-1/2} \sim 10^{-2}$] and three [$(10^{4})^{-2/3}\sim 10^{-3}$] orders of magnitude for the not heralded and heralded schemes, respectively. Reaching an infidelity of the order $10^{-2}$ requires a cooperativity of the order $10^5$ and $10^4$ for the not heralded and heralded schemes, ignoring other error sources.

\section{Discussion}\label{sec:discuss}

It is well-known that Grover's algorithm achieves a quadratic speedup in searching for an unknown element in a database \cite{Grover1997}. Grover's algorithm offers the same quadratic speedup in state preparation compared to probabilistic schemes that use projective measurements. In the cavity carving schemes, e.g., \cite{Chen2015carving, ramette2024counterfactual}, the atom-cavity interaction with light acts as a projective measurement, projecting an initial product state into a Dicke state $\ket{m}$. The success probability of such projection is the overlap (square of the inner product) of the initial state with the Dicke state. To maximize the probability, one can choose to rotate the initial product state to have the largest overlap possible with $\ket{m}$, which will be given by $\sim 1/m^{1/2}$ (and $1/N^{1/2}$ for $m=N/2)$. To make the state preparation ``deterministic'', the cavity carving protocol merely needs to be repeated, on average, $m^{1/2}$ times. On the other hand, Grover's algorithm only requires $k \sim m^{1/4}$ steps for state preparation, which is a quadratic speedup over probabilistic carving schemes. The quadratic speedup comes precisely from the fact that $k$ depends on the overlap amplitude (not its square) in Grover's algorithm. 

In the algorithm part of this work, it was shown that Grover's algorithm can always prepare a Dicke state $\ket{m}$ perfectly in $O(m^{1/4})$ steps.  In particular, the Dicke states $m=N/2$ are the most expensive to prepare, requiring $O(N^{1/4})$ steps. Due to this favorable scaling, it was shown that any Dicke state with $3\le N \le 500$ qubits can be prepared within four steps. The W state, in particular, can always be prepared in a single step, independent of the qubit number. These conclusions and scalings are consistent with findings from other proposals \cite{Cirac_Dicke_2024}. We have also presented a two-step Grover's algorithm to efficiently prepare the GHZ states in $O(N^{1/4})$ steps, starting from a Dicke state. Our work shows that the resource for GHZ state preparation is comparable to that of the Dicke states. An approximate two-step Grover's algorithm for Cat states has also been presented. We leave it for future work to prove the scaling of the steps for Cat states with the number of qubits. A key idea in all the proposed protocols above is the identification of trivial product states, prepared by global rotations, that allows perfect transformation to interesting quantum states by a small integer number of single photon scattering events. In these cases, there is only a single parameter, the global rotation angle $\phi$, that needs to be optimized in contrast to previous amplitude amplification schemes that employ $O(N)$ optimization parameters \cite{Grover_geometric_2020,Grover_2023}. 

The favorable $O(N^{1/4})$ scaling of Grover's algorithm only applies if we have a  physical mechanism that realizes each Grover iteration with resources that do not grow with the qubit number. We have identified cavity-meditated interactions between atomic ensembles and a single photon as a promising candidate. The conditional change of sign in Grover's algorithm is realized by the phase shift the atoms and cavity impart on the photon and {\it vice versa} due to the product state character of each quantum state component. Crucially, the number of photon reflections per single Grover step is independent of the qubit number, i.e., it is two and three photons for Dicke and GHZ states, respectively. This means that only a few $O(N^{1/4})$ photon reflections, interleaved with global rotations, are required to prepare Dicke and GHZ states. Thus, the proposed physical mechanism requires resources scaling favorably as $O(N^{1/4})$, compared to previous schemes that scale (super) linearly. 

We have identified the dominant sources of error and  provided analytical and numerical error analyses of the proposed physical implementation. Since it relies on reflection, the protocol is sensitive to spatial mode matching between the cavity and the photonic modes. Achieving a fidelity beyond $F \ge 99 \%$ would require a mode matching efficiency higher than $ 99 \%$. Another requirement is that the bandwidth of the photon wavepacket needs to be small relative to that of the cavity, which is achievable in typical experiments \cite{Reiserer_2014}. For given cavity parameters, the fidelity decreases with $m$. This implies that the present scheme is able to prepare Dicke states with small $m$ (e.g., the W state) with relatively higher fidelity, compared to Dicke states with larger $m$ or the GHZ state. The scaling of the errors in the protocol is $O(C^{-1/2})$, which could be improved to $O(C^{-2/3})$ by heralding on detection of the reflected photon. For current state-of-the art optical cavities $C=20-100$ \cite{Lukin_carve_2024,High_C}, we obtain fidelities in the range of $70-80 \%$ and $80-90 \%$ for the not heralded and heralded cases, for Dicke states with small $m$.  Achieving a significantly higher fidelity, i.e., $F \ge 99 \%$, requires high cooperativity in the order of $C=10^{4}-10^{5}$. While experimentally demanding, cooperativities in the order of $10^{3}-10^{4}$ should be achievable \cite{Very_High_C_2010}. In the absence of pure single-photon sources, the protocol can still be implemented using weak coherent pulses and heralding on detection of a single reflected photon. 

The proposed physical implementation of Grover's algorithm is not restricted to atom-cavity systems. It can be realized in other physical systems with a Hamiltonian formally equivalent to the Jaynes/Tavis-Cummings model, e.g., an ensemble of Rydberg atoms \cite{Rydberg_optimal_control_2016,Grover_Klaus_Mark_2011,Klaus_toffoli_2020,Dicke_partition_2021}, superconducting qubits in circuit QED \cite{Circuit_QED_2009,Toffoli_superconducting_2020}, and trapped ions \cite{Multiqubit_Ion_Grover_2001, Ion_tavis_2007}. An interesting future research direction is to investigate the scaling and behavior of errors in these different physical systems.

The present work begs an interesting research question on the algorithm and physical implementation sides. On the algorithm side, it would be worth investigating the possibility to efficiently prepare other classes of entangled states by replacing the phase inversion operation ($\chi_m$) with some other non-linear phase gate \cite{Grover_geometric_2020,Grover_2023}. Specifically, which non-linear phase gates, interleaved with global rotations, can prepare which class of entangled states efficiently, as a function of the number of qubits? Once such nonlinear phase gates are identified, finding physical platforms and mechanisms to implement them in a resource-efficient way would have important theoretical and practical consequences for quantum state preparation and beyond.

\begin{acknowledgments}

This material is based upon work supported by the U.S.
Department of Energy Office of Science National Quantum
Information Science Research Centers as part of the Q-NEXT
center, as well as support from NSF Grant No. 2016136 for the
QLCI center Hybrid Quantum Architectures and Networks,
and NSF Grant No. 2228725.  KM acknowledges funding from the Carlsberg Foundation
through the “Semper Ardens” Research Project QCooL and from
the Danish National Research Foundation (Center of Excellence “Hy-Q,” grant number DNRF139).
\end{acknowledgments}

\bibliography{carve_PRA,qc_refs}

\appendix

\section{Grover's algorithm for Dicke states}

\subsection{Analysis of the Dicke  states preparation}\label{appendix:Dicke_prep}

We wish to calculate the fidelity of the Dicke states prepared by the modified Grover's algorithm in Sec. \ref{subsec:modified_grover_Dicke} in the limit of many qubits. $\ket{\psi_i}=R(\phi)^{\otimes N}\ket{0}^{\otimes N}$ can be decomposed as $\sin (\theta/2) \ket{m} +  \cos (\theta/2)   \ket{m_{\perp}}$ where the overlap between the initial state and $\ket{m}$ is
\begin{equation} \label{overlap_max}
\sin (\theta/2)=\bigg(1-\dfrac{m}{N} \bigg)^{(N-m)/2} \bigg(\dfrac{m}{N}\bigg)^{m/2}\sqrt{{N\choose m}}
\end{equation}
After applying $k$ iterations, the fidelity becomes $F_{\rm Grover}=\sin ([2k+1] \theta/2)^2$, i.e.,
\begin{align}
&F_{\rm Grover}=\\ & \nonumber \sin^2\bigg([2k+1] \arcsin \bigg \{ \sqrt{\bigg(1-\dfrac{m}{N} \bigg)^{N-m} \bigg(\dfrac{m}{N}\bigg)^{m}{N\choose m}  } \bigg \} \bigg)
\end{align}
Imposing the condition $\sin ([2k+1] \theta/2)=1$ on the equation above and solving for $k$ gives:
\begin{align}\label{k_exact_dicke_modified}
k=   \dfrac{\pi}{  4\arcsin  \sqrt{\bigg(1-\dfrac{m}{N} \bigg)^{N-m} \bigg(\dfrac{m}{N}\bigg)^{m}{N\choose m}  }     }  -\dfrac{1}{2}
\end{align}
This is an exact result for any $N$ and $m$. Next, we find the scaling in the limit of many qubits, i.e., $N \rightarrow \infty$. Using the identity, $\lim_{N \to \infty} {N\choose m}/N^m \rightarrow 1/m!$, $\sin( \theta/2)$ becomes
\begin{equation}
\sin( \theta/2)= \bigg(1-\dfrac{m}{N} \bigg)^{(N-m)/2} m^{m/2} \sqrt{\dfrac{1}{m!}}  
\end{equation}
Moreover,
\begin{equation}
\lim_{N \to \infty} \bigg(1-\dfrac{m}{N} \bigg)^{(N-m)/2}= e^{-m/2}
\end{equation}
Therefore, the fidelity becomes:
\begin{equation}\label{fidelity_dicke_large_m}
F_{\rm Grover}=\sin^2\bigg([2k+1] \arcsin \bigg \{ e^{-m/2} m^{m/2} \sqrt{\dfrac{1}{m!}}  \bigg \} \bigg)
\end{equation}
For $m \gg 1$, this can be approximated by the Stirling formula $m! \approx e^{-m}m^m \sqrt{2 \pi m}$:
\begin{equation}
F_{\rm Grover}=\sin^2\bigg([2k+1] \arcsin \bigg ( \dfrac{1}{2 \pi m}  \bigg )^{1/4} \bigg)   
\end{equation}
Using Eq. \eqref{k}, the number of steps required to prepare $\ket{m}$ becomes
\begin{equation}
 k= \dfrac{\pi}{ 4 \arcsin [(1/2 \pi m)^{1/4}]}-\dfrac{1}{2}, \ \ N \gg 1,  \ m \ll N/2 
 \end{equation}
\subsection{Condition to prepare a Dicke state exactly in a few steps}\label{overlap_appendix}

Here it is shown that that the overlap condition
\begin{equation}\label{overlap}
  \braket{m|\psi_i}=  \sqrt{{N\choose m}} \cos^{N-m} (\phi/2) \sin^{m} (\phi/2)=x
\end{equation}
has a solution for a given $N,m,$ and $x$ if they satisfy the condition:
\begin{equation}\label{overlap_sol}
{N\choose m} \bigg(1-\dfrac{m}{N}\bigg)^{N-m}\bigg(\dfrac{m}{N}\bigg)^m\ge x^2
\end{equation}
Observe that the LHS of the overlap condition \eqref{overlap} is a product of sines and cosines. For a given $N$ and $m$, its maximum value is
\begin{equation}
\sqrt{{N\choose m} \bigg(1-\dfrac{m}{N}\bigg)^{N-m}\bigg(\dfrac{m}{N}\bigg)^m}
\end{equation}
Therefore, a solution exists if the maximum value of the LHS is equal to or greater than the constant $x$ (graphically, this means the LHS plotted as a function of $\phi$ intersects the constant function $x$): 
\begin{equation}
\sqrt{{N\choose m} \bigg(1-\dfrac{m}{N}\bigg)^{N-m}\bigg(\dfrac{m}{N}\bigg)^m}\ge x
\end{equation}
Squaring this equation readily leads to Eq. \eqref{overlap_sol}.

\subsection{The $m=1$ Dicke state can always be prepared in one step for any qubit number $N$}\label{W_one_step_appendix}

A Dicke state $m$ with $N$ qubits can be prepared in one step if $N$ and $m$ satisfy Eq. \eqref{Dicke_condition_solution} with $k=1$:
\begin{equation}
{N\choose m} \bigg(1-\dfrac{m}{N}\bigg)^{N-m}\bigg(\dfrac{m}{N}\bigg)^m\ge \dfrac{1}{4}
\end{equation}
For $m=1$, this reduces to:
\begin{equation}
 \bigg(1-\dfrac{1}{N}\bigg)^{N-1}\ge \dfrac{1}{4}
\end{equation}
The LHS a monotonically decreasing function of $N$ taking the values from $4/9=0.\periodfl{4}$ to $1/e \approx 0.368$ for $N \in [3,\infty)$. Therefore, the condition above is satisfied for all $N$, which concludes the proof.

\subsection{The minimum number of steps to prepare the Dicke state $m=N/2$}\label{N_over_2_steps_appendix}

A Dicke state $m$ with $N$ qubits can be prepared exactly in $k$ steps if $N$ and $m$ satisfy
\begin{equation}
{N\choose m} \bigg(1-\dfrac{m}{N}\bigg)^{N-m}\bigg(\dfrac{m}{N}\bigg)^m\ge \sin^2 \bigg( \dfrac{\pi}{2(2k+1) }\bigg)
\end{equation}
For $m=N/2$ this becomes:
\begin{equation}
\dfrac{1}{2^N} {N\choose N/2} \ge \sin^2 \bigg( \dfrac{\pi}{2(2k+1) }\bigg)
\end{equation}
Solving for $k$, we get
\begin{equation}
k \ge \dfrac{\pi}{4 \arcsin(\sqrt{{N\choose N/2}}/2^{N/2})}-\dfrac{1}{2}
\end{equation}
or approximately
\begin{equation}
k \ge 0.88N^{1/4}-\dfrac{1}{2}
\end{equation}
This is precisely the scaling obtained by applying the ``normal'' Grover iteration from Subsec. \ref{subsec:normal_grover} to prepare $\ket{m=N/2}$.

\section{Grover's algorithm for the GHZ and Cat states}\label{GHZ_appendix}

\subsection{Analysis of the GHZ state preparation}\label{GHZx_appendix}

First, we analyze the Grover's algorithm to prepare the GHZ state, starting from the Dicke state in the $x$-basis $\ket{N/2}_x$. The overlap between the GHZ state and a Dicke state in the $x$-basis is given by
\begin{equation}
\braket{m|H^{\otimes N}| {\rm GHZ}}=\dfrac{1}{2^{(N-1)/2}}\sqrt{{N\choose m}}
\end{equation}
Where $m$ needs to be even here to have a finite overlap with the GHZ state. Here, we take $m=N/2$ as our initial state. If $m=N/2$ is not even then we choose an even $m$ close to $N/2$. After $k$ steps of applying the Grover iteration $G=H^{\otimes N}\chi_{N/2}H^{\otimes N}\chi_0 \chi_N$ on the initial state $\ket{N/2}$, the fidelity becomes [Eq. \eqref{fidelity_k}]
\begin{equation}
F_{\rm Grover}=\sin^2\bigg[(2k+1)\arcsin\bigg(\dfrac{\sqrt{{N\choose N/2}}}{2^{(N-1)/2}}\bigg)\bigg]
\end{equation}
Using Eq. \eqref{k}, the number of steps required to prepare the GHZ state becomes
\begin{equation}
k=\dfrac{\pi}{4 \arcsin(\sqrt{{N\choose N/2}}/2^{(N-1)/2})}-\dfrac{1}{2}
\end{equation}
which for large $N$ is approximately given by
\begin{equation}
F_{\rm Grover}=\sin^2\bigg[(2k+1)\arcsin\bigg(\dfrac{8}{ \pi N}\bigg)^{1/4}\bigg]
\end{equation}
and
\begin{equation}
 k= \dfrac{\pi}{ 4 \arcsin [(8/\pi N)^{1/4}]}-\dfrac{1}{2} \approx 0.62 N^{1/4}-\dfrac{1}{2}, \ \ N \gg 1  
 \end{equation}

\subsection{Preparing the GHZ state using Dicke states in the $y$-basis}\label{GHZyz_appendix}

Alternatively, we can start with a Dicke state in the $y$-basis as our initial state, i.e., $\ket{\psi_i}=R(-\pi/2)^{\otimes N}\ket{N/2}$. Applying the following Grover iteration repeatedly on that state would yield the GHZ state  
\begin{align}
&\ket{\psi_i}=R(-\pi/2)^{\otimes N}\ket{N/2}=\ket{N/2}_y,\\ &  \nonumber \ket{\psi_t}=\ket{\rm GHZ},
\\ & \nonumber G=R(-\pi/2)^{\otimes N}\chi_{N/2} R(\pi/2)^{\otimes N}\chi_0 \chi_N
\end{align}
where $\chi_0 \chi_N$ flips the phase of the GHZ state, and $R(-\pi/2)^{\otimes N}\chi_{N/2} R(\pi/2)^{\otimes N}$ flips the phase of the initial state. Note that for even (odd) $N$, the GHZ state has finite overlap with the even (odd) $m$ Dicke states in the $y$-basis. Thus, for an even (odd) $N$, we need to choose an even (odd) $m$ close to $N/2$ as the initial state. To prepare the GHZ perfectly in integer steps, the phase inversion operations need to be modified by a phase $\alpha$ as described in Subsec. \ref{subsec:mod_phase}.

\subsection{Overlap condition for preparing the GHZ state exactly}\label{GHZ_condition_solution_appendix}

Next, we derive the overlap condition to prepare the GHZ state exactly in $k$ steps exactly without using a modified phase $\alpha$. The overlap between a Dicke state rotated by $R(-\phi)^{\otimes N}$ and the GHZ state is
\begin{align}
&\braket{m|R(\phi)^{\otimes N}|\rm GHZ}= \dfrac{1}{\sqrt{2}} \sqrt{{N\choose m}} \times \\ & \nonumber  \bigg(  \cos^{N-m} (\phi/2) \sin^{m} (\phi/2)+ \\ & \nonumber (-1)^{N-m} \sin^{N-m} (\phi/2) \cos^{m} (\phi/2) \bigg) 
\end{align}
To end up with the GHZ state, we need to choose as our initial state an $m$ that has the same overlap with the two components $\ket{0}^{\otimes N}$ and $\ket{1}^{\otimes N}$:
\begin{align}
&\braket{m|R(\phi)^{\otimes N}|0}^{\otimes N}=  \\ & \nonumber \dfrac{1}{\sqrt{2}} \sqrt{{N\choose m}}    \cos^{N-m} (\phi/2) \sin^{m} (\phi/2) 
\end{align}
\begin{align}
&\braket{m|R(\phi)^{\otimes N}|1}^{\otimes N}=  \\ & \nonumber \dfrac{1}{\sqrt{2}} \sqrt{{N\choose m}}    (-1)^{N-m} \sin^{N-m} (\phi/2) \cos^{m} (\phi/2) 
\end{align}
This is achieved by choosing an even $m=N/2$. Therefore, the overlap condition to prepare the GHZ state exactly after $k$ steps, starting from $\ket{N/2}$, becomes
\begin{align}
&\braket{N/2, -\phi|\rm GHZ}= \\& \nonumber \sqrt{2{N\choose N/2}}  \cos^{N/2} (\phi/2) \sin^{N/2} (\phi/2)=  \sin \bigg( \dfrac{\pi}{2(2k+1) }\bigg)
\end{align}
We claim a solution exists if $N$ satisfies
\begin{equation}
\dfrac{1}{2^{N-1}}{N\choose N/2} \ge  \sin^2 \bigg( \dfrac{\pi}{2(2k+1) }\bigg)
\end{equation}
The LHS of $\braket{N/2, -\phi|\rm GHZ}$ achieves a maximum at $\phi=\pi/2$ with the value $ \sqrt{{N\choose N/2}}/2^{(N-1)/2}$. Therefore, a solution exists if the maximum value of the LHS is equal to or greater than the right hand side   $\sin( \frac{\pi}{2(2k+1) })$:
\begin{equation}
\sqrt{\dfrac{1}{2^{N-1}}{N\choose N/2} }\ge  \sin \bigg( \dfrac{\pi}{2(2k+1) }\bigg)
\end{equation}
Squaring  both sides readily gives the condition for a solution. 

\subsection{Analysis of the Cat states preparation}\label{Cat_appendix}

Starting from the Cat states
\begin{equation}
\ket{{\rm cat \pm},\phi}=\dfrac{1}{\sqrt{2 \pm 2\braket{\phi|-\phi}}}(\ket{\phi}^{\otimes N} \pm \ket{-\phi}^{\otimes N})
\end{equation}
where the CSS states $\ket{\phi}^{\otimes N}=R(\phi)^{\otimes N}\ket{m=0}$ and $\ket{-\phi}^{\otimes N}=R(-\phi)^{\otimes N}\ket{m=0}$ are given by
\begin{align}
&\ket{\phi}^{\otimes N}=\bigg(\cos(\phi/2) \ket{0}+\sin (\phi/2) \ket{1}\bigg)^{\otimes N}\\&= \nonumber \sum_{m=0}^{N} \sqrt{{N\choose m}} \cos^{N-m} (\phi/2) \sin^{m} (\phi/2) \ket{m} 
\end{align}
\begin{align}
&\ket{-\phi}^{\otimes N}=\bigg(\cos(\phi/2)\ket{0}-\sin (\phi/2) \ket{1}\bigg)^{\otimes N}\\&= \nonumber \sum_{m=0}^{N} (-1)^m  \sqrt{{N\choose m}} \cos^{N-m} (\phi/2) \sin^{m} (\phi/2) \ket{m} 
\end{align}
Using $\braket{\phi|-\phi}=\cos^N( \phi)$, it follows that the Cat states are given by
\begin{align}
&\ket{{\rm cat \pm },\phi}=\dfrac{2}{\sqrt{2 \pm 2\cos^N (\phi)}}\\&\nonumber \times \sum_{m={\rm even/odd}}\sqrt{{N\choose m}} \cos^{N-m} (\phi/2) \sin^{m} (\phi/2) \ket{m} 
\end{align}
Using the equation above, to prepare $\ket{{\rm cat \pm},\phi}$ in $k$ steps, the overlap condition between the initial state $\ket{m}$ and $\ket{{\rm cat \pm},\phi}$ needs to satisfy
\begin{align}
&\braket{m|{\rm cat \pm},\phi}\nonumber=\dfrac{2}{\sqrt{2 \pm 2\cos^N( \phi)}} \\& \times  \sqrt{{N\choose m}} \cos^{N-m} (\phi/2) \sin^{m} (\phi/2)= \sin \bigg( \dfrac{\pi}{2(2k+1) }\bigg)
\end{align}
\section{Previous cavity carving schemes }\label{appendix:cavity_carve}
\rsub{We give a brief overview of previous probabilistic carving schemes, as exemplified by Ref. \cite{Chen2015carving}. We assume an initial product state, expressed as a superposition of Dicke states [Eq. \eqref{psi_initial}]. Consider a symmetric two-sided cavity that transmits photons on resonance with the cavity and reflects ones that are off-resonance. We have the same energy level scheme in Fig. \ref{fig:energy_scheme}(a) and the same Hamiltonian $H=\hbar \Omega \hat{m} \hat{n}_c $ in the dispersive regime, with $\Omega \ll \Delta$. Thus, the shifted cavity resonance frequency, $\omega_m=\omega_0+\delta\omega_m=\omega_0+m \Omega$, depends on the number of atoms coupled to the cavity. If the initial state is $\ket{\psi_i}=\sum_{n}c_n\ket{n}$, and the transmission amplitude when there are $n$ atoms coupled is $t_n(\delta\omega_m)$ [Eq. \eqref{t_n}], then the projected atomic state conditioned on photon transmission is }\rsub{
\begin{equation}
\ket{\psi_t}=\dfrac{1}{\sqrt{\sum_n |t_n|^2 |c_n|^2}}\sum_{n=0}^{N}t_n(\delta \omega_m)c_n \ket{n}
\end{equation}}
\rsub{where the prefactor is a normalization constant. In the ideal case there is only transmission when the atomic state is $\ket{m}$ while the photon is reflected back for all other Dicke states, i.e., $|t_n(\delta \omega_m)|=\delta_{m,n}$. In practice, for any nonideal cavity, there is non-zero probability of transmission for the other Dicke states. The resulting infidelity can be characterized in terms of the atom-cavity cooperativity $C=g^2/\kappa \gamma$ as (cf. the supplement section ``Factual Carving Infidelity'' in \cite{ramette2024counterfactual})}\rsub{
\begin{equation}
1-F=1-|\braket{m|\psi_t}|^2 \sim \dfrac{m}{C}, \ C \gg 1
\end{equation}
}
\rsub{where $\kappa$ and $\gamma$ are the cavity and atomic decay rates, and $g$ is the atom-cavity coupling constant. This is an inherently probabilistic scheme, where the success probability is the overlap between the initial state and the Dicke state $\ket{m}$. For an initial CSS the success probability is maximized by choosing the initial state $R^{\otimes N}(\phi)\ket{0}^{\otimes N}$ with $\phi=\arccos[(N-2m)/N]$. As calculated in Appendix \ref{appendix:Dicke_prep}, the overlap squared for that state is $\sim 1/m^{1/2}$. Thus the theoretical success probability to carve a Dicke state $\ket{m}$ with $N$ qubits is}\rsub{
\begin{align}
&P_{\rm succ}(m)\sim \dfrac{1}{m^{1/2}}, \ m \ll N/2, \  N \gg 1\\
&P_{\rm succ}(m=N/2)\sim \dfrac{1}{N^{1/2}}
\end{align}
}
\rsub{i.e., on average $O(m^{1/2}) [$ $O(N^{1/2})$] trials are required to succeed.}

\rsub{We note that a single-sided cavity can implement this projective carving scheme by using a polarizing beam splitter (PBS) and an incoming photon in a superposition of two orthogonal polarizations $\ket{\sigma_{\pm}}$, where one polarization, e.g., $\ket{\sigma_{+}}$, couples to the atom-cavity system while the other polarization $\ket{\sigma_{-}}$ is always decoupled \cite{Welte2017cavity}. A polarization flip of the reflected photon, e.g., from horizontal to vertical, heralds the projective measurement in this case.}
\section{Effect of the spatial mode mismatch}\label{appendix:mismatch}

Here, an error model is developed to study the effect of spatial mode mismatch on the performance of Grover's algorithm. We have shown [see Sec. \ref{sec:physical_implement}] that the phase inversion operator $\chi$ acting on the atomic qubits $\rho$ corresponds to reflecting a photon off the cavity. If the spatial mode matching efficiency is $\zeta$, then the mismatched part of the photon mode $1-\zeta$ is reflected off the cavity without interacting with the atomic qubits, i.e., after reflection the atomic density matrix becomes
\begin{equation}\label{mismatch}
 \rho \rightarrow  \zeta \chi \rho  \chi^{\dag}+(1-\zeta)\rho
\end{equation}
Thus, spatial mode mismatch acts as a depolarizing channel. Note that $\zeta=1$ corresponds to no mismatch. For the Dicke states preparation, there are two spatial mode mismatch events per Grover step coming from $\chi_0$ and $\chi_m$. Using $G=\chi_{i}\chi_{t}$, $\chi_m=\chi_t$, $\chi_i=R(\phi)^{\otimes N}\chi_0R(-\phi)^{\otimes N}$, and Eq. \eqref{mismatch}, the relation between the output $\rho^{(1)}_{\rm out}$ after a single Grover step and input $\rho_i$ atomic states becomes
\begin{align}
&\rho^{(1)}_{\rm out}= \\ &  \nonumber \zeta^2 G\rho_i G^{\dag}+\zeta(1-\zeta)\chi_i \rho_i \chi_i^{\dag}+\zeta(1-\zeta)\chi_t \rho_i \chi_t^{\dag}+(1-\zeta)^2\rho_i
\end{align}
To get the output after $k$ steps $\rho^{(k)}_{\rm out}$, we feed $\rho^{(k-1)}_{\rm out}$ as input in the right hand side of the equation above. By recursive application of this rule, $\rho^{(k)}_{\rm out}$ can be explicitly expressed in terms of $\rho_i, \zeta, \chi_i,$ and $\chi_t$. The initial state before applying Grover's algorithm is  $\rho_i= \ket{\psi_i} \bra{\psi_i}$ with
\begin{equation}
 \ket{\psi_i}=   \sin (\theta/2) \ket{\psi_t} +  \cos(\theta/2) \ket{\psi_{t,\perp}}
\end{equation}
The fidelity after $k$ steps then becomes 
\begin{equation}
F_{\rm Grover}(k)= \bra{\psi_t}\rho^{(k)}_{\rm out}\ket{\psi_t} 
\end{equation}
To prepare the target perfectly in $k$ steps, the overlap condition requires $\theta/2=\frac{\pi}{2(2k+1)}$ (see Sec. \ref{sec:exact_prep}). Consider the four initial states with $\theta/2=\pi/6,\ \pi/10, \ \pi/14,$ and $\pi/18$. After applying the corresponding (ideal) Grover iteration on these four states, the target state is prepared exactly after one, two, three, and four steps, respectively. For the nonideal case, using the equations in this section and Eqs. \eqref{chi_t} and \eqref{chi_i}, we have analytically calculated the fidelity as a function of $\zeta$ given these four input states:
\begin{widetext}
\begin{equation}\label{F_mismatch1}
F(k=1)=\dfrac{1}{4}(1+3 \zeta^2), \ \ \dfrac{\theta}{2}=\dfrac{\pi}{6}
\end{equation}
\begin{equation}\label{F_mismatch2}
F(k=2)=\dfrac{1}{8} \bigg([5+3 \sqrt{5}] \zeta ^4-8 \sqrt{5} \zeta ^3+6 \sqrt{5} \zeta ^2-\sqrt{5}+3 \bigg),  \ \ \dfrac{\theta}{2}=\dfrac{\pi}{10}
\end{equation}
\begin{align}\label{F_mismatch3}
&F(k=3)= \sin^2 \bigg( \frac{\pi }{14}\bigg) \bigg \{ 2 \zeta ^6  \bigg(10+4 \sin \frac{\pi }{14}+13 \sin \frac{3 \pi
   }{14}+19 \cos \frac{\pi }{7}\bigg)  -16 \zeta ^5  \bigg(3+\sin
   \frac{\pi }{14}+4 \sin \frac{3 \pi }{14}+6 \cos \frac{\pi }{7}\bigg) \\ & \nonumber +\zeta ^4  \bigg(54+10 \sin \frac{\pi }{14}+64 \sin \frac{3 \pi }{14}+108
   \cos \frac{\pi }{7}\bigg)-32 \zeta ^3  \bigg(1+\sin \frac{3 \pi
   }{14}+2 \cos \frac{\pi }{7}\bigg)  +12 \zeta ^2  \bigg(1+\sin
   \frac{3 \pi }{14}+2 \cos \frac{\pi }{7}\bigg)  \\ & \nonumber +1\bigg\} , \ \ \dfrac{\theta}{2}=\dfrac{\pi}{14}
\end{align}
\begin{align}\label{F_mismatch4}
&F(k=4)=\sin ^2 \bigg(\frac{\pi
   }{18}\bigg) \bigg \{\zeta ^8 \bigg(104+28 \sin \frac{\pi }{18}+138 \cos \frac{\pi
   }{9}+108 \cos \frac{2 \pi }{9}\bigg) \\ & \nonumber +\zeta ^7  \bigg(-348-96 \sin
   \frac{\pi }{18}-480 \cos \frac{\pi }{9}-360 \cos \frac{2 \pi }{9}\bigg)  \\ & \nonumber  +\zeta ^6
   \bigg(523+136 \sin \frac{\pi }{18}+760 \cos \frac{\pi
   }{9}+530 \cos \frac{2 \pi }{9} \bigg) \\ & \nonumber +\zeta ^5  \bigg(-448-96 \sin \frac{\pi }{18}-704 \cos \frac{\pi }{9}-448 \cos \frac{2 \pi }{9} \bigg)  \\ & \nonumber +\zeta ^4
 \bigg (240+30 \sin \frac{\pi }{18}+420 \cos \frac{\pi
   }{9}+240 \cos \frac{2 \pi }{9}\bigg) \\ & \nonumber +\zeta ^3 \bigg(-80-160 \cos \frac{\pi }{9}-80 \cos \frac{2 \pi }{9}\bigg) \\ & \nonumber +\zeta ^2 
   \bigg(20+40 \cos \frac{\pi }{9}+20 \cos \frac{2 \pi }{9}\bigg)+1 \bigg\} ,  \ \ \dfrac{\theta}{2}=\dfrac{\pi}{18}
\end{align}
For $|\zeta-1| \ll 1$, the infidelity after applying $k$ steps, to the lowest order, is found by Taylor expanding the equations above:
\begin{equation}
1-F(k) \approx \dfrac{2k+1}{2}(1-\zeta), \ \ \dfrac{\theta}{2}=\frac{\pi}{2(2k+1)}
\end{equation}
More generally, for an arbitrary initial state with $\theta$, $F(k)$ is given by
\begin{equation}
F(k=1)=\frac{1}{2} \zeta ^2 (\cos \theta-\cos 3 \theta)+\frac{1}{2} (1-\cos \theta )
\end{equation}
\begin{align}
&F(k=2)= \frac{1}{2} \zeta ^4 (2 \cos \theta-2 \cos \theta \cos 4\theta)+\frac{1}{2}
   \zeta ^3 (8 \cos \theta \cos 2\theta-8 \cos \theta)\\&  \nonumber+\frac{1}{2} \zeta ^2
   (6 \cos \theta-6 \cos \theta \cos 2\theta)+\frac{1}{2} (1-\cos \theta)
\end{align}
\begin{align}
&F(k=3)=\frac{1}{2} \zeta ^6 (5 \cos \theta-\cos 3\theta-3 \cos 5\theta-\cos 7\theta) +\frac{1}{2} \zeta ^5 (-16 \cos \theta+8 \cos 3\theta+8 \cos 5\theta ) \\ & \nonumber +\frac{1}{2} \zeta ^4 (22 \cos \theta-17 \cos 3\theta-5 \cos 5\theta)+\frac{1}{2} \zeta ^3 (16 \cos 3\theta-16 \cos \theta)+\frac{1}{2}
   \zeta ^2 (6 \cos \theta-6 \cos 3\theta)+\frac{1}{2} (1-\cos \theta)
\end{align}
\begin{align}
&F(k=4)=\frac{1}{2} \zeta ^8 (14 \cos \theta-8 \cos 5\theta-5 \cos 7\theta-\cos 9\theta )+\frac{1}{2} \zeta ^7 (-60 \cos \theta+12 \cos 3\theta+36 \cos 5\theta +12 \cos 7\theta)\\ & \nonumber +\frac{1}{2} \zeta ^6 (115 \cos \theta-47 \cos 3\theta -61 \cos 5\theta-7 \cos 7\theta)+\frac{1}{2} \zeta ^5 (-128 \cos
   \theta+80 \cos 3\theta+48 \cos 5\theta)\\ & \nonumber +\frac{1}{2} \zeta ^4 (90 \cos
   \theta-75 \cos 3\theta-15 \cos 5\theta)+\frac{1}{2} \zeta ^3 (40 \cos 3\theta -40 \cos \theta)+\frac{1}{2} \zeta ^2 (10 \cos \theta-10 \cos 3\theta  )+\frac{1}{2} (1-\cos \theta)
\end{align}
\end{widetext}
This error model, which introduces two spatial mode mismatch events per Grover step, is for the Dicke states. A similar analysis and formulas can be calculated for the GHZ and Cat states by introducing three spatial mode mismatch events per Grover step.

\section{Kraus operators description of cavity QED experiments}\label{appendix:kraus}

Consider $N$ qubits in a cavity in an initial state $\rho_{\rm in}$. After the atom-cavity system interacts with light, the atomic state becomes $\rho_{\rm out}$. Given that this is an open quantum system, we wish to find a Kraus operator description that maps the atomic input to the atomic output \cite{Kraus_2012}:
\begin{equation}\label{Kraus_definition}
\rho_{\rm out}=\sum_iK_i\rho_{\rm in}K^{\dag}_i
\end{equation}
where $K_i$ are the Kraus operators that capture the cavity QED interaction of the atomic qubits and light. The Kraus operators, by construction, conserve probability, i.e.,
\begin{equation}\label{Kraus_normalization}
\sum_iK^{\dag}_iK_i=\mathbb{1}
\end{equation}
The Kraus operators should only depend on the scattering amplitudes of the cavity (reflection, transmission, spontaneous emission, cavity losses) and the state of the light (e.g., the average photon number if the light is a coherent source). If we can find these Kraus operators, then we can describe a variety of cavity QED experiments in a straightforward way. In particular, we want to find the Kraus operators describing our cavity QED implementation of the phase inversion $\chi_m$ and the Grover iteration $G$. Once found, we can model how $\chi_m$ and $G$ will act under many different experimental settings.

We follow the formalism in Ref. \cite{Beukers2024Remote}, extending it to arbitrary number of qubits $N$ and more than one loss mode. An arbitrary input atomic state is given by:
\begin{eqnarray}
\rho_{\rm in}=\sum_{k,k'}C_{kk'}\ket{k}\bra{k'}
\end{eqnarray}
This is in either the computational or the Dicke basis. We wish to model the interaction of that input state with an initial arbitrary state of light $\ket{\psi}_{\rm p}$, which is taken to be a Fock state presently. To model the open nature of the quantum system, we introduce an auxiliary loss photonic mode. Before interaction, we have the initial atom-light state:
\begin{eqnarray}
\sigma_{\rm in}=\rho_{\rm in} \otimes \ket{\psi}_{\rm p}  \bra{\psi}_{\rm p}\otimes\ket{0}_{\rm L}\bra{0}_{\rm L}
\end{eqnarray}
i.e., the atomic $\rho_{\rm in}$ and light $\ket{\psi}_{\rm p}$ states are separable and there are no photons in the loss mode (later we will introduce more than one loss mode). The cavity QED interaction is described by the following unitary:
\begin{equation}
\chi=\sum_k \ket{k}\bra{k} \otimes e^{i\arg(r_k) a^{\dag}a+i\arg(l_k) a_L^{\dag}a_L}e^{i \theta_k(a^{\dag}a_L+aa_L^{\dag})}
\end{equation}
where $a$ and $a_L$ are the annihilation operators for the photonic and loss modes, respectively. $r_k$ and $l_k$ are the complex reflection and loss amplitudes the light and the atom experience when the atom is in $\ket{k}$. Note that $|r_k|^2+|l_k|^2=1$, by probability conservation. $e^{i\theta_k(a^{\dag}a_L+aa_L^{\dag})}$ is a beam-splitter unitary that describes the photonic mode leaking into the loss mode. It mixes the photonic mode $\ket{\psi}_{\rm p}$ and the loss mode $\ket{0}_L$ with an angle $\theta_k$ depending on the photon loss, where $\theta_k=\arcsin |l_k|$. After the interaction, the atoms in the state $\ket{k}$ acquire a phase shift $e^{i\arg r_k a^{\dag}a+i\arg l_k a_L^{\dag}a_L}$. Observe how both the photon loss and phase change depend on $\ket{k}$, making this unitary a highly entangling operation. To obtain the output atomic state $\rho_{\rm out}$, we apply $\chi$ on $\sigma_{\rm in}$ and trace out the loss mode completely and the photonic mode, either partially or completely.

The expressions in this appendix are for monochromatic light. For a photon with a finite bandwidth, the resulting expressions involving the coefficients of reflection, transmission, spontaneous emission, and mirror scattering should be integrated over the wavepacket $|\Phi(\omega)|^2$ with respect to $\omega$. The case of finite bandwidth is tackled in Appendix \ref{appendix:error_analysis}.

\subsection{Interaction with a single photon}

For a single incoming photon $\ket{1}_{\rm p}$ interacting with the atom-cavity system, the initial total state is:
\begin{align}
&\sigma_{\rm in}=\rho_{\rm in} \otimes \ket{1}_{\rm p}  \bra{1}_{\rm p}\otimes\ket{0}_{\rm L}\bra{0}_{\rm L}\nonumber\\&=\sum_k C_{kk'}\ket{k}\bra{k'} \otimes\ket{1}_{\rm p}  \bra{1}_{\rm p}\otimes\ket{0}_{\rm L}\bra{0}_{\rm L}
\end{align}
First, let's apply the beam splitter operator $e^{i\theta_k(a^{\dag}a_L+aa_L^{\dag})}$ on the state of the light. This gives:
\begin{equation}
e^{i\theta_k(a^{\dag}a_L+aa_L^{\dag})}\ket{1}_{\rm p}\ket{0}_{\rm L}=\cos (\theta_k)\ket{1}_{\rm p}\ket{0}_{\rm L}+i\sin (\theta_k)\ket{0}_{\rm p}\ket{1}_{\rm L}
\end{equation}
Observing that $\cos (\theta_k)=|r_k|$ and $\sin (\theta_k)=|l_k|$:
\begin{equation}
e^{i\theta_k(a^{\dag}a_L+aa_L^{\dag})}\ket{1}_{\rm p}\ket{0}_{\rm L}=|r_k|\ket{1}_{\rm p}\ket{0}_{\rm L}+i|l_k|\ket{0}_{\rm p}\ket{1}_{\rm L}
\end{equation}
The action of the phase shift operator is:
\begin{align}
&e^{i\arg(r_k) a^{\dag}a+i\arg(l_k) a_L^{\dag}a_L}(|r_k|\ket{1}_{\rm p}\ket{0}_{\rm L}+i|l_k|\ket{0}_{\rm p}\ket{1}_{\rm L})\nonumber\\ &=e^{i\arg(r_k)}|r_k|\ket{1}_{\rm p}\ket{0}_{\rm L}+ie^{i\arg(l_k)}|l_k|\ket{0}_{\rm p}\ket{1}_{\rm L}
\end{align}
Using $r_k=e^{i\arg(r_k)}|r_k|$ and $l_k=e^{i\arg(l_k)}|l_k|$ to get:
\begin{align}
&e^{i[\arg(r_k) a^{\dag}a+\arg(l_k) a_L^{\dag}a_L]}(|r_k|\ket{1}_{\rm p}\ket{0}_{\rm L}+i|l_k|\ket{0}_{\rm p}\ket{1}_{\rm L})\nonumber\\ &=r_k\ket{1}_{\rm p}\ket{0}_{\rm L}+il_k\ket{0}_{\rm p}\ket{1}_{\rm L}
\end{align}
This describes the action of $\chi$. So we have

\begin{widetext}
\bea
\chi \sigma_{\rm in}\chi^{\dag}&=&\sum_{kk'}C_{kk'}\ket{k}\bra{k'}\bigg\{ r_k r_{k'}^*\ket{1}_{\rm p}\bra{1}_{\rm p} \otimes \ket{0}_{\rm L}\bra{0}_{\rm L}-i r_k l^*_{k'}\ket{1}_{\rm p}\bra{0}_{\rm p} \otimes \ket{0}_{\rm L}\bra{1}_{\rm L}+il_{k} r^*_{k'} \ket{0}_{\rm p}\bra{1}_{\rm p} \otimes \ket{1}_{\rm L}\bra{0}_{\rm L}\nonumber\\
 &&~~~~~~~~~~~~~~~~~~~~+  l_k l_{k'}^*\ket{0}_{\rm p}\bra{0}_{\rm p} \otimes \ket{1}_{\rm L}\bra{1}_{\rm L}\bigg \}
\eea

Tracing over the loss mode we get the atom-light state:
\begin{align}
{\rm Tr}_{\rm L}(\chi \sigma_{\rm in}\chi^{\dag})=\sum_{kk'}C_{kk'}\ket{k}\bra{k'}\bigg\{r_k r_{k'}^*\ket{1}_{\rm p}\bra{1}_{\rm p}+l_k l_{k'}^*\ket{0}_{\rm p}\bra{0}_{\rm p}\bigg \}
\end{align}
\end{widetext}

We are interested in the output atomic state under two different settings: 1) conditioned on detecting a photon reflecting back and 2) unconditioned on the photon being detected. 

\subsubsection{Heralding on detecting the photon}

In this case, the (unnormalized) atomic state $\rho_{\rm out}$ is given by:
\begin{equation}
\rho_{\rm out}=\bra{1}_{\rm p}{\rm Tr}_{\rm L}(\chi \sigma_{\rm in}\chi^{\dag})\ket{1}_{\rm p}=\sum_{kk'}C_{kk'}r_k r^*_{k'}\ket{k}\bra{k'}
\end{equation}
Define the reflection Kraus operator:
\begin{equation}
K_r=\sum_k r_k \ket{k}\bra{k}
\end{equation}
Then from the expression for $\rho_{\rm out}$ above, the input and the output are related as:
\begin{equation}
\rho_{\rm out}=K_r\rho_{\rm in}K_r^{\dag}
\end{equation}
Note that $K_r$ is diagonal. For a one-sided cavity, $K_r$ enacts an entangling conditional phase gate on the atomic qubits. For a double-sided cavity, it enacts an atomic projection operator (i.e., reflection or transmission of light projects the atomic qubit into different orthogonal states). The success probability of this operation is given by the norm of $\rho_{\rm out}$:
\begin{equation}
P_{\rm succ}={\rm Tr}(\rho_{\rm out})=\sum_k C_{kk}|r_k|^2
\end{equation}
\subsubsection{Unconditioned on the photon detection}

If we want to find $\rho_{\rm out}$ unconditioned on detecting the photon, then we trace over all  modes of light:
\begin{align}
&\rho_{\rm out}={\rm Tr}_{\rm L, p}(\chi \sigma_{\rm in}\chi^{\dag})=\sum_{kk'}C_{kk'}(r_k r^*_{k'}+l_k l^*_{k'})\ket{k}\bra{k'}
\end{align}
This is a normalized state. The loss mode can be decomposed into three modes: transmission $t_k$, spontaneous emission $a_k$, and mirror scattering $m_k$. Thus, by generalization, the output becomes: 
\begin{align}
&\rho_{\rm out}=\sum_{kk'}C_{kk'}(r_k r^*_{k'}+t_k t^*_{k'}+a_k a^*_{k'}+m_k m^*_{k'})\ket{k}\bra{k'}
\end{align}
Defining the Kraus operators for reflection, transmission, spontaneous emission, and mirror scattering: 
\begin{subequations}
\begin{align}
K_r(\omega)=&\sum_k r_k(\omega) \ket{k}\bra{k} \\
K_t(\omega)=&\sum_k t_k(\omega) \ket{k}\bra{k}\\
K_a(\omega)=&\sum_k a_k(\omega) \ket{k}\bra{k}\\
K_m(\omega)=&\sum_k m_k(\omega) \ket{k}\bra{k}
\end{align}
\end{subequations}
Then the atomic input and the output are related by:
\begin{equation}\label{appendix_rho_out_mono}
\rho_{\rm out}=K_r\rho_{\rm in}K_r^{\dag}+K_t\rho_{\rm in}K_t^{\dag}+K_a\rho_{\rm in}K_a^{\dag}+K_m\rho_{\rm in}K_m^{\dag}
\end{equation}
Observe that these Kraus operators satisfy the normalization condition [Eq. \eqref{Kraus_normalization}] as expected:
\begin{equation}
K_r^{\dag}K_r+K_t^{\dag} K_t +K_a^{\dag}K_a+K_m^{\dag}K_m=\mathbb{1}
\end{equation}
which is equivalent to $|r_k|^2+|t_k|^2+|a_k|^2+|m_k|^2=1$. To conclude, we have found the Kraus operators to describe the interaction of single photons with an arbitrary atomic state in a cavity, provided that we have analytic expressions for the scattering amplitudes. In the present work, we use Eqs. \eqref{r_n}--\eqref{m_n} for the scattering amplitudes.

\begin{widetext}

\section{Error analysis of the phase inversion and Grover operators}\label{appendix:error_analysis}

In this section we describe the error in implementing the phase inversion $\chi_m$ and Grover iteration $G$, due to a wavepacket with finite bandwidth interacting with a nonideal cavity. We build on the results from appendix \ref{appendix:kraus}.

\subsection{Phase inversion operator}

\subsubsection{Not heralding on the photon reflecting back}

Given an input atomic state $\rho_{\rm in}$ and a wavepacket $|\Phi(\omega)|^2$, the atomic output state $\rho_{\rm out}$ after the photon reflection, for the case of not heralding, is:
\begin{equation}\label{rho_out_unherald}
\rho_{\rm out}=\int_{-\infty}^{\infty} d \omega |\Phi(\omega)|^2 \bigg ( K_r(\omega)\rho_{\rm in}K_r^{\dag}(\omega)+K_t(\omega)\rho_{\rm in}K_t^{\dag}(\omega)+K_a(\omega)\rho_{\rm in}K_a^{\dag}(\omega)+K_m(\omega)\rho_{\rm in}K_m^{\dag}(\omega)\bigg )
\end{equation}
I.e., the atomic state is in a mixed state of the cases where the photon reflected, transmitted, scattered with the atom(s) or the cavity mirror. The Kraus operators are given by Eqs \eqref{K_r}--\eqref{K_m} and Eqs. \eqref{r_n}--\eqref{m_n}. This equation is the generalization of Eq. \eqref{appendix_rho_out_mono} when the photon is not monochromatic but has a finite bandwidth \cite{Klaus_Cohen_network_2018}, where the integrations comes from tracing over the photon wavepacket. Next, we will compute the output atomic state for an arbitrary input Gaussian wavepacket and arbitrary input state. First, we reshape our density matrix $\rho$ into a vector $\bm{\rho}$. A vectorization ${\rm vec}(\rho)$ maps the density matrix $\rho=\sum_{ij} \rho_{ij} \ket{i} \otimes \bra{j}$ into the following column vector \cite{optically_pumped_atoms}:  
\begin{equation}
\bm{\rho}={\rm vec}(\rho)= \sum_{ij} \rho_{ij} \ket{i}\otimes\ket{j}
\end{equation} 
Applying the vectorization identity ${\rm vec}(ABC)=(A \otimes C^{\rm T}){\rm vec}(B)$ on Eq. \eqref{rho_out_unherald} gives the vectorized equation:
\begin{equation}
\bm{\rho}_{\rm out}= \bigg[ \int d \omega |\Phi(\omega)|^2\bigg (K_r(\omega)\otimes K_r^{*}(\omega)+K_t(\omega)\otimes K_t^{*}(\omega)+K_a(\omega)\otimes K_a^{*}(\omega)+K_m(\omega) \otimes K_m^{*}(\omega)\bigg ) \bigg] \bm{\rho}_{\rm in}
\end{equation}
Now the input $\bm{\rho}_{\rm in}$ has factored out from the integration over the Kraus operators. By evaluating the integral in the brackets, we can compute the output for any input atomic state and wavepacket. First, Let's focus on the first term. $K_r(\omega)\otimes K_r^{*}(\omega)$, being the tensor product of two diagonal matrices, is nothing but a diagonal matrix itself:
\begin{equation} 
K_r(\omega)\otimes K_r^{*}(\omega)= \sum_{m,m'}^{N,N}r_m(\omega)r^*_{m'}(\omega) \big(\ket{m}\bra{m}\otimes\ket{m'} \bra{m'}\big) \
\end{equation}
Therefore, the problem essentially boils down to computing terms of the form
\begin{equation}
\int_{-\infty}^{\infty} d\omega  |\Phi(\omega)|^2 r^*_m(\omega)r_{n}(\omega)
\end{equation}
and similarly for the other terms $K_t(\omega)\otimes K_t^{*}(\omega),K_a(\omega)\otimes K_a^{*}(\omega)$, and $K_m(\omega)\otimes K_m^{*}(\omega)$. For concreteness, consider a Gaussian wavepacket
\begin{equation}\label{gaussian}
|\Phi(\omega,\Omega,\sigma)|^2=\dfrac{1}{\sqrt{2 \pi} \sigma}\exp \bigg(-\dfrac{(\omega-\Omega_c)^2}{2 \sigma^2} \bigg)
\end{equation}
where $\Omega_c$ is the central frequency and $\sigma$ is the width of the packet. Then the integral becomes
\begin{equation}
\int d\omega  |\Phi(\omega,0,\sigma)|^2 r^*_m(\omega+\Omega_c)r_{n}(\omega+\Omega_c)
\end{equation}
where we have shifted the integration variable $\omega \rightarrow \omega+\Omega_c$ so that the Gaussian is centered around $\omega=0$. \rsub{Thus, the diagonal entries of the averaged Kraus superoperators associated with reflection, transmission, mirror scattering, and  spontaneous emission are}
\begin{align}\label{Kr_avg}
&\mathcal{K}^{nm}_{\rm r,avg}=\int d\omega  |\Phi(\omega,0,\sigma)|^2 r^*_m(\omega+\Omega_c)r_{n}(\omega+\Omega_c) 
\end{align}
\begin{align}\label{Kt_avg}
\mathcal{K}^{nm}_{\rm t,avg}&=\int d\omega  |\Phi(\omega,0,\sigma)|^2 t^*_m(\omega+\Omega_c)t_{n}(\omega+\Omega_c)
\end{align}
\begin{align}\label{Km_avg}
&\mathcal{K}^{nm}_{\rm m,avg}=\int d\omega  |\Phi(\omega,0,\sigma)|^2 m^*_m(\omega+\Omega_c)m_{n}(\omega+\Omega_c)
\end{align}
\begin{align}\label{Ka_avg}
&\mathcal{K}^{nm}_{\rm a,avg}=\int d\omega  |\Phi(\omega,0,\sigma)|^2 a^*_m(\omega+\Omega_c)a_{n}(\omega+\Omega_c)
\end{align}\rsub{
where the scattering amplitudes are given by \cite{Daiss2019Single} 
\begin{subequations}
\begin{align}
r_n(\omega)=&1-\dfrac{2 \kappa_r(i \Delta+i\omega + \gamma)}{n g^2+(i \Delta+i\omega + \gamma)(i \omega + \kappa)} \label{r_n_appendix} \\ 
t_n(\omega)=&\dfrac{2 \sqrt{\kappa_r \kappa_t }(i \Delta+i\omega + \gamma)}{n g^2+(i \Delta+i\omega  + \gamma)(i \omega + \kappa)}  \label{t_n_appendix} \\
a_n(\omega)=&\dfrac{2 \sqrt{\kappa_r \gamma }\sqrt{n}g}{n g^2+(i \Delta+i\omega  + \gamma)(i \omega + \kappa)} \label{a_n_appendix}\\
m_n(\omega)=&\dfrac{2 \sqrt{\kappa_r \kappa_m }(i \Delta+i\omega  + \gamma)}{n g^2+(i \Delta+i\omega  + \gamma)(i \omega + \kappa)}  \label{m_n_appendix}
\end{align}
\end{subequations}}
Under this superoperator approach, the error analysis becomes straightforward. Let's denote the averaged Kraus superoperators as
\begin{equation}\label{K_avg_appendix}
\mathcal{K}_{\rm avg}(\Omega_c)= \bigg[ \int d \omega |\Phi(\omega)|^2\bigg (K_r(\omega)\otimes K_r^{*}(\omega)+K_t(\omega)\otimes K_t^{*}(\omega)+K_a(\omega)\otimes K_a^{*}(\omega)+K_m(\omega) \otimes K_m^{*}(\omega)\bigg ) \bigg]
\end{equation}
i.e., the matrix elements of $\mathcal{K}_{\rm avg}(\Omega_c)=\mathcal{K}_{\rm r,avg}+\mathcal{K}_{\rm t,avg}+\mathcal{K}_{\rm a,avg}+\mathcal{K}_{\rm m,avg}$ are given by Eqs. \eqref{Kr_avg}, \eqref{Kt_avg}, \eqref{Km_avg}, and \eqref{Ka_avg}, which can be computed numerically. Then the output atomic state [Eq. \eqref{rho_out_unherald}], for an arbitrary input atomic state and wavepacket, simply becomes:
\begin{equation}\label{rho_out_vector}
\bm{\rho}_{\rm out}= \mathcal{K}_{\rm avg}(\Omega_c)\bm{\rho}_{\rm in}
\end{equation}
\rsub{ To implement $\chi_m$, we choose the central frequency $\Omega_c$ of the wavepacket that achieves phase inversion when there are $m$ atoms coupled. That is, we choose $\Omega_c$ to match the shifted cavity resonance frequency $\delta \omega_m$ in the presence of $m$ coupled atoms (in $\ket{1}$). Solving for $r_m(\Omega_c)=-1$ [Eq. \eqref{r_n_appendix}] gives two frequencies associated with the shifted resonance frequencies of the atoms and the cavity. The one associated with the shifted cavity resonance frequency is:}
\rsub{\begin{equation}\label{cavity_shift_freq_appendix}
\Omega_c = \delta \omega_m= {\rm Re} \bigg \{ \dfrac{1}{2}\Bigl(
  i\gamma - \Delta 
  + i(\kappa - \kappa_r)
  + \sqrt{4g^{2}m - \bigl(\gamma + i\Delta - \kappa + \kappa_r\bigr)^{2}}
\Bigr) \bigg \}
\end{equation}}
\rsub{where the presence of imaginary part is due to dissipation and implies that the resonance condition is only approximately observed as $r_m(\Omega_c)\approx-1$. In the cavity dispersive regime, $mg^2/\Delta^2 \ll 1$, we have $\Omega_c \approx m g^2/\Delta$ to a leading order.} 

\rsub{Approximate analytical expressions for the averaged Kraus superoperators can be calculated in certain limiting cases of the cavity dispersive regime. We assume that 1) the photon wavepacket central frequency $\Omega_c$ and its width $\sigma$ are close to resonance with that of the bare and shifted cavity resonances $\omega_m=\omega_0+m\Omega$,  2) the atomic frequency is far detuned from that of the cavity such that the photon does not excite the atomic resonance, and  3) the shifted cavity resonances are small relative to the cavity-atom detuning, i.e., $m \Omega \ll \Delta$ or $mg^2/\Delta^2 \ll 1$. Under these assumptions, the atomic resonance can be ignored, and we can approximately replace $i \Delta+i\omega$ by $i \Delta$ in the expressions for the scattering amplitudes:}
\rsub{\begin{subequations}
\begin{align}
r_n(\omega)\approx&1-\dfrac{2 \kappa_r(i \Delta + \gamma)}{n g^2+(i \Delta + \gamma)(i \omega + \kappa)} \label{r_n_approx} \\ 
t_n(\omega)\approx&\dfrac{2 \sqrt{\kappa_r \kappa_t }(i \Delta+ \gamma)}{n g^2+(i \Delta  + \gamma)(i \omega + \kappa)}  \label{t_n_approx} \\
a_n(\omega)\approx&\dfrac{2 \sqrt{\kappa_r \gamma }\sqrt{n}g}{n g^2+(i \Delta + \gamma)(i \omega + \kappa)} \label{a_n_approx}\\
m_n(\omega)\approx&\dfrac{2 \sqrt{\kappa_r \kappa_m }(i \Delta  + \gamma)}{n g^2+(i \Delta + \gamma)(i \omega + \kappa)}  \label{m_n_approx}
\end{align}
\end{subequations}}
\rsub{These approximate expressions effectively ignore the atomic resonance in the scattering amplitudes and only focus on the interaction of the photon wavepacket with the (shifted) cavity resonances. Integrating Eqs. \eqref{Kr_avg}-\eqref{Ka_avg} by using these approximate scattering amplitudes, this evaluates to}
\begin{align}\label{Kr_avg_approx}
&\mathcal{K}^{nm}_{\rm r,avg}\approx 1+\left[e^{f_1^2(m)}    \text{erfc}\left(f_1(m)\right)  + e^{(f_1^*(n))^2} \text{erfc}\left(f_1^*(n)\right) \right] f_2
\end{align}
\begin{align}\label{Kt_avg_approx}
\mathcal{K}^{nm}_{\rm t,avg}\approx
\left[ e^{f_1^2(m)}   \text{erfc}\left(f_1(m)\right)  + e^{(f_1^*(n))^2}\text{erfc}\left(f_1^*(n)\right)\right] f_3(\kappa_t) 
\end{align}
\begin{align}\label{Km_avg_approx}
&\mathcal{K}^{nm}_{\rm m,avg}\approx \left[ e^{f_1^2(m)}   \text{erfc}\left(f_1(m)\right)  + e^{(f_1^*(n))^2}\text{erfc}\left(f_1^*(n)\right)\right] f_3(\kappa_m) 
\end{align}
\begin{align}\label{Ka_avg_approx}
&\mathcal{K}^{nm}_{\rm a,avg}\approx \left[ e^{f_1^2(m)}   \text{erfc}\left(f_1(m)\right)  + e^{(f_1^*(n))^2}\text{erfc}\left(f_1^*(n)\right)\right] f_3(\sqrt{mn}g^2\gamma/(\gamma^2+\Delta^2))
\end{align}
where $\text{erfc}(x)=1-\text{erf}(x)$ is the complementary error function and we have introduced the definitions 
\begin{eqnarray}
   f_1(x)&=& \dfrac{g^2 x+(\gamma -i \Delta ) (\kappa -i \Omega_c)}{\sqrt 2
   \sigma  (\gamma -i \Delta )}\nonumber\\
   f_2&=&-\frac{\sqrt{2 \pi } \kappa_r }{\sigma }+\dfrac{ 2 \sqrt{2 \pi } \kappa_r^2 \left(\gamma ^2+\Delta ^2\right)}{\sigma  \left(2 \kappa  \left(\gamma ^2+\Delta ^2\right)+g^2 [\gamma 
   (m+n)+i \Delta  (m-n)]\right)}\nonumber\\
   f_3(x)&=& \dfrac{ 2 \sqrt{2 \pi } \kappa_r x \left(\gamma ^2+\Delta ^2\right)}{\sigma  \left(2 \kappa  \left(\gamma ^2+\Delta ^2\right)+g^2 [\gamma 
   (m+n)+i \Delta  (m-n)]\right)}.
\end{eqnarray}
\rsub{The accuracy of these expressions depends on how well the above assumptions are satisfied, especially $mg^2/\Delta^2 \ll 1$}.

The ideal output after applying $\chi_m$ is given by:
\begin{equation}\label{rho_out_ideal}
\bm{\rho}_{\rm out}({\rm ideal})=(\chi_m \otimes \chi_m^{*})\bm{\rho}_{\rm in}
\end{equation}
The fidelity $F$ in implementing $\chi_m$ is the overlap between the ideal (pure) state and the actual mixed state. In the superoperator formalism, this overlap is simply given by the dot product between these two vectors \cite{optically_pumped_atoms}:
\begin{equation}\label{F}
F=\bm{\rho}^*_{\rm out}({\rm ideal})\cdot \bm{\rho}_{\rm out}=\bm{\rho}_{\rm out}({\rm ideal})\cdot \bm{\rho}_{\rm out}
\end{equation}
where the last equation is true since the ideal output, \rsub{a superposition of the Dicke states under phase inversion}, only has real amplitudes. It is numerically found that, with all the system parameters being fixed, $\chi_m$ achieves minimum fidelity for the state $\ket{\psi_i}=R^{N}(\phi)\ket{m=0}$ with $\phi=\arccos[(N-2m)/N]$. Carrying out the analysis using this state will give us a lowerbound on the fidelity with respect to all input product states of the form $R^{N}(\phi)\ket{m=0}$. That is, our initial vectorized density matrix is:
\begin{equation}\label{rho_in_min}
\bm{\rho}_{\rm in}= \ket{\psi_i}\otimes \ket{\psi_i}= \sum_{n,l}^{N,N} \sqrt{{N \choose{n}}} \sqrt{{N \choose{l}}} \left(\frac{m}{N}\right)^{n/2} \left(\frac{m}{N}\right)^{l/2} \left(1-\frac{m}{N}\right)^{(N-n)/2}  \left(1-\frac{m}{N}\right)^{(N-l)/2} \ket{n} \otimes  \ket{l}
\end{equation}
The ideal output after the phase inversion $\mathcal{X}_m=\chi_m \otimes \chi_m^{*}$ is
\begin{equation}
 \bm{\rho}_{\rm out}({\rm ideal})=\sum_{n,l}^{N,N} (1-2 \delta_{mn}) (1-2 \delta_{ml}) \sqrt{{N \choose{n}}} \sqrt{{N \choose{l}}} \left(\frac{m}{N}\right)^{n/2} \left(\frac{m}{N}\right)^{l/2} \left(1-\frac{m}{N}\right)^{(N-n)/2}  \left(1-\frac{m}{N}\right)^{(N-l)/2} \ket{n}\otimes  \ket{l}
\end{equation}
While the nonideal output is \rsub{$ \bm{\rho}_{\rm out}({\rm nonideal})=\mathcal{K}_{\rm avg}(\delta \omega_m) \bm{\rho}_{\rm in}$}:

\begin{equation}
 \bm{\rho}_{\rm out}=\sum_{n,l}^{N,N} \mathcal{K}_{nl} \sqrt{{N \choose{n}}} \sqrt{{N \choose{l}}} \left(\frac{m}{N}\right)^{n/2} \left(\frac{m}{N}\right)^{l/2} \left(1-\frac{m}{N}\right)^{(N-n)/2}  \left(1-\frac{m}{N}\right)^{(N-l)/2} \ket{n}\otimes  \ket{l}
\end{equation}
where the matrix elements $\mathcal{K}_{nl}$ are given by the sum of Eqs. \eqref{Kr_avg}, \eqref{Kt_avg}, \eqref{Km_avg}, and \eqref{Ka_avg}. The fidelity $F=\bm{\rho}_{\rm out}({\rm ideal})\cdot \bm{\rho}_{\rm out}$ is then given by:
\begin{equation}\label{F_chi}
F(\chi_m)=\bm{\rho}_{\rm out}({\rm ideal}) \cdot \bm{\rho}_{\rm out}({\rm nonideal})=\sum_{n,l}^{N,N} \mathcal{K}_{nl} (1-2 \delta_{mn}) (1-2 \delta_{ml}) {N \choose{n}} {N \choose{l}} \left(\frac{m}{N}\right)^{n+l} \left(1-\frac{m}{N}\right)^{2N-n-l}  
\end{equation}
This expression is confirmed by our numerical simulations. 

\subsubsection{Heralding on detection of the reflected photon}

This case is given by only considering the reflection Kraus operator (cf. Appendix \ref{appendix:kraus}) and discarding the rest of the Kraus operators. The (unnormalized) atomic output is
\rsub{
\begin{equation}
\bm{\rho}_{\rm out}= \bigg(\int d\omega  |\Phi(\omega)|^2  [K_r(\omega)\otimes K_r^{*}(\omega)] \bigg) \bm{\rho}_{\rm in}=\mathcal{K}_{\rm r, avg}(\delta \omega_n)\bm{\rho}_{\rm in}
\end{equation}}
where we have already evaluated the matrix elements for $\mathcal{K}_{\rm r, avg}$ in Eq. \eqref{Kr_avg}. We can get a lowerbound on the fidelity by choosing the initial state given by Eq. \eqref{rho_in_min}. The fidelity is the overlap of the ideal output with the actual output, after normalizing, i.e.,
\begin{equation}
F=\dfrac{\bm{\rho}_{\rm out}({\rm ideal})\cdot \bm{\rho}_{\rm out}}{{\rm Tr}(\rho_{\rm out})}=\dfrac{\bm{\rho}_{\rm out}({\rm ideal})\cdot \bm{\rho}_{\rm out}}{{\rm vec}(\mathbb{1})\cdot \bm{\rho}_{\rm out}} 
\end{equation}
where the trace in the superoperator formalism is computed by taking the dot product of $\bm{\rho}_{\rm out}$ and the vectorization of the identity matrix \cite{optically_pumped_atoms}.
Carrying out the same analysis as before, we get
\begin{equation}
F(\chi_m)=\dfrac{1}{{\rm vec}(\mathbb{1})\cdot \bm{\rho}_{\rm out} }\sum_{n,l}^{N,N} \mathcal{K}_{\rm r,avg}^{nl} (1-2 \delta_{mn}) (1-2 \delta_{ml}) {N \choose{n}} {N \choose{l}} \left(\frac{m}{N}\right)^{n+l} \left(1-\frac{m}{N}\right)^{2N-n-l}  
\end{equation}
and 
\begin{equation}
{\rm Tr}(\rho_{\rm out})={\rm vec}(\mathbb{1})\cdot \bm{\rho}_{\rm out}=\sum_{n}^{N} \mathcal{K}_{\rm r,avg}^{nn} {N \choose{n}} \left(\frac{m}{N}\right)^{n} \left(1-\frac{m}{N}\right)^{N-n}  
\end{equation}

\subsection{Scaling of the fidelity of the phase inversion operator with the cavity parameters}

Here, we compute the dependence of the fidelity in implementing $\chi_m$ on the cavity parameters.

\subsubsection{Not heralding on detection of the reflected photon}

We first analyze $\chi_m$ for $m \neq 0$. For an initial state $\ket{\psi_i}=\sum_{n=0}^{N}c_n\ket{n}$, the fidelity in implementing $\chi_m$ is found by applying Eqs. \eqref{rho_out_vector}, \eqref{rho_out_ideal}, and \eqref{F} on $\ket{\psi_i}$:
\begin{equation}\label{F_chi_general}
F(\chi_m)=\bm{\rho}_{\rm out}({\rm ideal}) \cdot \bm{\rho}_{\rm out}({\rm nonideal})=\sum_{n,l}^{N,N} |c_n|^2 |c_l|^2 (1-2 \delta_{mn}) (1-2 \delta_{ml}) \mathcal{K}_{nl} 
\end{equation}
To be able to obtain analytical estimates that explicitly depend on the cavity parameters, we make the following simplifying assumptions: we assume no mirror or transmission losses, i.e., $\kappa_t=\kappa_m=0$ so that $\kappa=\kappa_r$. This implies that $\mathcal{K}_{nl}(\delta \omega_m \approx mg^2/\Delta)$ is the the sum of the reflection [Eq. \eqref{Kr_avg}] and spontaneous emission [Eqs. \eqref{Ka_avg}] averaged Kraus superoperators only. We further assume that the wavepacket width relative to the cavity is negligible, i.e., $w=\sigma/\kappa \ll 1$. In this case, the infidelity will be given by:
\begin{equation}\label{1_Fchi_general}
1-F(\chi_m)=\sum_{n,l}^{N,N} |c_n|^2 |c_l|^2  \big \{ 1-  (1-2 \delta_{mn}) (1-2 \delta_{ml}) (r_n r^{*}_l+a_n a^{*}_l) \big  \}
\end{equation}
where $r_n(\omega)$ and $a_n(\omega)$ are given by Eqs. \eqref{r_n} and \eqref{a_n} with $\omega \approx m g^2/\Delta$. Observe that for $n=l$, the term in the curly brackets vanish (since $|r_n|^2+|a_n|^2=1$), and so it does not contribute to the infidelity. Only the off diagonal terms $n \neq l$ are going to contribute. The entire behavior of the system can be captured by the three dimensionless parameters: the cooperativity $C=g^2/\kappa \gamma$, $w=\sigma/\kappa$, and $d=\Omega/\kappa$, where $\Omega=g^2/\Delta$. $d$ measures the ability of the cavity to physically distinguish between two neighboring Dicke states $m$ and $m\pm1$, i.e., $d$ is the cavity ``resolution''. Re-expressing the equations above in terms of $C$ and $d$, plugging them in Eq. \eqref{F_chi}, and then doing an expansion, assuming $d \gg 1$ and $d^2/C \ll 1$, we find the terms with the leading-order contributions are
\begin{equation}
r_m r^{*}_n+r_n r^{*}_m \sim -2 +4m \dfrac{d^2}{C}+\dfrac{4}{d^2(m-n)^2}
\end{equation}
and we get $a_m a^{*}_n+a_n a^{*}_m \sim 1/C$ from the spontaneous emission terms. Plugging this into $1-F(\chi_m)$, gives the scaling:
\begin{equation}\label{F_unhearld_expand}
1-F(\chi_m) \sim a\dfrac{1}{d^2}+  a_m\dfrac{d^2}{C}, \ d \gg 1, \ d^2/C \ll 1
\end{equation}
where $a$ and $a_m$ are constants. The infidelity decreases monotonically with $C$, for a fixed $d$. For a given $C$, there is a tradeoff between the two error terms as a function of $d$, i.e., increasing (decreasing) $d$ decreases the first (second) error term and vice versa. This implies that the fidelity $F$ achieves a maximum at a certain $d$, namely $d \sim C^{1/4}$. Plugging in $d=C^{1/4}$ in Eq. \eqref{F_unhearld_expand}, we find that, to a leading order, the fidelity scales as
\begin{equation}
1-F(\chi_m) \sim \dfrac{1}{\sqrt{C}}, \ m \neq 0
\end{equation}
We have numerically found that $d=(C/m)^{1/4}$ gives a rough estimate to the exact $d_{\rm max}$ that maximizes the fidelity for the state given by Eq. \eqref{rho_in_min}. Finally, we point out that we computed $1-F(\chi_m)$ here up to the zeroth order in the small parameter $w=\sigma/\kappa$. We can keep the first nonvanishing term in $w$ by employing the expansion $e^{x^2}{\rm erfc}(x)\approx \frac{1}{\sqrt{\pi}x}-\frac{1}{\sqrt{2 \pi} x^3}$ in $\mathcal{K}_{nl}$ [Eq. \eqref{K_avg_appendix}]. Assuming that $w$ is much smaller than the other two parameters $C$ and $d$, then the fidelity will scale as
\begin{equation}
F(\chi_m)=1-a \dfrac{1}{d^2}- a_m \dfrac{d^2}{C}-b w^2, \ d \gg 1, \ d^2/C \ll 1, \ w \ll 1
\end{equation}
where $b$ is a constant. Thus, for a given $C$, the value of $d$ that maximizes the fidelity still does not depend on $w$, and it is still be given by $d \sim C^{1/4}$. Therefore, the infidelity will scale as $1/\sqrt{C}+w^2$.

Next, we analyze $\chi_0$. Carrying out the same analysis before with $m=0$ and $\Omega_c=0$, the leading-error terms in Eq. \eqref{1_Fchi_general} are of the form:
\begin{equation}
r_0 r^{*}_n+r_n r^{*}_0 \sim -2 +\dfrac{4}{d^2 n^2}  + \dfrac{4}{n C}
\end{equation}
and $a_m a^{*}_n+a_n a^{*}_n\sim 1/C$. Observe that, unlike the case of $m \neq 0$, there is no tradeoff between the two error terms by increasing $d$. I.e., the optimal choice that minimizes the infidelity is $d\rightarrow \infty$ or equivalently $\Delta=0$. To summarize, the case of not heralding on the photon reflection has the scaling 
\begin{align}
&1-F(\chi_m) \sim \dfrac{1}{\sqrt{C}}+w^2, \ m \neq 0, \ d\sim(C/m)^{1/4} \\ &
1-F(\chi_0) \sim \dfrac{1}{C}+w^2, \ d=\infty
\end{align}
\subsubsection{Heralding on detection of the reflected photon}

Here the fidelity is given by 
\begin{equation}
F=\dfrac{\bm{\rho}_{\rm out}({\rm ideal})\cdot \bm{\rho}_{\rm out}}{{\rm vec}(\mathbb{1})\cdot \bm{\rho}_{\rm out}} 
\end{equation}
or
\begin{equation}
F(\chi_m)=\dfrac{1}{\sum_{n}|c_n|^2|r_n|^2}\sum_{n,l}^{N,N} |c_n|^2 |c_l|^2 (1-2 \delta_{mn}) (1-2 \delta_{ml})r_n r^{*}_l 
\end{equation}
There are no spontaneous emission terms $a_na^{*}_l$ here since we herald on the photon being reflected, and the prefactor ensures renormalization of the final atomic state. We employ the same approximations as before in the previous section: $\kappa_t=\kappa_m=0$ so that $\kappa=\kappa_r$ and $w=\sigma/\kappa \ll 1$. Rewriting $1-F(\chi_m)$ above to get
\begin{equation}
1-F(\chi_m)=\dfrac{1}{\sum_{n}|c_n|^2|r_n|^2}\sum_{n,l}^{N,N} |c_n|^2 |c_l|^2  \big \{ |r_n|^2-  (1-2 \delta_{mn}) (1-2 \delta_{ml})r_n r^{*}_l \big  \}
\end{equation}
Again, only terms with $n \neq l$ contributes to the infidelity. Expanding the infidelity in terms of $d$ and $C$, assuming $d \gg 1$ and $d^2/C \ll 1$, the leading-order terms are of the form:
\begin{equation}
|r_n|^2+|r_m|^2+(r_m r^{*}_n+r_n r^{*}_m) \sim  4\dfrac{d^4 m^2}{C^2}+\dfrac{4}{d^2(m-n)^2}
\end{equation}
Therefore, the infidelity will scale as:
\begin{equation}
F(\chi_m)=1-a \dfrac{1}{d^2}-b\dfrac{d^4}{C^2}, \ d \gg 1, \ d^2/C \ll 1
\end{equation}
For a given $C$, the fidelity is maximized by choosing $d \sim C^{1/3}$. Plugging in $d=C^{1/3}$ in $F(\chi_m)$, we get the scaling:
\begin{equation}
1-F(\chi_m)\sim  \dfrac{1}{C^{2/3}}, \ m \neq 0
\end{equation}
It is numerically found that $d \approx (C/m)^{1/3}$ gives a rough estimate to the exact $d_{\rm max}$ that maximizes the fidelity for the state given by Eq. \eqref{rho_in_min}. Observe that this is a better scaling than the not heralded case with $1/\sqrt{C}$.

Carrying out the same kind of analysis for $\chi_0$, we find the leading-order error terms scale like:
\begin{equation}
|r_0|^2+|r_n|^2+(r_0 r^{*}_n+r_n r^{*}_0)\sim \dfrac{4}{n^2C^2}+\dfrac{4}{n^2d^2}
\end{equation}
Again, \rsub{just like in the unheralded case above,} we find no tradeoff between the two error terms, so the ideal $d$ value is $d=\infty(\Delta=0)$, and we get $1-F(\chi_0) \sim 1/C^2$. To summarize, the heralding case gives the scaling:
\begin{align}
&1-F(\chi_m) \sim \dfrac{1}{C^{2/3}}+w^2, \ m \neq 0 , \ d\sim(C/m)^{1/3} \\ &
1-F(\chi_0) \sim \dfrac{1}{C^2}+w^2, \ d=\infty
\end{align}
Next, we estimate the success probability in implementing $\chi_m$ ($m\neq 0$) for the case of heralding, which is given by the probability that the photon is reflected back after the atom-cavity interaction, i.e.,
\begin{equation}
P_{\rm success}={\rm vec}(\mathbb{1})\cdot \bm{\rho}_{\rm out}=\sum_{n=0}^{N}|c_n|^2|r_n|^2
\end{equation}
Expanding $|r_n|^2$ in terms of $d$ and $C$, and assuming $d \gg 1$ and $d^2/C \ll 1$, the leading-order terms are
\begin{equation}
P_{\rm success}=\sum_{n\neq m}|c_n|^2\left(1-\dfrac{4n}{C (m-n)^2}\right)+|c_m|^2\left(1-4\dfrac{md^2}{C}\right)
\end{equation}
Using conservation of the norm of the atomic state, i.e., $\sum_n |c_n|^2=1$, and plugging in the optimal value $d\sim C^{1/3}$ that maximizes the fidelity, we get
\begin{equation}
P_{\rm success}\sim 1-O(C^{-1/3})
\end{equation}
\subsection{Grover iteration}

\subsubsection{Not heralding on detection of the reflected  photon}

The effect of one Grover step, $G=R(\phi)^{\otimes N}\chi_0R(-\phi)^{\otimes N}\chi_n$, under this description would involve hitting the cavity with two photon wavepackets and not heralding on detection of the reflected photons. Therefore, the output after one Grover step would be:
\begin{equation} \label{rho_out_grover_1}
\rho^{(1)}_{\rm out}= \int d \omega_0 |\Phi(\omega_0)|^2  \int  d \omega_n  |\Phi(\omega_n)|^2 \bigg ( R(\phi)^{\otimes N}K_r(\omega_0)R(-\phi)^{\otimes N}K_r(\omega_n)\rho_{\rm in}(...)^{\dag}\bigg )
\end{equation}
Where $(...)^{\dag}$ denotes the adjoint of the operator to the left of $\rho_{\rm in}$. The first wavepacket has a central frequency \rsub{$\Omega_c=\delta \omega_n$} while the second has $\Omega_c=0$. Rewriting this equation in the superoperator form using the identity ${\rm vec}(ABC)=(A \otimes C^{\rm T}){\rm vec}(B)$:
\begin{align} 
\nonumber \bm{\rho}^{(1)}_{\rm out}= \bigg( R(\phi)^{\otimes N} \otimes R(\phi)^{\otimes N} \bigg) \left(\int d \omega_0 |\Phi(\omega_0)|^2 K_r(\omega_0) \otimes K_r^{*}(\omega_0) \right)  \bigg(R(-\phi)^{\otimes N} \otimes R(-\phi)^{\otimes N} \bigg) \\ \times \left( \int  d \omega_n   |\Phi(\omega_n)|^2 K_r(\omega_n) \otimes K_r^{*}(\omega_n) \right)  \bm{\rho}_{\rm in}
\end{align}
and defining the rotation superoperator
\begin{equation}
\mathcal{R}(\phi)=R(\phi)^{\otimes N} \otimes R(\phi)^{\otimes N}
\end{equation}
where the global rotation matrix elements in the Dicke basis is given by the Wigner small $d$-matrix, i.e., $\bra{m}R(\phi)^{\otimes N} \ket{n}=d^{j=N/2}_{m-N/2,n-N/2}(\phi)$. Here we adopt the sign convention where the Wigner matrix for $N=1$ qubit is given by Eq. \eqref{R_phi}. After vectorization, the two integrations separate and give \rsub{$\mathcal{K}_{\rm avg}(\delta \omega_n)$} and $\mathcal{K}_{\rm avg}(0)$, independent of the initial state or the rotation angle $\phi$. Then one Grover iteration in the superoperator form becomes:\rsub{
\begin{equation}\label{G_physical_appendix}
\mathcal{G}=  \mathcal{R}(\phi)  \mathcal{K}_{\rm avg}(0) \mathcal{R}(-\phi) \mathcal{K}_{\rm avg}(\delta \omega_n)
\end{equation}}
The output after $k$ steps then becomes
\begin{equation}
\bm{\rho}^{(k)}_{\rm out}=\mathcal{G}^k\bm{\rho}_{\rm in} 
\end{equation}
It is important to emphasize that everything in $\mathcal{G}$ is already precomputed and is independent of $\bm{\rho}_{\rm in}$. 

\subsubsection{Heralding on detection of the reflected photon}

For that case, the Grover iteration would be given by the averaged reflection Kraus superoperator:\rsub{
\begin{equation}
\mathcal{G}=  \mathcal{R}(\phi)  \mathcal{K}_{\rm r, avg}(0) \mathcal{R}(-\phi) \mathcal{K}_{\rm r, avg}(\delta \omega_n)
\end{equation}}
with the (unnormalized) output atomic state after $k$ steps being $\mathcal{G}^k\bm{\rho}_{\rm in}$. The norm of this state, ${\rm vec}(\mathbb{1})\cdot \bm{\rho}^{(k)}_{\rm out}$, gives the efficiency or success probability of implementing $k$ Grover steps without photon losses.

\subsection{Numerical simulation}
We outline the procedure used in this work to numerically simulate the effects of the physical phase inversion operator [Eq. \eqref{K_avg_appendix}] and the Grover iteration [Eq. \eqref{G_physical_appendix}].

For $N$ qubits, the corresponding operators/matrices are $(N+1) \times (N+1)$ in the Dicke basis. After vectorization, the matrices become $(N+1)^2 \times (N+1)^2$. The Kraus superoperators are sparse matrices (since they were originally diagonal) while the rotation superoperators are not (since the Wigner $d$-matrix is not sparse). The cost of matrix multiplication for an $N \times N$ matrix is $O(N^3)$ operations. For vectorized operators, this scales as $O(N^6)$ operations, which is costly for simulation of many qubits. Based on the preceding remarks, we present the following approach for numerical simulation, which is more efficient. First, observe that the superoperator $\mathcal{K}_{\text{avg}}$ is a diagonal matrix with diagonal entries $(\mathcal{K}_1 \; \mathcal{K}_2 \; \ldots \; \mathcal{K}_{(N+1)^2})$ acting on the vectorized density matrix $\bm{\rho} = \text{vec}(\rho) = (\rho_1 \; \rho_2 \; \ldots \; \rho_{(N+1)^2})^{\rm T}$, which is a column vector of size $(N+1)^2$. The action of the Kraus superoperator on the atomic state is:
\begin{equation}
(\mathcal{K}_1 \rho_1 \;\; \mathcal{K}_2 \rho_2 \;\; \ldots \;\; \mathcal{K}_{(N+1)^2} \rho_{(N+1)^2}),
\end{equation}
i.e., it is the element-wise multiplication of these two vectors. This corresponds to $O(N^2)$ operations, which is an improvement over $O(N^4)$ that would result from the normal matrix multiplication between the Kraus superoperator and the vectorized density matrix. Inspired by the previous comments, we model the action of $\mathcal{K}_{\text{avg}}$ through the following procedure:
\begin{enumerate}
    \item  Construct the diagonal entries of $\mathcal{K}_{\text{avg}}$ as a vector $(\mathcal{K}_1 \; \mathcal{K}_2 \; \ldots \; \mathcal{K}_{(N+1)^2})$.
    \item Implement its action on $\rho$ as the elementwise multiplication 
$(\mathcal{K}_1 \rho_1 \;\; \mathcal{K}_2 \rho_2 \;\; \ldots \;\; \mathcal{K}_{(N+1)^2} \rho_{(N+1)^2})$.
\end{enumerate}
This approach can be extended to model the physical Grover iteration:\rsub{
\begin{equation}
\mathcal{G}=\mathcal{R}(\phi)\mathcal{K}_{\text{avg}}(0)\mathcal{R}(-\phi)\mathcal{K}_{\text{avg}}\left(\delta \omega_m\right),
\end{equation}}
as follows:
\begin{enumerate}
    \item Implement the action of $\mathcal{K}_{\text{avg}}$ as discussed above.
  \item   To model the action of the rotation superoperator $\mathcal{R}(-\phi)$ after the action of $\mathcal{K}_{\text{avg}}$, reshape
\begin{equation}
\bm{\rho}= (\mathcal{K}_1 \rho_1 \;\; \mathcal{K}_2 \rho_2 \;\; \ldots \;\; \mathcal{K}_{(N+1)^2} \rho_{(N+1)^2})
\end{equation}
into the original density matrix format, $\rho$, then compute
\begin{equation}
R^N(-\phi) \rho R^N(\phi).
\end{equation}
which costs $O(N^3)$ operations.
\item  To compute $\mathcal{R}(\phi)\mathcal{K}_{\text{avg}}(0)$, repeat steps 1 and 2.  
\end{enumerate} 
Therefore, the total computational cost of this approach is $O(N^3)$ operations. We note that further optimizations, based on the symmetry properties of the operators and the density matrix, could be possible.

\end{widetext}

\end{document}